\documentclass[12pt]{article}
\usepackage{amsmath}
\usepackage{graphicx}
\usepackage{enumerate}
\usepackage{natbib}
\usepackage{url} 
\usepackage{amsfonts,amsmath,verbatim,amssymb,epsfig,enumitem,url,amsbsy,enumitem,color,natbib,graphicx}
\usepackage{algorithm, algorithmic}
\usepackage{bm,ulem,courier}
\usepackage{subfigure}
\usepackage{float}

\newcommand{\blind}{1}

  \renewenvironment{thebibliography}[1]{
    \begin{oldthebibliography}{#1}
      \setlength{\parskip}{0ex}
      \setlength{\itemsep}{0.ex}}
  {\end{oldthebibliography}}

\newcommand{\myvec}[1]%
{\stackrel{\raisebox{-2pt}[0pt][0pt]{\tiny$\rightharpoonup$}}{#1}}

\makeatletter
\newenvironment{breakablealgorithm}
  {
   \begin{center}
     \refstepcounter{algorithm}
     \hrule height.8pt depth0pt \kern2pt
     \renewcommand{\caption}[2][\relax]{
       {\raggedright\textbf{\ALG@name~\thealgorithm} ##2\par}%
       \ifx\relax##1\relax 
         \addcontentsline{loa}{algorithm}{\protect\numberline{\thealgorithm}##2}%
       \else 
         \addcontentsline{loa}{algorithm}{\protect\numberline{\thealgorithm}##1}%
       \fi
       \kern2pt\hrule\kern2pt
     }
  }{
     \kern2pt\hrule\relax
   \end{center}
  }
\makeatother

\newcommand{\oldchange}[1]{\textcolor{black}{#1}}

\newcommand{\change}[1]{\textcolor{black}{#1}}

\newtheorem{thm}{Theorem}
\newtheorem{cor}{Corollary}
\newtheorem{prop}{Proposition}
\newtheorem{lemma}{Lemma}
\newtheorem{assump}{Assumption}

\newtheorem{emp}{Example}
\def\argmin{\mathop{\rm argmin}}
\def\rank{\mathop{\rm rank}}

\def\myvec{\mathop{\rm vec}}
\def\Tr{\operatorname{Tr}}

\newenvironment{pf}{{\noindent\it Proof}\ }{\hfill $\square$\par}

\usepackage{xr-hyper}
\usepackage{hyperref}
\makeatletter
\newcommand*{\addFileDependency}[1]{
\typeout{(#1)}
\@addtofilelist{#1}
\IfFileExists{#1}{}{\typeout{No file #1.}}
}
\makeatother

\begin{document}

\def\spacingset#1{\renewcommand{\baselinestretch}%
{#1}\small\normalsize} \spacingset{1}


\if1\blind
{
  \title{\bf Dynamic Matrix Recovery}
  \author{
  Ziyuan Chen\\    
    School of Mathematical Sciences, Center for Statistical Science, \\
    Peking University, Beijing, China\\ 
  Ying Yang\\
    Academy of Mathematics and Systems Science, \\ 
    Chinese Academy of Sciences, Beijing, China\\
    Fang Yao
     \thanks{
    Ziyuan Chen and Ying Yang are the co-first authors, and contribute equally to this work. Fang Yao is the corresponding author, E-mail: fyaomath.pku.edu.cn.  Fang Yao's research is partially supported by the National Key R\&D Program of China (No. 2022YFA1003801, 2020YFE0204200), the National Natural Science Foundation of China (No. 12292981, 11931001), the LMAM and the Fundamental Research Funds for the Central Universities and the LMEQF.  Ying Yang's research is partially supported by China Postdoctoral Science Foundation (No. 2022TQ0360, 2022M723334), the National Natural Science Foundation of China (No. 71988101) and the Guozhi Xu Posdoctoral Research Foundation.
       }\hspace{.2cm} 
    \\
    School of Mathematical Sciences, Center for Statistical Science, \\
    Peking University, Beijing, China\\ 
  }  
\date{}
\maketitle
} \fi

\if0\blind
{
  \bigskip
  \bigskip
  \bigskip
  \begin{center}
    {\LARGE\bf Dynamic Matrix Recovery}
\end{center}
\date{}
\medskip
} \fi

\bigskip
\begin{abstract}
Matrix recovery from sparse observations is an extensively studied topic emerging in various applications, such as recommendation system and signal processing, which includes the matrix completion and compressed sensing models as special cases.
In this work, we propose a general framework for dynamic matrix recovery of low-rank matrices that evolve smoothly over time. We start from the setting that the observations are independent across time, then extend to the setting that both the design matrix and noise possess certain temporal correlation via modified concentration inequalities. By pooling neighboring observations, we obtain sharp estimation error bounds of both settings, showing the influence of the underlying smoothness, the dependence and effective samples. We propose a dynamic fast iterative shrinkage-thresholding algorithm that is computationally efficient, and characterize the interplay between algorithmic and statistical convergence. Simulated and real data examples are provided to support such findings.
\end{abstract}

\noindent%
\vfill

\newpage
\spacingset{1.9} 
\section{Introduction}
\label{intro}
  Matrix recovery from sparse observations has been extensively studied in a wide range of problems, such as signal processing \citep{li2019survey}, recommendation system \citep{koren2009matrix} and the quantum state tomography \citep{gross2010quantum}. The classical model is to link the response $Y$ with the random matrix $X\in\mathbb{R}^{m_1\times m_2}$ via
  \begin{equation}\label{eq:static model 0}
      Y = \Tr(X^\top M^0) + \xi
  \end{equation}
  to recover the low-rank matrix $M^0\in\mathbb{R}^{m_1\times m_2}$ from the independent observations $\{(X_i,Y_i)\}_{i=1}^n$, where $\Tr(\cdot)$ is the trace of a matrix, $\xi$ is a mean zero random error and $n$ denotes the sample size, which is typically much smaller than $m_1 m_2$ (i.e., $n \ll m_1 m_2$). By imposing different assumptions
on $X$, the model includes a variety of matrix recovery models, in which matrix completion and compressed sensing are two important cases.
  Most existing works assume that the target matrix $M^0$ is static \citep{klopp2011rank,koltchinskii2015optimal,fan2021shrinkage,xia_statistical_2021}. 
  However, there are various scenarios where $M^0$ is dynamic and observed at multiple time points,
  such as in time-varying user interests in recommendation systems \citep{xu2020contextual}, temporally-changed graph models \citep{qiu2017time} and dynamic quantum dot matrices \citep{csurgay2000signal}. 
  To formulate a general framework accounting the dynamic changes in numerous applications, we propose investigating the recovery of the dynamic matrix as follows,
  \begin{equation}\label{eq:dynamic model 0}
      Y_t = \Tr(X_t^\top M_t^0) + \xi_t,\ t=1,\ldots,T,
  \end{equation}
  where $t$ is the timestamp, $\{Y_t,X_t,\xi_t\}$ is the tuple of the response, the design matrix and the noise at time $t$, respectively, and $M_t^0$ is a dynamic matrix which is assumed smooth across time $t$ with a low-rank structure. 

A prevalent line of research for solving the classical static model \eqref{eq:static model 0} is to impose strict low-rank constraints based on matrix factorization. For instance, \cite{keshavan2009matrix} proposed singular value decomposition to make the recovery, \cite{zhao2015nonconvex} used QR decomposition instead, and \cite{zheng2016convergence} adopted Burer-Monteiro factorization for rectangle matrix completion. 
  Another stream is the convex approaches, thanks to \cite{candes2009exact} that first relaxed the non-convex rank restriction to nuclear norm for efficient optimization. 
  A variety of convex algorithms and theoretical results are further established \citep{recht_guaranteed_2010,candes2011tight
  ,agarwal2012noisy}. 
  Though $M_t^0$ in \eqref{eq:dynamic model 0} can be recovered by adopting the static matrix recovery at each time $t$, these {\it single-stage} estimates are less desirable, as they only use data at time $t$ to reconstruct $M_t^0$ and fail to take advantage of the neighboring observations and smoothness of the underlying matrix. 
Related work also includes tensor regression to rebuild $M_t^0$, which views the dynamic model \eqref{eq:dynamic model 0} from a static perspective by letting $N\in\mathbb{R}^{T\times m_1\times m_2}$ be a three order tensor whose $(t,i,j)$th element is the $(i,j)$th element of $M_t^0$. 
 \cite{gandy2011tensor} and \cite{liu2012tensor} 
  proposed algorithms for tensor completion and 
  \cite{zhang2020islet} considered the low-rank tensor trace regression via the importance sketching algorithm. 
We mention that their low-rank assumption of $N$ is different from model \eqref{eq:dynamic model 0} that $M_t^0$ is dynamically and smoothly evolving along time $t$. Consequently, one needs to process the whole data to recover $M_t$ in tensor regression setting, which is time- and space-consuming. 

A more efficient approach is to adopt the dynamic modeling/algorithm that avoids computing the whole data for update and has been adopted in some relevant problems. 
\cite{lois2015online} designed a memory-saving algorithm for batch robust principle component analysis; \cite{gao2018dynamic} studied dynamic robust principal component analysis and proposed a fast iterative algorithm; \cite{hu2021dynamic} proposed a unified framework to directly estimate dynamic principal subspace in high dimension. The dynamic algorithm for matrix recovery has been explored in \cite{xu2016dynamic}, which assumed that $\xi_t$ are independently and identically distributed (i.i.d.) zero-mean Gaussian error and imposed a parametric assumption on structure of the underlying matrix, $M_t^0=M_{t-1}^0+\varepsilon_t$, where $\varepsilon_t$ are i.i.d. zero-mean Gaussian noises.
\oldchange{Another line of work, namely the dynamic tensors, also finds extensive applications in various domains, and several works have been noted in this direction. For instance, \cite{zhang2021dynamic} proposed a time-varying coefficient model for temporal tensor factorization, employing a polynomial spline approximation, and \cite{zhou2013tensor} introduced a family of rank-R generalized linear tensor regression models. Most of the current research on dynamic tensors relies on tensor decomposition techniques and assumes that only certain decomposition factors or coefficients are changing \citep{zhang2021dynamic,bi2021tensors,koren2009collaborative,wang2016recommending}.}
  
\oldchange{Different from the existing work, we consider the general scenario  where the dynamics are not limited to decomposition components or a parametric structure {\it priori}.}
There exist two major challenges. First, one needs to maintain the low-rank structure in the whole procedure of dynamic algorithm. Second, the combination of neighboring information often inflates the estimation error when the observations are correlated along time.
To resolve these, we utilize local smoothing which has the advantages that the manifold structure of low-dimensional matrices can be approximately retained and temporal dependence has less influence when $T$ increases. 
Based on these considerations, we propose a dynamic matrix recovery algorithm without imposing structural assumption besides smoothness and allowing the measurement error $\xi_t$ and/or design matrix $X_t$ to be possibly dependent across time.
 The idea is to pool together the observations that lie in the local window to estimate $M_t^0$. Noting that $M_t^0$ is low-rank, we set the target function composed of a locally weighted loss function and the nuclear norm penalty of $M_t^0$ and propose a Dynamic Fast Iterative Shrinkage-Thresholding Algorithm (DFISTA) for optimization. 
 
  The main contributions of this work are summarized as follows.
  First, we establish a general framework for dynamic matrix recovery with theoretical guarantees. We allow the underlying matrix varying smoothly under the low-rank constraint and allow the observations to be dependent across time, which has more flexibility and applicability. While existing works only study the estimation error bounds for specific problems such as matrix completion and compressed sensing \citep{koltchinskii_nuclear-norm_2011,negahban2011estimation,xu2016dynamic,fan2021shrinkage}, we derive the error bounds under this general framework, which can be readily adopted to such matrix recovery problems.
  Second, the proposed method attains a faster convergence rate than classical static methods in the following sense. Take the matrix completion for instance. The ratio of error bounds of the proposed estimator compared to the classical one (i.e., single-stage estimates at each $t$) is $O_p\big(\{m\log m/n_t\}^{-1/10}T^{-2/5}\big)$, where $n_t$ is the number of observations at time $t$ and $m=m_1\vee m_2$ ("$\vee$" indicates the greater of two quantities). When the observations are sparse in the sense that $n_t\ll m\log m$, the ratio becomes $o_p\big(T^{-2/5}\big)$. 
  Specifically, the dynamic method uses only $n'=\big(m\log m/n\big)^{-1/4}n/T=o(n/T)$ observations to attain the same convergence rate as the static method using $n$ observations.
  Last but not least, a computationally efficient algorithm is devised for dynamic matrix recovery. This algorithm iterates only a subset of the data to update the estimate at each step. By using the estimate of the last step as the initial of the current optimization procedure, the algorithm converges faster than the single-stage recovery. Specifically, the ratio of computational complexities of our algorithm compared to the single-stage recovery is $O(T^{-3/5}\vee n^{-1/2})$.

  The rest of the paper is organized as follows. In Section \ref{method}, we present the proposed method and algorithm for dynamic matrix recovery, and establish theoretical guarantees in Section \ref{theo} for both independent and dependent observations with algorithmic analysis. 
  Finally, we conduct simulation and real data experiments in Section \ref{sec:simu} and \ref{sec:data} to show the favorable performance of the proposed dynamic method/algorithm compared with other existing methods. 
  Additional theoretical results and technical proofs are deferred to an online Supplementary Material due for space economy.



\section{Methodology and Algorithm}
\label{method}
We consider the general framework in the form of dynamic trace regression whose coefficient matrix is changing smoothly along time. Specifically, at time $t=1,\ldots,T$, we observe $(Y_{ti},X_{ti})$ following model (\ref{eq:dynamic model 0}), 
  \begin{equation*}
  Y_{t i} = \Tr(X_{t i}^\top M_t^0) +\xi_{t i},\quad i=1,\ldots,n_t,
  \end{equation*}
  where $M_t^0$ is the time-varying coefficient matrix, $X_{ti}$ and $\xi_{ti}$ are design matrices and zero-mean noises, respectively, $\{X_{ti},\xi_{ti}\}_i$ (given $t$) are mutually independent across subject and $\{X_{ti},\xi_{ti}\}_t$ (given $i$) may be correlated across time.
 Parallel to the static model \eqref{eq:static model 0}, the dynamic trace regression includes several problems such as dynamic matrix completion and dynamic compressed sensing that can be applied to dynamic recommendation systems and video signal processing as illustrated in numerical experiments of Section \ref{sec:data}.

  \begin{emp}\label{emp:mc} (Dynamic matrix completion). For each $t$, suppose that the design matrices $X_{ti}$ are i.i.d. uniformly distributed on the set
  \begin{equation} \label{eq:dmc}
  \mathcal{E}=\{e_j(m_1)e_k^\top(m_2):1\le j\le m_1, 1\le k\le m_2\},
  \end{equation}
  where $e_j(m)$ are the canonical basis vectors in $\mathbb{R}^m$ and then $\mathcal{E}$ forms an orthonormal basis in the space $\mathbb{R}^{m_1\times m_2}$. 
  Then estimating $M_t^0$ is equivalent to the problem of matrix completion under uniform sampling at random (USR). 
  \end{emp} 

  \begin{emp}\label{emp:cs} (Dynamic compressed sensing).  For each $t$, suppose that the design matrices $X_{ti}$ are i.i.d. replicates of a random matrix $X_t$ such that $\langle M,X_t\rangle$ is a sub-gaussian random variable for any $M\in\mathbb{R}^{m_1\times m_2}$. 
  In this work, we focus on the concrete example that each design matrix $X_{ti}$ is a random matrix whose elements are independent mean-zero sub-gaussian random variables with variance $\sigma_{X_t}^2$. Moreover, if the design $X_t$ is homogeneous across $t$, we denote the variance as $\sigma_{X}^2$. 
  \end{emp} 

  We first briefly review the classical static method in the context of dynamic modeling, i.e., adopting the static method at each time $t$, which is referred to as the single-stage estimates in the sequel. 
  Define the matrix inner product by $\langle M,X\rangle = \Tr(M^\top X)$. 
  \cite{koltchinskii_nuclear-norm_2011} proposed to incorporate the knowledge of the distribution of $X_{ti}$ and estimate $M_t^0$ by
  \begin{equation}\label{eq_classical_estimator}
  \widehat{M}_t^\lambda=\argmin_{M\in\mathbb{M}}\left[\frac{1}{n_t}\sum_{i=1}^{n_t}\mathbb{E}(\langle M,X_{ti} \rangle)^2-\left\langle\frac{2}{n_t}\sum_{i=1}^{n_t} Y_{ti}X_{ti},M\right\rangle+\lambda\|M\|_1\right],
  \end{equation}
  where $\mathbb{M}\subset\mathbb{R}^{m_1\times m_2}$ is a convex set of matrices, $\lambda>0$ is a regularization parameter and $\|M\|_1$ is the nuclear norm of $M$. 
  They proved that in the setting of matrix completion in Example \ref{emp:mc}, 
  if $\lambda$ is appropriately set, with probability at least $1-3/(m_1+m_2)$,
  \begin{equation*}
  (m_1m_2)^{-1/2}\|\widehat{M}_t^\lambda-M_t^0\|_2\le C\left\{\frac{\log(m_1+m_2)\max(m_1,m_2)r_t}{n_t}\right\}^{1/2}, 
  \end{equation*}
  where $r_t=\rank(M_t^0)$ and $\|M\|_2$ is the Frobenius norm of matrix $M$.
  Though they have proved that the rate is optimal up to logarithmic factors comparing to the minimax lower bounds in classical trace regression,
 this approach fails to incorporate temporal information and hence is not efficient in multi-stage case.

  A direct modification is to utilize the {\it two-step} procedure that first obtains the single-stage estimates $\widehat{M}^\lambda_t$ and then applies local smoothing to $\{\widehat{M}_t^\lambda\}_t$ to achieve the final estimate as follows,
   \begin{equation}
   \label{eq:twostep}
      \widehat{M}_{t}^{\text{t-s}} = \sum_{j=1}^T \omega_h(j-t)\widehat{M}_j^\lambda,
   \end{equation}
 where the local smoothing weights 
  $\omega_h(j-t)=K_{Th}\left(j-t\right)/\sum_{k=1}^T K_{Th}\left(k-t\right)$
  with the kernel function $K(x)$ and bandwidth $h$. 
  The two-step estimation approach computes a weighted average of low-rank matrices that have been estimated at distinct time points. However, this method encounters a significant challenge when dealing with sparse observations. \change{In such scenarios, the estimated matrix for each time point could potentially disrupt the true low-rank structure, displaying significant deviation from the actual value. Unfortunately, this disparity cannot be corrected through smoothing techniques.}
  
   \change{Motivated by the observation that the manifold structure of low-rank matrices can be approximately preserved locally  without matrix factorization, we design the dynamic target function with nuclear norm penalty, i.e.,}
  \begin{equation}
  \label{eq_local_estimator}
  \widetilde{M}_t^{\lambda}=\argmin_{M\in\mathbb{M}}\sum_{j=1}^T\omega_h(j-t)\left[\frac{1}{n_j}\sum_{i=1}^{n_j}\mathbb{E}(\left\langle M,X_{ji}\right\rangle)^2 - 2\left\langle\frac{1}{n_j}\sum_{i=1}^{n_j}Y_{ji}X_{ji}, M \right\rangle\right] +\lambda \|M\|_1,
  \end{equation}
  \change{which combines the observations across time and employs local smoothing for incorporating the temporal information.}
 It is worth noting that in many applications including our Examples \ref{emp:mc} and \ref{emp:cs},
the distribution of $X_t$ is known, and yet this information has not been exploited in empirical risk. Thus we may utilize the distribution knowledge by substituting the empirical form with true values $\mathbb{E}\langle M, X_{ji}\rangle$.
The temporal smoothness of $M_t^0$ guarantees that information in adjacent time points can be utilized to reduce the estimation error.
  We mention that the classical estimate $\widehat{M}_t^\lambda$ in \eqref{eq_classical_estimator} can be regarded as a special case of the proposed local estimate $\widetilde{M}_t^\lambda$ in \eqref{eq_local_estimator} as 
  the bandwidth $h\rightarrow 0$. 
   By choosing $h$ appropriately, our estimate pools more observations to reduce the estimation variance while controlling the bias caused by the varying of $M_t^0$, as discussed in  detail in Section \ref{theo}.

  For implementation, we adopt the Fast Iterative Shrinkage-Thresholding Algorithm (FISTA) which is an accelerated proximal method proposed by \cite{beck2009fast}. It improves traditional proximal methods and terminates in $\mathcal{O}(\varepsilon^{-1/2})$ iterations with a $\varepsilon$-optimal solution. 
  To handle with the trace norm penalty optimization problem \eqref{eq_local_estimator}, we use the matrix form FISTA established by \cite{ji2009accelerated} and \cite{toh2010accelerated}. 
  Define the objective function in \eqref{eq_local_estimator} as 
  $F_t(M) = f_t(M)+g(M)$, where 
  \begin{align*}
    f_t(M) = \sum_{j=1}^T \omega_h(j-t) \left[\frac{1}{n_j}\sum_{i=1}^{n_j}\mathbb{E}(\left\langle M,X_{ji}\right\rangle)^2 - 2\left\langle\frac{1}{n_j}\sum_{i=1}^{n_j}Y_{ji}X_{ji}, M \right\rangle\right].
  \end{align*}
  We mention that the gradient function of $f_t$ is equipped with Lipschitz constant $L_f$ presented in Section \ref{algorithm}, and that $g(M) = \lambda \|M\|_1$ is a continuous convex function. 
  Then the matrix form FISTA can be applied to solve the dynamic trace regression problem $$\widetilde{M}^\lambda_t = \arg\min_{M\in \mathbb{M}} f_t(M)+g(M).$$ 
  
 Note that $\widetilde{M}^\lambda_t$ need to be calculated for a sequence $t=1,2,\cdots,T$. The algorithm is accelerated at time $t$ by utilizing the estimate from the timestamp $(t-1)$. Specifically, suppose that at time $t$ we apply the matrix form FISTA for $k_{t}$ iteration times to numerically approximate the estimate $\widetilde{M}_{t}^\lambda = \arg\min_{M\in\mathbb{M}} f_{t}(M)+g(M)$ denoted by $M_{t}^{(k_{t})}$. Then input $M_t^{(k_t)}$ as the initial matrix in the procedure of estimating $\widetilde{M}_{t+1}^\lambda = \arg\min_{M\in\mathbb{M}} f_{t+1}(M)+g(M)$ at time $(t+1)$. Such initialization method accelerates the algorithm, as demonstrated in Section \ref{algorithm}. We summarize the implementations of the proposed DFISTA in Algorithm \ref{algo}. \oldchange{In the algorithm, ``svd'' means calculating the single value decomposition components $U,D,V$ for a matrix $G \in \mathbb{R}^{m_1\times m_2}$ 
 and ``$(\cdot)_{+}$'' means the threshold operator such as $
     (D - \gamma)_{+} = \operatorname{diag}((\lambda_1-\gamma)\vee 0,\cdots,(\lambda_{m_1}-\gamma)\vee 0)$ where $D = \operatorname{diag}(\lambda_1,\cdots,\lambda_{m_1})$. The learning rate $s_k$ is updated according to \cite{nesterov1983method}.}

  \begin{breakablealgorithm}\label{algo}
  \renewcommand{\algorithmicrequire}{\textbf{Input:}}
  \renewcommand{\algorithmicensure}{\textbf{Output:}}
  \caption{Dynamic Low-Rank Matrix Recovery}\label{alg1}
  \begin{algorithmic}[1]
  \REQUIRE $\sum_{t=1}^{T} n_t$ matrices $X_{t i}\in \mathbb{R}^{m_1\times m_2}$ and $\sum_{t=1}^{T} n_t$ scalars $Y_{t i}$, $t=1,2,\dots,T$, $i=1,2,\cdots,n_j$, max iteration steps $K$ and torrent {\tt tor}.
  \ENSURE $T$ matrices $M_t\in\mathbb{R}^{m_1\times m_2},\ t=1,2,\dots,T$.
  \STATE Choose a initial matrix $M_0 \in \mathbb{R}^{m_1\times m_2}$.
  \WHILE {$t\le T$}
  \STATE  $M_t^{(-1)} \leftarrow M_{t-1},\ M_t^{(0)} \leftarrow M_{t-1}$;
  \STATE  $s_{0}\leftarrow 1,\ s_{-1}\leftarrow 1$;
  \FOR{ $k=0,1,\ldots,K$}
  \STATE $N_t^{(k)} \leftarrow M_t^{(k)} + \frac{s_{k-1}-1}{s_{k}}\left(M_t^{(k)}-M_t^{(k-1)}\right)$;
  \STATE $ G_t^{(k)} \leftarrow N_t^{(k)} - L_f^{-1}\sum_{j=(t-Th/2)\vee 0}^{(t+Th/2)\wedge T} \frac{w_{h}(j-t)}{n_j} \sum_{i=1}^{n_j}X_{ji}\left(\langle X_{ji},N_t^{(k)}\rangle -Y_j\right)$;
  \STATE $U,D,V \leftarrow$ {\tt svd}($G_t^{(k)}$);
  \STATE $M_t^{(k+1)} \leftarrow U(D-2\lambda/L_f)_{+}V^T$;
  \STATE $s_{k+1} \leftarrow \frac{1+\sqrt{1+4s_{k}^2}}{2}$;
  \IF{$|F_t(M_t^{(k+1)})-F_t(M_t^{(k)})|\le {\tt tor}$} 
  \STATE {$\operatorname{break}$}
  \ENDIF
  \ENDFOR
  \STATE $M_t = M_t^{(k+1)}$;
  \ENDWHILE
  \RETURN $M_t,t=1,2,\cdots,T$ 
  \end{algorithmic}  
  \end{breakablealgorithm}
  
\change{Unlike most factorization based algorithms that are challenging to analyze for their convergence properties, we are able to analyze the error bounds of the algorithm's results with respect to the true value. This analysis bridges the gap between the statistical estimator and the estimator obtained from the algorithm, as discussed in detail in Section \ref{algorithm}.}
\section{Theoretical Guarantees}
\label{theo}
  In this section, we study both the statistical properties of the estimation error $\|\widetilde{M}_t^\lambda - M_t^0\|_2$ and the algorithmic convergence of the proposed method. We first present the error bound of the estimate $\widetilde{M}_t^\lambda$ for general dynamic trace regressions. Then we derive the explicit expressions of the upper bound of $\|\widetilde{M}_t^\lambda -M_t^0\|_2$ when $\{X_t\}_t$ and $\{\xi_t\}_t$ are independent across $t$ and when $\{X_t\}_t$ and $\{\xi_t\}_t$ are $\phi$-mixing processes, respectively. Noting that $X_t$ and/or $\xi_t$ may be  unbounded, the assumptions for classical concentration inequalities are violated \citep{merlevede2009bernstein, merlevede2011bernstein,hang2017bernstein}. We thus modify the Talagrand's inequalities and adopt  truncation technique appropriately to derive the convergence rate  for the dependent setting.  In both scenarios, we shall explain how the number of time points $T$, the smoothness of $M_t^0$ and the dependence of $X_t$ and $\xi_t$ across time influence the estimation error explicitly.



  In the sequel, we denote $\|M\|_0$, $\|M\|_1$, $\|M\|_2$, $\|M\|_{\infty}$ and $\langle M,N\rangle$ as the rank, nuclear norm, Frobenius norm, maximum singular value of $M$ and the trace of $M^\top N$, respectively.

 \subsection{General error bounds}




  Let $\myvec(X)\in \mathbb{R}^{m_1m_2}$ be the vectorization of $X \in \mathbb{R}^{m_1\times m_2}$ and define the second moment matrix of $X_{t}$ by $\Sigma_{t}\in \mathbb{S}^{m_1m_2\times m_1m_2}$ such that
  \begin{equation*}
  \Sigma_{t} = \frac{1}{n_t}\sum_{i=1}^{n_t}\mathbb{E} [\myvec({X}_{ti})\myvec({X}_{ti})^\top].
  \end{equation*}
  It is necesary to guarantee the identifiability of matrix recovery that the positive semi-definite matrix $\Sigma_{t}$ is invertible.
  To simplify, we consider the homogeneous scenario such that at each time point $t$, the design $\{X_t\}_t$ have the same second moment matrix as stated below.
  \begin{assump}
  \label{assump_Sigma}
  There exists a positive definite matrix $\Sigma \in \mathbb{S}_{+}^{m_1m_2\times m_1m_2}$ such that $\Sigma_j=\Sigma,\ j=1,2,\dots,T$ with smallest eigenvalue $\mu>0$ and bounded condition number $\kappa_{\Sigma}$.
  \end{assump}

  We remark that in the heterogeneous situation, the convergence rate can be derived similarly, which is at the same order as the homogeneous setting with a more tedious presentation. 
  An important indication of Assumption \ref{assump_Sigma} is that, for all $M\in \mathbb{R}^{m_1\times m_2}$ and $j=1,2,\cdots,T$, 
  \begin{equation*}
  \frac{1}{n_j}\sum_{i=1}^{n_j} \mathbb{E}(\langle M,X_{j i}\rangle )^2 = \myvec({M})^\top \Sigma \myvec({M}) \ge \mu\|M\|_2^2,
  \end{equation*}
  which is a standard assumption in matrix recovery.
The interpretation is that the smallest eigenvalue $\mu>0$ determines the information contained in the design matrix $X_t$. Intuitively, the larger $\mu$ is, the more information is revealed by data, thus the more precise the estimator $\widetilde{M}_t^\lambda$ is.
  In most cases, $\mu$ can be calculated directly. For instances,
  in matrix completion of Example \ref{emp:mc}, the smallest eigenvalue of $\Sigma$ is $\mu = 1/m_1m_2$, and in compressed sensing of Example \ref{emp:cs}, $\mu$ is the smallest variance of elements in $X_{j i}$, which is equal to $\sigma_X^2$ under Assumption \ref{assump_Sigma}. 

  For brevity, we further denote $\Delta_j =1/n_j \sum_{i=1}^{n_j}\left(Y_{j i}X_{j i} - \mathbb{E}(Y_{j i}X_{j i})\right)$ and 
  \begin{equation*}
    \delta_h M_{t} = \left\|M_t^0-\sum_{j=1}^T \omega_h(j-t) M_j^0\right\|_2,\quad \mathcal{W}_h\Delta_{t} =\left\| \sum_{j=1}^T \omega_h(j-t)\Delta_j\right\|_{\infty}.
  \end{equation*}
  The following theorem gives the general upper bound of estimation error $\|\widetilde{M}_t^\lambda - M^0_t\|_2$. 

  \begin{thm}
  \label{thm_error_bound}
  Under Assumption \ref{assump_Sigma}, if $\lambda \ge 2\mathcal{W}_h\Delta_{t}$, then
  \begin{equation}
   \label{eq_thm_error_bound}
  \left\|\widetilde{M}_t^\lambda - M_t^0\right\|_2 \le\delta_h M_t +\min\left\{ \left((\delta_h M_t)^2 + \frac{2\lambda}{\mu}\left\|M_t^0\right\|_1\right)^{1/2}, \frac{1+\sqrt{2}}{2} \frac{\lambda}{\mu}\sqrt{r_t}\right\}.
  \end{equation}
  When selecting $\lambda=2\mathcal{W}_h\Delta_{t}$, we have
  \begin{equation}
  \label{proposed_bound}
  \left\|\widetilde{M}_t^\lambda - M_t^0\right\|_2 \le\delta_h M_{t} + \min\left\{\left((\delta_h M_{t})^2 + 2\mu^{-1}\mathcal{W}_h\Delta_{t}\left\|M_t^0\right\|_1\right)^{1/2},{(1+\sqrt{2})}{\mu^{-1}}{\mathcal{W}_h\Delta_{t}}\sqrt{r_t}\right\}.
  \end{equation}
  \end{thm}

  The term $\delta_h M_t$ corresponds to the bias caused by kernel smoothing which is affected by kernel $K(x)$ and the bandwidth $h$, and $\mathcal{W}_h\Delta_{t}$ corresponds to the variance term caused by the effective samples in the local window and the measurement error $\xi$. 
  When $h=0$, the result in Theorem \ref{thm_error_bound} coincides with the classical result of Corollary 1 in \cite{koltchinskii_nuclear-norm_2011} such that when $\lambda = 2\|\Delta_t\|_{\infty}$,
  \begin{equation}\label{baseline_bound}
  \left\|\widehat{M}_t^\lambda - M_t^0 \right\|_2 \le \min\left\{\left(2\mu^{-1}\left\|\Delta_t\right\|_{\infty}\left\|M_t^0\right\|_1\right)^{1/2},{(1+\sqrt{2})}{\mu^{-1}}{\|\Delta_t\|_{\infty}}\sqrt{r_t}\right\}.
  \end{equation} 
  Compared to \eqref{baseline_bound}, the  bound \eqref{proposed_bound} replaces the error term $\Delta_t$ by the smoothed version $\mathcal{W}_h\Delta_t$ at the cost of the bias $\delta_hM_t$. 

  To obtain an explicit expression of the error bound that may be used for bandwidth selection, we further investigate the properties of $\delta_hM_t$ and $\mathcal{W}_h\Delta_t$. 
  Assumption \ref{assump_M} and \ref{assum_kernel} are standard conditions for temporal smoothness and the kernel function.

  \begin{assump}
  \label{assump_M}
  There exists a matrix function $M(t)$ supported on $[0,1]$, satisfying that
  \begin{itemize}
  \item[1.] for any $j=1,2,\dots,T$, $M_j^0 = M(j/T)$ and $\|M_j^0\|_1\le C_*,\ \max_{j,s,l}|[M_j^0]_{s,l}|\le C_M$, for some constants $C_*$ and $C_M$;
  \item[2.] the first and second derivatives $\nabla M(t),\nabla^2 M(t)$ are continuous and bounded by some constants $D_1$ and $D_2$, respectively.
  \end{itemize}
  \end{assump}

  \begin{assump}
  \label{assum_kernel} 
  The kernel function $K(x)$ is a symmetric probability density function on [-1, 1] and satisfies that
  \begin{eqnarray}\label{eq:kernel quantity}
  \alpha(K)=\int x^2K(x) \ dx<\infty,\quad R(K)=\int K(x)^2 \ dx<\infty.
  \end{eqnarray}
  \end{assump}

  Define the $\psi-$norm $\|Z\|_{\psi(\eta)}$ for random variable $Z$ with $0< \eta < \infty$ as 
  \begin{equation*}
  \|Z\|_{\psi(\eta)} =\inf\left\{t>0,\mathbb{E}\left[\exp\left(\frac{\|Z\|_\infty^\eta}{t^\eta}\right)\right]\le 2
  \right\}
  \end{equation*}
  and 
$
    \|Z\|_{\psi(\infty)}=\inf\left\{t > 0,\mathbb{E} \mathbf{1}_{\left\{\|Z\|_{\infty} > t\right\}}=0\right\}.
 $
  Then we impose the following assumption on the distributions of $\xi_{j i}$ and $X_{j i}$.

  \begin{assump} 
  \label{assump_distri}
  The designs and noises at the same time point $\{X_{ji}, \xi_{ji}\},\ i=1,2,\dots,n_j$ are mutually independent for each fixed $j$ and there exists constants $K_1,K_2>0,\ \alpha\ge 1,\ \beta\ge 2$ that for any $j=1,2\dots,T,\ i=1,2,\dots,n_j$
  \begin{itemize}
  \item[1.] $\xi_{ji}$ is sub-exponential distribution with $\|\xi_{ji}\|_{\psi(\alpha)} \le K_1$;
  \item[2.] $X_{ji}$ is sub-gaussian distribution with $\|X_{ji}\|_{\psi(\beta)} \le K_2$,
  \item[3.] There exists $\gamma \ge 1$ such that $1/\alpha+ 1/\beta= 1/\gamma$.
  \end{itemize}
  \end{assump}

  The tail distributions of $\xi_{j i}X_{j i}$ and $\langle M, X_{j i}\rangle X_{j i}$ are both sub-exponential when $X_{ji},\xi_{ji}$ satisfy Assumption \ref{assump_distri}. Denote $\mathcal{D}_{\alpha,\beta,K_1,K_2}$ as the distribution set of $\{X_{ji},\xi_{ji}\}_{ji}$ satisfying Assumption \ref{assump_distri}. We remark that $\mathcal{D}_{\alpha,\beta,K_1,K_2}$ is indeed general enough for applications. In the matrix completion problem of Example \ref{emp:mc}, $X_{ji}$ are bounded with $\|X_{ji}\|_{\psi(\infty)} = 1$. Then the distributions of $X_{ji},\xi_{ji}$ lie in $ \mathcal{D}_{1,\infty,K_1,1}$ as long as the noises $\xi_{ji}$ follow sub-exponential distributions with $\alpha\ge 1$. In the compressed sensing problem of Example \ref{emp:cs}, from Theorem 4.4.5 in \cite{vershynin2018high}, we can set $X_{ji}$ to be sub-guassian distributed with $\beta=2$ and $K_2 \asymp (m_1\vee m_2)^{1/2}\sigma_X$. Then the distributions of $X_{ji},\xi_{ji}$ lie in $\mathcal{D}_{2,2,K_1,\sigma_X}$ if noises $\xi_{ji}$ follow a sub-guassian distributions with $\alpha=2$.
  
Next we consider the temporal dependence between $\{X_{ji},\xi_{ji}\},i=1,2,\dots,n_j$ and $\{X_{si},\xi_{si}\},i=1,2,\dots,n_s$ that decays when the distance between time points $j$ and $s$ increases. 
  To characterize the structure of such dependence, we impose the strongly dependent assumption that $X_{j i}$ and $\xi_{j i}$ are $\phi$-mixing processes across $j$.
  Our definition of $\phi$-mixing is the same as in \cite{doukhan2012mixing}, and the detail definition is given in Supplementary Material S.4. In the following subsections, we first derive the estimation error bound in the ideal case that $\xi_{ji}$ and $X_{ji}$ are independent across time $j$, and then extend to the dependent case. These two settings are described in the following assumptions, respectively.

  \begin{assump}\label{assump_indpdt}
  (Independent case) The designs and noises $\{X_{ji},\xi_{ji}\}$ are mutually independent across $j$ for each $i$, $ j=1,2,\dots,T,\ i=1,2,\dots,N_j$.
  \end{assump}

  \begin{assump}
  \label{assump_phi_mixing}
  ($\phi$-mixing processes)
  There exist two independent sequences of $\sigma$-fields $\mathcal{X}_j$ and $\mathcal{Y}_j,j\in \mathbb{N}^*$ such that $$\sigma((X_{j1},X_{j2},\dots,X_{jn_j})^\top) \subseteq \mathcal{X}_j,\quad  \sigma((\xi_{j1},\xi_{j2},\dots,\xi_{jn_j})^\top)\subseteq\mathcal{Y}_j.$$
  The $\sigma$-fields  $\mathcal{X}_j$ and $\mathcal{Y}_j$, $j\in \mathbb{N}^*$ are both $\phi$-mixing with coefficients $\phi_{\mathcal{X}}(k)$ and $\phi_{\mathcal{Y}}(k)$, $k\in \mathbb{N}$ satisfying that  $$\Phi_\mathcal{X}  \overset{\triangle}{=} \sum_{k=0}^\infty \sqrt{\phi_\mathcal{X}(k)}< \infty,\quad \Phi_\mathcal{Y}\overset{\triangle}{=} \sum_{k=0}^\infty \sqrt{\phi_\mathcal{Y}(k)}< \infty.$$
  \end{assump}

\subsection{Independent case}
\label{sec:3.2}
  The theorem below states the upper bound of $\|\widetilde{M}_t^\lambda - M_t^0\|_2$ under the independent case when the distributions of $X_{ji},\xi_{ji}$ belong to $\mathcal{D}_{\alpha,\beta,K_1,K_2}$. For conciseness, we assume that the numbers of observations $n_j,j=1,2,\cdots,T$ have the same order denoted by $n$, i.e., $n_j\asymp n$ for $j=1,2,\cdots,T$. The conclusions for $n_j$ of different orders can be derived similarly. In the following, the notations $a_n\ll b_n$, $a_n\gg b_n$ and $a_n\asymp b_n$ mean that the ratio $a_n/b_n$ approaches to 0, infinity and  a constant, respectively,  and ``$\lceil \cdot \rceil$'' denotes the ceiling function.

  \begin{thm}\label{thm_total}
  Under Assumption \ref{assump_Sigma}--\ref{assump_indpdt}, let $n \lceil T h\rceil  \gg(K_*/\sigma_*)^2 \log(m_1+m_2)$, $h\rightarrow0$ and \\$(m_1m_2)^{1/2}\lceil T^2h^2\rceil T^{-1} \rightarrow\infty$ as $n,m_1,m_2,T\rightarrow\infty$, when
  $$\lambda = 2C_1\sigma_*\sqrt{\frac{\log(m_1+m_2)}{n \lceil Th\rceil}}, $$
  where $C_1$ and $\sigma_*$ are defined in Lemma S.4 of Supplementary Material, then 
  with probability at least $1-3/(m_1+m_2)$,
  \begin{equation} \label{err_bound_indpdt}   
  \begin{aligned}
  (m_1m_2)^{-1/2}\left\|\widetilde{M}_t^\lambda - M_t^0\right\|_2 \le \frac{1}{2}\alpha(K)D_2h^2 +\left(1+\sqrt{2}\right)C_1\sigma_*\left(\frac{r_t\log(m_1+m_2)}{\mu^2m_1m_2 n \lceil Th \rceil}\right)^{1/2}+o(h^2),
  \end{aligned}
  \end{equation}
  where $\alpha(K)$ is defined in \eqref{eq:kernel quantity} and $D_2$ in Assumption \ref{assump_M}.
  \end{thm}

  We remark that $n \lceil T h \rceil$ is the number of samples used in a local window and the condition $n \lceil T h\rceil  \gg (K_*/\sigma_*)^2 \log(m_1+m_2)$ is the requirement for the effective sample size to guarantee the upper bound \eqref{err_bound_indpdt} tending to zero.  
  The error bound \eqref{err_bound_indpdt} is a trade-off between the bias and variance which is controlled by the bandwidth $h$.
  \oldchange{When $n\mu^2 m_1m_2/(T^4\sigma_{*}^2r_t\log(m_1+m_2)) = o(1)$, we can choose the optimal bandwidth
  \begin{equation}
  \label{opt_h}
  h=C_h\left(\frac{\sigma_*^2r_t\log(m_1+m_2)}{\mu^2m_1m_2 n T}\right)^{1/5},
  \end{equation}
  where $C_h=\left[(2+2\sqrt{2})C_1/(\alpha(K)D_2)\right]^{2/5}$ is a constant. When $n\mu^2 m_1m_2/(T^4\sigma_{*}^2r_t\log(m_1+m_2)) = \mathcal{O}(1)$, we choose $h = 0$ which is degenerate to the classical situation with the convergence rate $n^{-1/2}$.}

  \oldchange{We focus on a large enough value for $T$ in the subsequent discussion, which allows us to omit the ceiling operator $\lceil\cdot \rceil$. For situations where $T$ is small, we can simply choose $h=0$, which yields the same results as static methods.}
  With an appropriately selected $h$, we have the following corollary.
  \begin{cor}
  \label{cor_general_result}
  Under assumptions of Theorem \ref{thm_total}, when 
  $$ nT\gg\max\{K_*^{5/2}\sigma_*^{-3}(\mu^{2}m_1m_2)^{1/4}\log(m_1+m_2),\ \sigma_*^2(\mu^{2}m_1m_2)^{-1}\log(m_1+m_2)\},$$
  with probability at least $1-3/(m_1+m_2)$,
  \begin{equation}\label{eq_general_bound}
  (m_1m_2)^{-1/2}\left\|\widetilde{M}_t^\lambda - M_t^0\right\|_2 \le C_2\left(\frac{\sigma_*^2r_t\log(m_1+m_2)}{\mu^2m_1m_2 n T}\right)^{2/5},
  \end{equation}
  where $C_2 =1/2\left[(2+2\sqrt{2})C_1\right]^{4/5}\left[\alpha(K)D_2\right]^{1/5}$.
  \end{cor}

  The bound requirement for $nT$ is composed of two terms: the first is to make $n T h \gg (K_*/\sigma_*)^2 $ $\log(m_1+m_2)$ hold when $h$ takes the form \eqref{opt_h}, and the second is to guarantee the optimal bandwidth \eqref{opt_h} tends to 0 as $n,T\rightarrow\infty$.
  In \eqref{eq_general_bound}, the upper bound is proportional to $(nT)^{-2/5}$ where $n T$ is the total sample size. This indicates that our dynamic method makes efficient use of information from adjacent time points to improve the estimation quality. 
  The number of observations in a single time point $n$ and the number of time points $T$ complement each other to reduce the upper bound \eqref{eq_general_bound}. 
  When the samples are extremely sparse at each time point, it is infleasible for the classical static method to reconstruct the eigenvalues and eigenvectors of $M_t^0$. However, the proposed method can still control the estimation error $\|\widetilde{M}^\lambda_t-M^0_t\|_2$ to a desirable level, as long as there are sufficient time points. We also mention that the two-step smoothing method performs poorly, because the first-step estimate $\widehat{M}_t^\lambda$ and the reconstruction of eigen-subspace at each time point are far from satisfaction, which cannot be fixed by the second smoothing step regardless how dense the time points are. We emphasize that Theorem \ref{thm_total} and Corollary \ref{cor_general_result} can be applied directly to any dynamic trace regression problems as long as $X_{ji}$ and $\xi_{ji}$ lie in $\mathcal{D}_{\alpha,\beta,K_1,K_2}$. Taking dynamic matrix completion for example, we have the following corollary. The corresponding result for dynamic compressed sensing can be derived similarly, which is presented in Corollary S.3 of the Supplement.

  \begin{cor}
  \label{cor_matrix_completion}
  Under Assumption \ref{assump_Sigma}--\ref{assump_indpdt}, when $X_{ji}$ are i.i.d. uniformly distributed on $\mathcal{E}$, $\xi_{ji}$ are independently follow sub-exponential mean-zero distributions, $n T h \gg (m_1\wedge m_2)\log^{1+2/\alpha} (m_1+m_2)$, and $(m_1m_2)^{1/2}Th^2\rightarrow\infty$ as $h\rightarrow 0,n,m_1,m_2,T\rightarrow\infty$,
  then with probability at least $1-3/(m_1+m_2)$,
  \begin{align*}
  &(m_1m_2)^{-1/2}\left\|\widetilde{M}_t^\lambda - M_t^0\right\|_2 \\&\le
  \frac{1}{2}\alpha(K)D_2h^2+
  (1+\sqrt{2})C_1 (C_M\vee \sigma_\xi)\left(\frac{r_t(m_1\vee m_2)\log (m_1+m_2)}{n Th}\right)^{1/2}+o(h^2).
  \end{align*}
  When $nT\gg(m_1\vee m_2)\log^{1+5/(2\alpha)}(m_1+m_2)$, let
  \begin{equation}
  \label{eq_select_h}
  h=C_{h}\left(\frac{(C_M\vee \sigma_\xi)^2r_t(m_1\vee m_2)\log(m_1+m_2)}{n T}\right)^{1/5},
  \end{equation}
  then
  \begin{equation}\label{eq_upper_bound_mc}
  (m_1m_2)^{-1/2}\left\|\widetilde{M}_t^\lambda - M_t^0\right\|_2 \le C_2(C_M\vee \sigma_\xi)^{4/5}\left(\frac{r_t(m_1\vee m_2)\log(m_1+m_2)}{n T}\right)^{2/5},
  \end{equation}
  where $C_1,C_h,C_2$ are the same constants as above.
  \end{cor}
  
  Recall that the classical static result for the error bound of matrix completion in \cite{koltchinskii_nuclear-norm_2011} is
  $$(m_1m_2)^{-1/2}\left\|\widehat{M}_t^\lambda-M_t^0\right\|_2\le (1+\sqrt{2})C_1(C_M\vee \sigma_\xi)\left(\frac{r_t(m_1\vee m_2)\log(m_1+m_2)}{n}\right)^{1/2}.$$
  The dynamic method provides a sharper upper bound than the classic one. In other words, when the sample size in each single time point is sparse, our method can borrow information from adjacent time points to improve the estimation, as long as the number of time points $T$ is sufficiently large.


\subsection{Dependent case}
\label{sec:dependent}
  In parallel to the independent case, we derive the bounds of  $\|\widetilde{M}_t^\lambda - M_t^0\|_2$ when $\xi_{ji},X_{ji}$ are $\phi$-mixing processes, as stated in Theorem \ref{thm_phi_mixing} below.

  \begin{thm}
  \label{thm_phi_mixing}
  With Assumption \ref{assump_Sigma}--\ref{assump_distri} and \ref{assump_phi_mixing}, when $Th\gg\log(m_1+m_2)$, $nTh \gg \left(K_*/\sigma_*\right)^2\log^3 (m_1+m_2)$, $h\rightarrow0$ and $(m_1m_2)^{1/2}Th^2\rightarrow\infty$ as $n,m_1,m_2,T\rightarrow\infty$,
  with probability at least $1-3/(m_1+m_2)$,
  \begin{equation}\label{eq_phi_mixing}
  \begin{aligned}
  (m_1m_2)^{-1/2}\left\|\widetilde{M}_t^\lambda - M_t^0\right\|_2 \le \frac{1}{2}\alpha(K) D_2 h^2 + (1+\sqrt{2})\mathcal{C}_1(\Phi_{\mathcal{X}}\vee \Phi_{\mathcal{Y}}) \sigma_*\left(\frac{r_t\log(m_1+m_2)}{\mu^2m_1m_2nTh}\right)^{1/2}+o(h^2)
  \end{aligned}
  \end{equation}
  where $\mathcal{C}_1>0$ is a constant defined in Lemma S.5 of Supplementary Material.
  \end{thm}

  We use $\Phi_\mathcal{X}$ and $\Phi_\mathcal{Y}$ to measure the temporal dependence of design matrices $X_{ji}$ and noises $\xi_{ji}$, respectively, where larger $\Phi_\mathcal{X}$ and $\Phi_{\mathcal{Y}}$ mean stronger dependence. When $\Phi_{\mathcal{X}}=\Phi_{\mathcal{Y}}=1$, it indicates that $X_{ji},\xi_{ji}$ are independent across $j$ and the bound \eqref{eq_phi_mixing} is the same as \eqref{err_bound_indpdt} in Theorem \ref{thm_total} in the independent case.
  When $nT\gg (\mu^2m_1m_2)^{-1}\sigma_*^2\log(m_1+m_2)$,
  if we choose
  \begin{equation}
  \label{eq_bandwidth}
  h=\mathcal{C}_h\left(\frac{\sigma_*^2(\Phi_{\mathcal{X}}\vee \Phi_{\mathcal{Y}})^2r_t\log(m_1+m_2)}{\mu^2m_1m_2nT}\right)^{1/5},
  \end{equation}
  where $\mathcal{C}_h = \left[(2+\sqrt{2})\mathcal{C}_1/(\alpha(K)D_2)\right]^{2/5}$, then we have the following corollary.
  \begin{cor}
  \label{cor_phi_mixing}
  Under assumptions of Theorem \ref{thm_phi_mixing}, when 
  $$n^{-1}T^4\gg\sigma_*^{-2}\mu^2m_1m_2\log^4(m_1+m_2) $$
  and
  $$ nT\gg\max\{K_*^{5/2}\sigma_*^{-3}(\mu^{2}m_1m_2)^{1/4}\log(m_1+m_2),\ \sigma_*^2(\mu^{2}m_1m_2)^{-1}\log(m_1+m_2)\},$$
  with probability at least $1-3/(m_1+m_2)$,
  \begin{equation}\label{eq_phi_mixing_general_bound}
  (m_1m_2)^{-1/2}\left\|\widetilde{M}_t^\lambda - M_t^0\right\|_2 \le \mathcal{C}_2\left(\frac{\sigma_*^2(\Phi_\mathcal{X}\vee \Phi_\mathcal{Y})^2r_t\log(m_1+m_2)}{\mu^2m_1m_2 n T}\right)^{2/5},
  \end{equation}
  where $\mathcal{C}_2 =1/2\left[(2+\sqrt{2})\mathcal{C}_1\right]^{4/5}\left[\alpha(K)D_2\right]^{1/5}$.
  \end{cor}

  When $X_{ji}$ and/or $\xi_{ji}$ have strong long-term dependence across $j$, $\Phi_\mathcal{X}\vee \Phi_{\mathcal{Y}}$ becomes large and inflates the bound \eqref{eq_phi_mixing_general_bound}, which means that the convergence rate of $\widetilde{M}_t^\lambda$ is usually slower than that in independent case. 
  Note that the bound for the single-stage estimates at each $t$ using static method can also be derived from Theorem \ref{thm_total} by using a degenerated kernel $\delta_0(x)$, given by
  \begin{equation*}
  (m_1m_2)^{-1/2}\left\|\widehat{M}_t^\lambda - M_t^0\right\|_2 \le (1+\sqrt{2})C_1\sigma_*\left(\frac{r_t\log(m_1+m_2)}{\mu^2m_1m_2n}\right)^{1/2}.
  \end{equation*}
  Then when 
  \begin{equation}
  \label{eq_condition_sharper}
  \Phi_{\mathcal{X}}\vee \Phi_{\mathcal{Y}} \ll T^{1/2}\left(\frac{\sigma^{*2}r_t\log(m_1+m_2)}{\mu^2 m_1 m_2 n}\right)^{1/8},
  \end{equation} 
  our dynamic method yields a sharper bound than the static one though strong dependence across time exists. With the expression of $h$ in \eqref{eq_bandwidth}, the condition \eqref{eq_condition_sharper} is equal to
  $T h \gg (\Phi_{\mathcal{X}}\vee \Phi_{\mathcal{Y}})^2$. 
  Note that $Th$ corresponds to the effective time points used in estimation at each $t$. This implies that as the number of effective time points increases, the negative effect caused by temporal correlation would be offset to some extent. 
For space economy, the applications of Theorem \ref{thm_phi_mixing} and Corollary \ref{cor_phi_mixing} to dynamic matrix completion and dynamic compressed sensing are presented as Corollary S.2 and Corollary S.4 in Supplementary Material.
  
\subsection{Empirical error and complexity analysis}
\label{algorithm}

Recall that the sequence generated by the iteration of Algorithm \ref{alg1} at time $t$ are denoted by $\{M^{(k)}_t\}_k,k\in \mathbb{N}^*$, where $M^{(k)}_t$ is the output matrix after $k$-th iteration and $M_t^{(0)}$ is the initial input matrix. 
In practice, we use the output $M^{(k)}_t$ to approximate $M_t^0$ for a certain $k$. 
Thus, we derive the upper bound of the empirical error $\big\|M_t^{(k)} - M_t^0\big\|_2$ in the following theorem, which takes both the algorithmic error between $M^{(k)}_t$ and $\widetilde{M}_t^\lambda$ and the estimation error between $\widetilde{M}_t^\lambda$ and $M_t^0$ into account.

\begin{thm}
	\label{thm_algorithm} Under assumptions of Theorem \ref{thm_error_bound},
	for any $k>1$ and $\lambda \ge 2\|\mathcal{W}_h\Delta_{t}\|_\infty$, we have
	\begin{equation*}
		\begin{aligned}
			\left\|M_t^{(k)} - M_t^0\right\|_2\le  \delta_h M_t +\min\left\{\left((\delta_h M_t)^2 +\frac{2\lambda}{\mu}\|M_t^0\|_1 + \frac{2L_f\gamma_t^2}{\mu(k+1)^2}\right)^{1/2},\right.&\\
  \left.\frac{1+\sqrt{2}}{2} \frac{\lambda}{\mu}\sqrt{r_t}+\left(\frac{2L_f \gamma_t^2}{\mu(k+1)^2}\right)^{1/2}\right\}& 
		\end{aligned}
	\end{equation*}
	where  $\gamma_t = \big\|M_t^{(0)}-\widetilde{M}^\lambda_t\big\|_2 $ and $L_f =2 \big\|\sum_{j=1}^T \omega_h(j-t)\cdot\frac{1}{n_j}\sum_{i=1}^{n_t} \left(\|X_{ji}\|_2X_{ji}\right)\big\|_2$
	 is the Lipschitz constant of gradient function $\nabla f_t(M)$ that is finite under Assumption \ref{assump_distri}.
\end{thm}

The empirical error bound in Theorem \ref{thm_algorithm} has an additional term $2L_f\gamma_t^2/\mu(k+1)^2$ compared to the estimation error bound \eqref{eq_thm_error_bound} in Theorem \ref{thm_error_bound}, which is the so-called algorithmic error. The optimal choice is to stop iterating when the algorithmic error attains the same order as the estimation error. Let $k_t$ be the minimal number of iteration steps needed to attain the desired order. From the expression of the algorithmic error, $k_t$ depends on the distance between the initial value $M_t^{(0)}$ and the target $\widetilde{M}^\lambda_t$, denoted by $\gamma_t$. Since less iterations are required for a smaller initial distance, the naive strategy to initialize $M_t^{(0)}$ randomly at each time $t$ would lead a large $\gamma_t$. Noting that $\widetilde{M}_t^\lambda$ is close to $\widetilde{M}_{t-1}^\lambda$, we thus make use of the neighboring information in the initialization to reduce computation cost.
Hence we propose to use the output matrix at time $(t-1)$ as the initial value for estimating $\widetilde{M}_t^\lambda$ at time $t$, i.e. $M_t^{(0)}=M_{t-1}^{(k_{t-1})}$. 
Take the dependent case in Section \ref{sec:dependent} for instance. Recall that the optimal estimation error bound is derived in Corollary \ref{cor_phi_mixing} with appropriate choice of the bandwidth $h$. 
Then desired empirical error bound for Theorem \ref{thm_algorithm} can be derived by
\begin{equation}\label{eq_bound_mixing_alg}
	(m_1m_2)^{-1/2}\left\|M_t^{(k_t)} - M_t^0\right\|_2 \le 2\mathcal{C}_2\left(\frac{\sigma_*^2(\Phi_\mathcal{X}\vee \Phi_\mathcal{Y})^2r_t\log(m_1+m_2)}{\mu^2m_1m_2 n T}\right)^{2/5}.
\end{equation}
To quantify the improvement of our initialization strategy, in Corollary \ref{cor_iteration_step_mixing}, we compute the total number of iteration steps to attain the bound \eqref{eq_bound_mixing_alg} at each $t$, and compare with that of the random initialization.
The result for independent case can be derived similarly, which is presented in Corollary S.5 of Supplementary Material for space economy.

 \begin{cor}
 \label{cor_iteration_step_mixing}
	Under the assumptions of Theorem \ref{thm_phi_mixing},
	when the total number of iterations of random initialization $K_0$ satisfies
	\begin{equation*}
		K_0 \ge \mathcal{C}_2^{-1}\left(\frac{\sigma_*^2(\Phi_\mathcal{X}\vee \Phi_\mathcal{Y})^2r_t\log(m_1+m_2)}{\mu^2m_1m_2}\right)^{-2/5} \left(\frac{2L_f}{\mu m_1m_2}\right)^{1/2}\gamma_1 n^{2/5}T^{7/5},
	\end{equation*}
	 or total number of iterations of the proposed initial strategy $K_1$ satisfies
	\begin{align*}
		K_1 \ge  &\mathcal{C}_2^{-1}\left(\frac{\sigma_*^2(\Phi_\mathcal{X}\vee \Phi_\mathcal{Y})^2r_t\log(m_1+m_2)}{\mu^2m_1m_2}\right)^{-2/5}\left(\frac{2L_f}{\mu m_1m_2}\right)^{1/2} (\gamma_1 +D_1) (nT)^{2/5}
         \\&+ 3\sqrt{2}\left(\frac{L_f}{\mu}\right)^{1/2}T.
	\end{align*}
      the empirical error is lower than the optimal upper bound \eqref{eq_bound_mixing_alg} for each $t=1,\ldots,T$.
   \end{cor}

From Corollary \ref{cor_iteration_step_mixing}, once the total number of iterations of random initialization attains the level $\mathcal{O}(n^{2/5}T^{7/5})$ and that of Algorithm \ref{alg1} attains $\mathcal{O}(n^{2/5}T^{2/5}+T)$, the error is bounded by \eqref{eq_bound_mixing_alg}. 
Then the ratio of computational cost of our proposed algorithm compared to the random initialization method is $\mathcal{O}(T^{-1}\vee n^{-2/5}T^{-2/5})$.
We finally mention that the classical static method need $\mathcal{O}(n^{1/2})$ iterations to attain the optimal  rate $\mathcal{O}(n^{-1/2})$ for each $t$, and hence the single-stage method (at all $T$ times) need $\mathcal{O}(n^{1/2}T)$ iterations in total. 
Therefore the ratio of computational cost of Algorithm \ref{alg1} compared to the single-stage method is $\mathcal{O}(T^{-3/5}\vee n^{-1/2})$, which leads to the conclusion that our algorithm is computationally more efficient than the random initial method and the classical static methods. 


\section{Simulation Studies}
\label{sec:simu}
  We take the dynamic matrix completion as the instance to verify the usefulness of our method.
  To demonstrate the advantages,  we compare the proposed estimates with three benchmark methods: (i) the classical static estimator $\widehat{M}_t^\lambda$ defined as in (\ref{eq_classical_estimator}), abbreviated as Static; (ii) the two-step smoothing estimator $\widehat{M}_t^{\text{t-s}}$ given in \eqref{eq:twostep}, abbreviated as TwoStep; (iii) the low-rank tensor completion estimator, abbreviated as Tensor,
  \begin{equation*}
  \widehat{N}^\lambda =  \argmin_{N\in \mathbb{R}^{T\times m_1\times m_2}} \|\mathcal{P}_\Omega(N)-Y\|_2^2 + \lambda \sum_{i=1}^3 \|N_{(i)}\|_1,
  \end{equation*}
  where $\mathcal{P}_\Omega$ is the projection to the observation subspace $\Omega$, Y is the observed data tensor with element 
  \begin{align*}[Y]_{j,s,l}=\left\{
		\begin{aligned}
			&\frac{1}{|\mathcal{A}_{j,s,l}|}\sum_{i\in \mathcal{A}_{j,s,l}}Y_{ji},\ \text{ if }  \mathcal{A}_{j,s,l}\neq \emptyset \\
			 &0,\ \text{ otherwise}
		\end{aligned}
		\right.
  \end{align*}
  for $\mathcal{A}_{j,s,l} = \left\{i\mid X_{ji} = e_s(m_1)e^\top_l(m_2),1\le i\le n_j\right\}$, and $N_{(i)}$ is the mode-$i$ unfolding of tensor $N$ which is a matrix that arranges the mode-$i$ fibers to be columns. We use matrix form FISTA algorithm in \cite{toh2010accelerated} to approximate $\widehat{M}_t^\lambda$, and $\widehat{M}^{\text{t-s}}_{t}$ has a closed form solution. In Tensor, we use AMD-TR(E) algorithm in \cite{gandy2011tensor} to numerically approximate $\widehat{N}^\lambda$.

  We set $m_1=500,m_2=300,r_t=10$ and the matrix function $M(t)$ in Assumption \ref{assump_M} is constructed by $M(t)=U(t)D(t)V^T(t)$ with
  \begin{align*}
  U(t) &= \cos\left(\frac{t\pi}{2}\right)U_0 + \sin\left(\frac{t\pi}{2}\right)U_1,\\
  D(t) &= 10 \left(\operatorname{diag}\left\{10^2,9^2,\dots,1\right\}+ t\operatorname{diag}\left\{10,9,\dots,1\right\}\right),\\
  V(t) &= \cos\left(\frac{t\pi}{2}\right)V_0+\sin\left(\frac{t\pi}{2}\right)V_1,
  \end{align*}
  where $U_0,U_1\in \mathbb{R}^{m_1\times r_t}$, $V_0,V_1\in \mathbb{R}^{m_2\times r_t}$ are composed of standard orthogonal basis. Here we consider the situation that each $n_t$ is the same, i.e., $n_t=n,t=1,2,\cdots,T$. Denote the rate of observation samples as $\rho= n/(m_1m_2)$ and the density of time points as $\tau=1/T$. We use a 5-fold cross validation to select $\lambda$ in the numerical experiments. \oldchange{Please refer to  Section S.9.4 of the Supplement for more details.}

\begin{figure}[!ht]
  \centering 
  \subfigbottomskip=2pt 
  \subfigcapskip=-5pt 
  \subfigure{
  \includegraphics[width=0.5\linewidth]{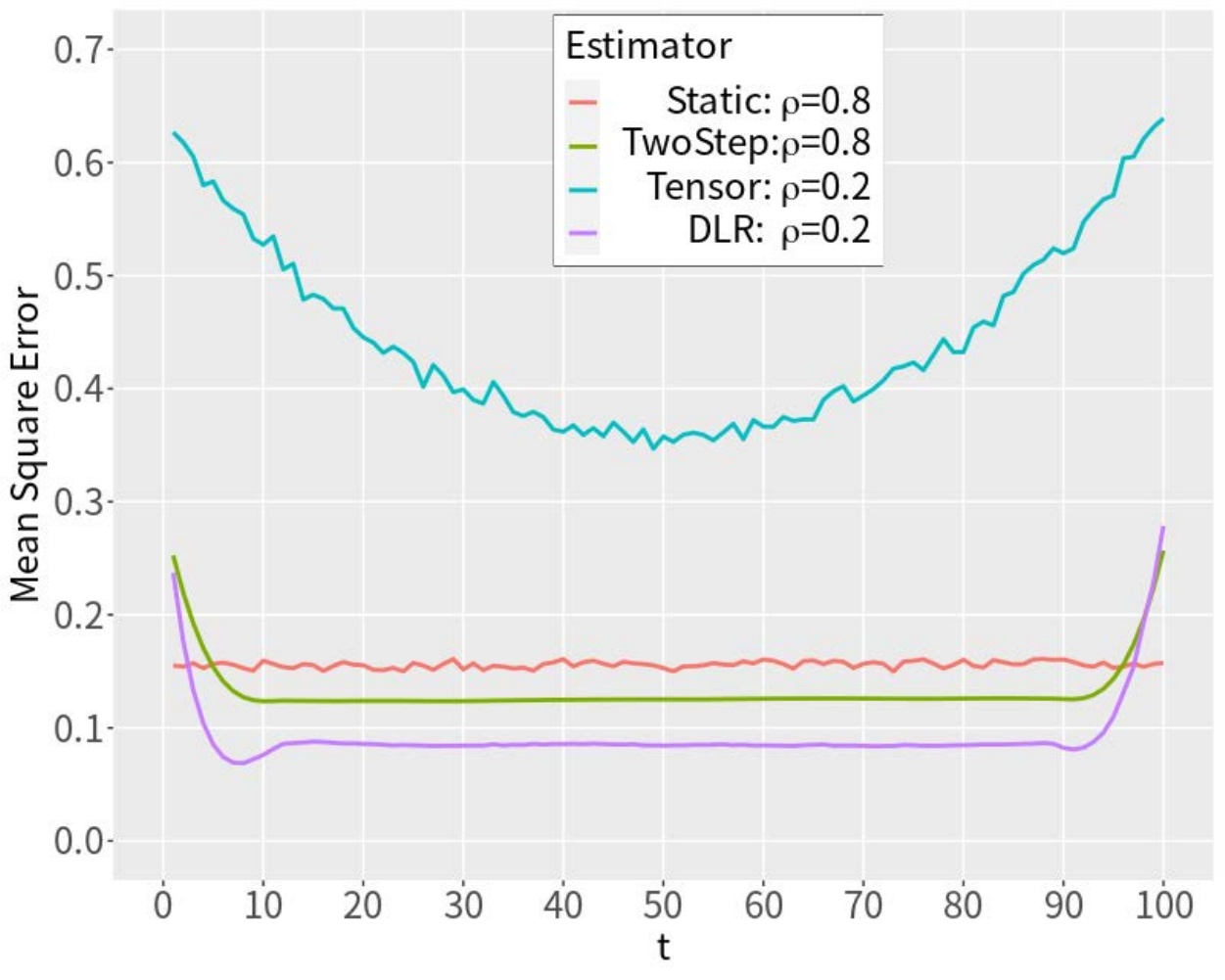}}
  \caption{The comparisons of MSE$_t$ for different methods. In DLR and Tensor, $\rho=0.2$, and in Static and TwoStep, $\rho=0.8$.}
  \label{fig_0}
  \end{figure}

  For the independent case, we sample $X_{ji},\ i=1,2,\dots,n$ from $\mathcal{E}$ defined in (\ref{eq:dmc}) independently with uniform probabilities, and let $\xi_{ji},j=1,2,\dots,T,\ i=1,2,\dots,n$ i.i.d follow the normal distribution $N(0,\sigma_\xi^2)$, where $T=100$ and $\sigma_\xi=1$. We use the Epanechnikov kernel and set the bandwidth $h$ by the optimal selection form \eqref{eq_select_h} in which we approximate $C_M\vee \sigma_\xi$ by mean of the top 10\% largest $Y_{j i}$ and $D_2$ by $ \sum_{j=2}^T |1/n_j\sum_{i=1}^{n_j} Y_{j i} - 1/n_{j-1}\sum_{i=1}^{n_{j-1}} Y_{j-1 i}|$. We set $\rho=0.2$ for our proposed dynamic low-rank (DLR) and Tensor methods. In Static and TwoStep, $\widehat{M}_t^\lambda$ and $\widehat{M}_{t}^{\text{t-s}}$ fail to recover the matrix structure under the same setting $\rho = 0.2$, so we set $\rho = 0.8$ for feasibility. Figure \ref{fig_0} illustrates the mean square errors 
  \begin{equation}\label{eq_mse_1}
  \mbox{MSE}_t= (m_1m_2)^{-1}\left\|\widehat{M}_t - M^0_t\right\|_2^2, 
  \end{equation}
  for $t=1,2,\cdots,T$. 
  For the proposed DLR and TwoStep methods, the MSE$_t$ floats up at the boundaries due to slight edge effect.  The average MSE, i.e., $T^{-1}\sum_{t=1}^T \mbox{MSE}_t$ for the DLR estimate is $0.085$ and for Tensor is $0.44$ when $\rho=0.2$. This implies that the tensor recovery is not effective at least when the temporal structure is smooth. For Static and TwoStep, when $\rho=0.8$, the average MSE is $0.16$ and $0.13$ respectively. That is, these two methods with four times data still perform worse than the proposed DLR method.  
  This verifies the theoretical finding that when the sample size $n$ at a single time point is small, the classical trace regression fails to estimate $M^0_t$, while our dynamic method can borrow adjacent information to improve the estimation quality. 

\begin{figure}[!ht]
  \centering 
  \subfigure{
  \includegraphics[width=0.48\linewidth]{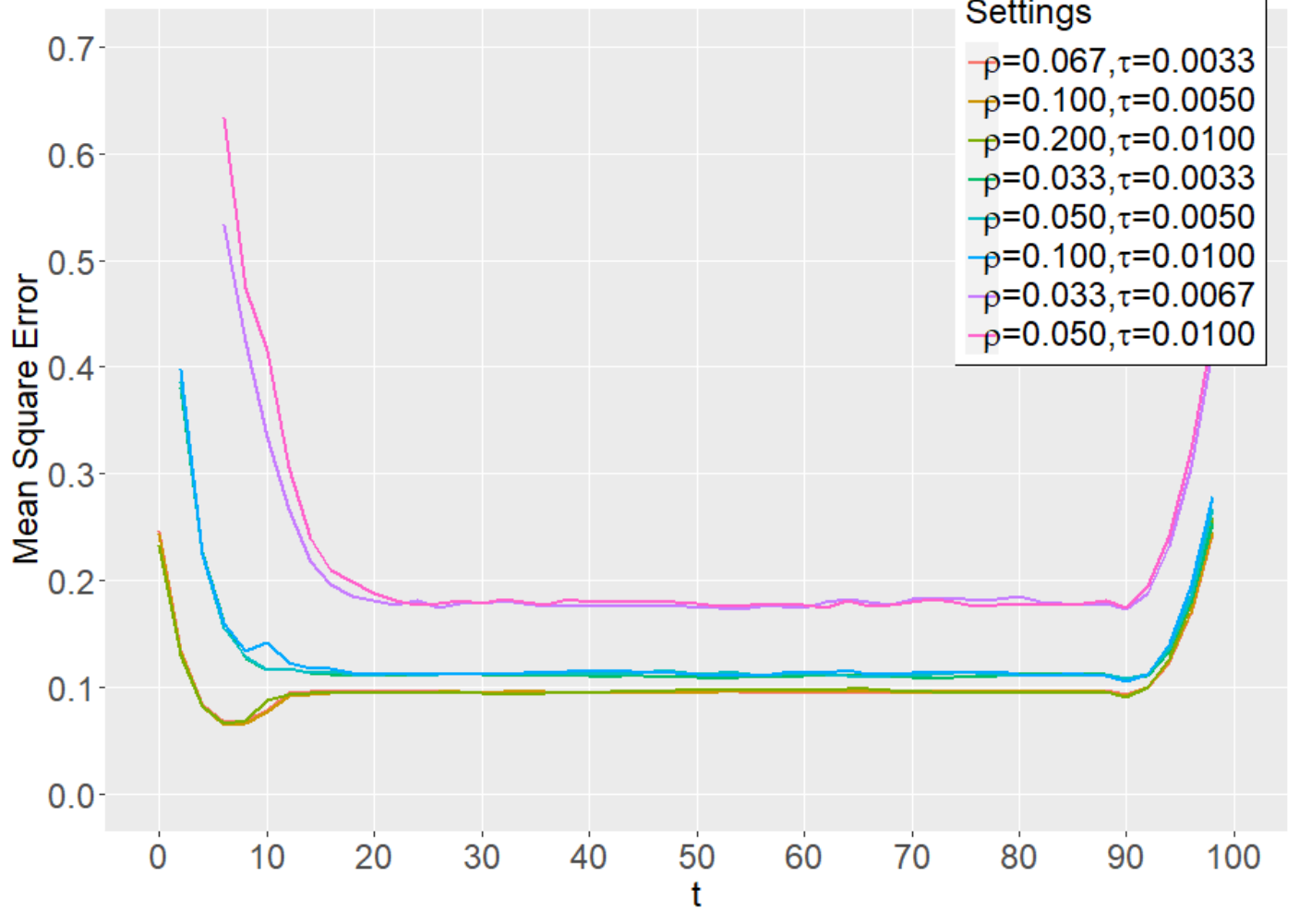}}
  \subfigure{
  \includegraphics[width=0.48\linewidth]{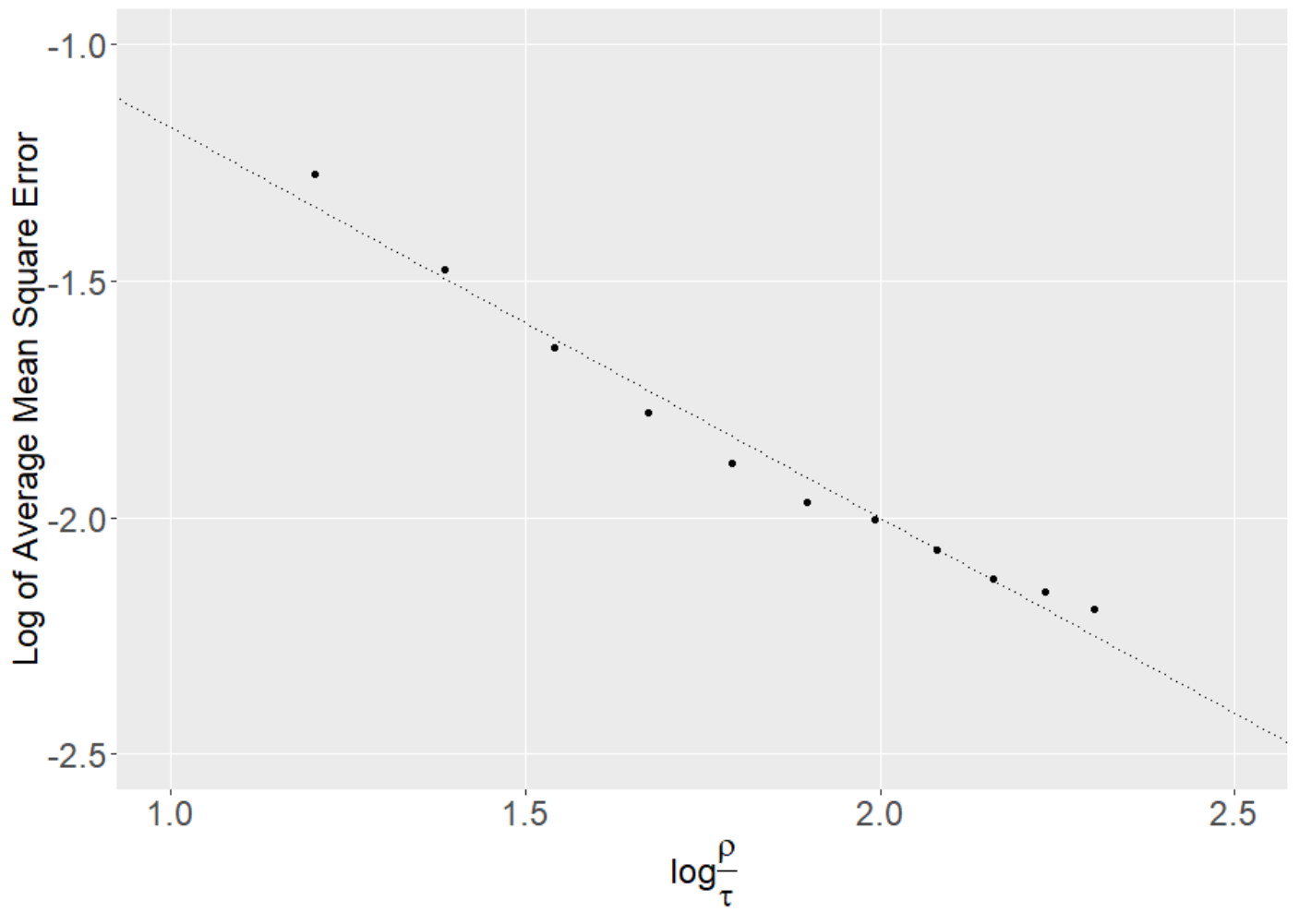}}
  \caption{The left panel corresponds to MSE$_t$ of DLR estimates under different settings of $\rho$ and $\tau$. The curves are clustered into three classes corresponding to $\rho/\tau=5,10,20$. The right panel shows the logarithm of average MSE versus the logarithm of $\rho/\tau$ showing a linear trend with the slope -4/5.}
  \label{fig_1}
  \end{figure}

  Now we study the empirical influence of $\rho$ and $\tau$ on the estimation results. One can see from \eqref{eq_upper_bound_mc} that given the underlying matrix, the bound of MSE is in proportion to $(\rho/\tau)^{-4/5}$.
  We set eight different values for $(\rho,\tau)$ which can be divided into three groups, i.e. $\rho/\tau=5,10,20$. The $MSE_t$ across $t$ under these settings are shown in the left panel of Figure \ref{fig_1}, which indicates that given the matrix dimension, the empirical MSE depends on $\rho/\tau$.
  We further plot the logarithm of average MSE across time versus the logarithm of $\rho/\tau$ in the right panel of Figure \ref{fig_1}, which reveals the linear relationship of the slope -4/5 between them. This further validates the conclusion \eqref{eq_upper_bound_mc}.

\begin{figure}[!ht]
  \centering 
  \subfigbottomskip=2pt 
  \subfigcapskip=-5pt 
  \subfigure{
  \includegraphics[width=0.48\linewidth]{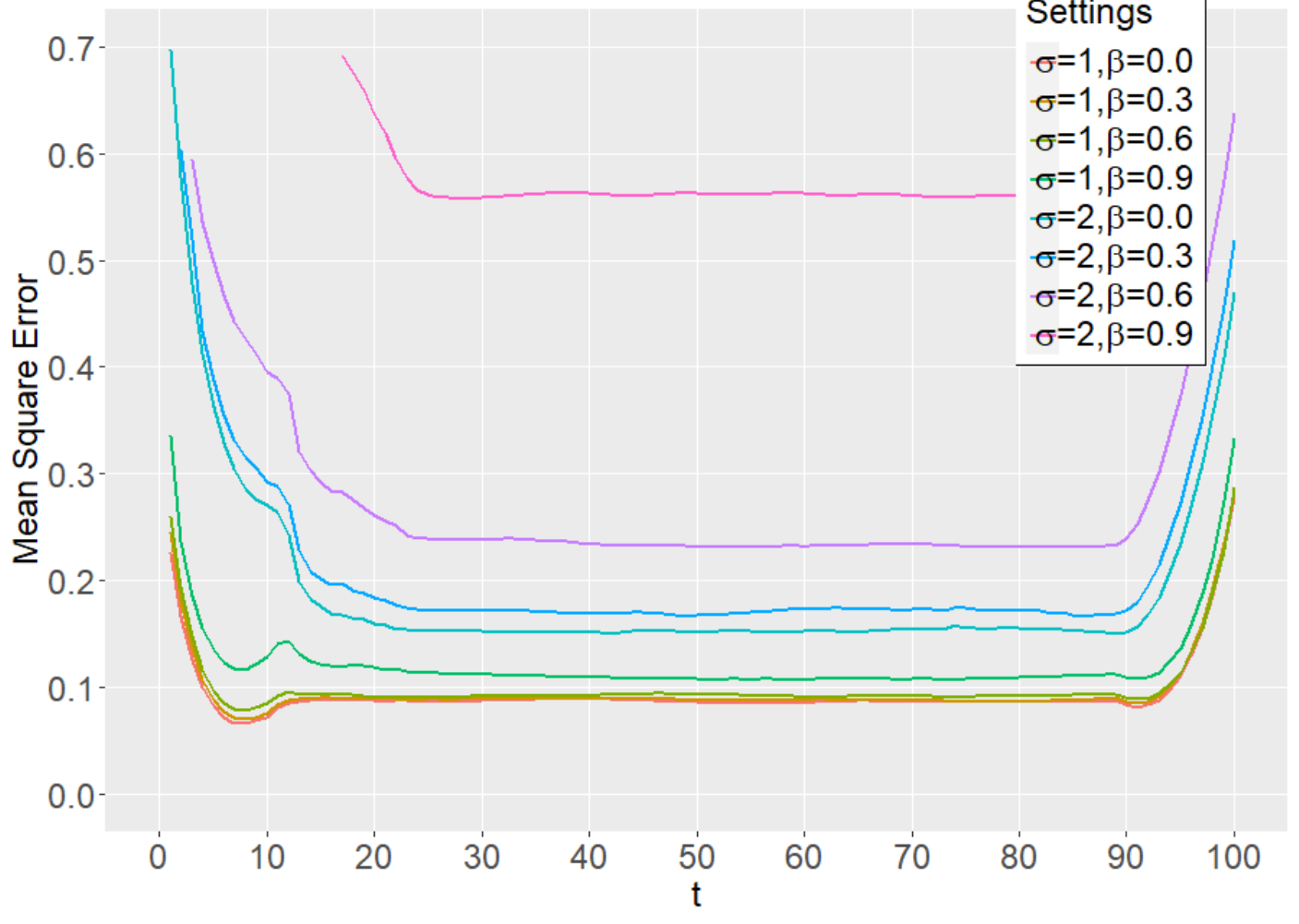}}
  \subfigure{
  \includegraphics[width=0.48\linewidth]{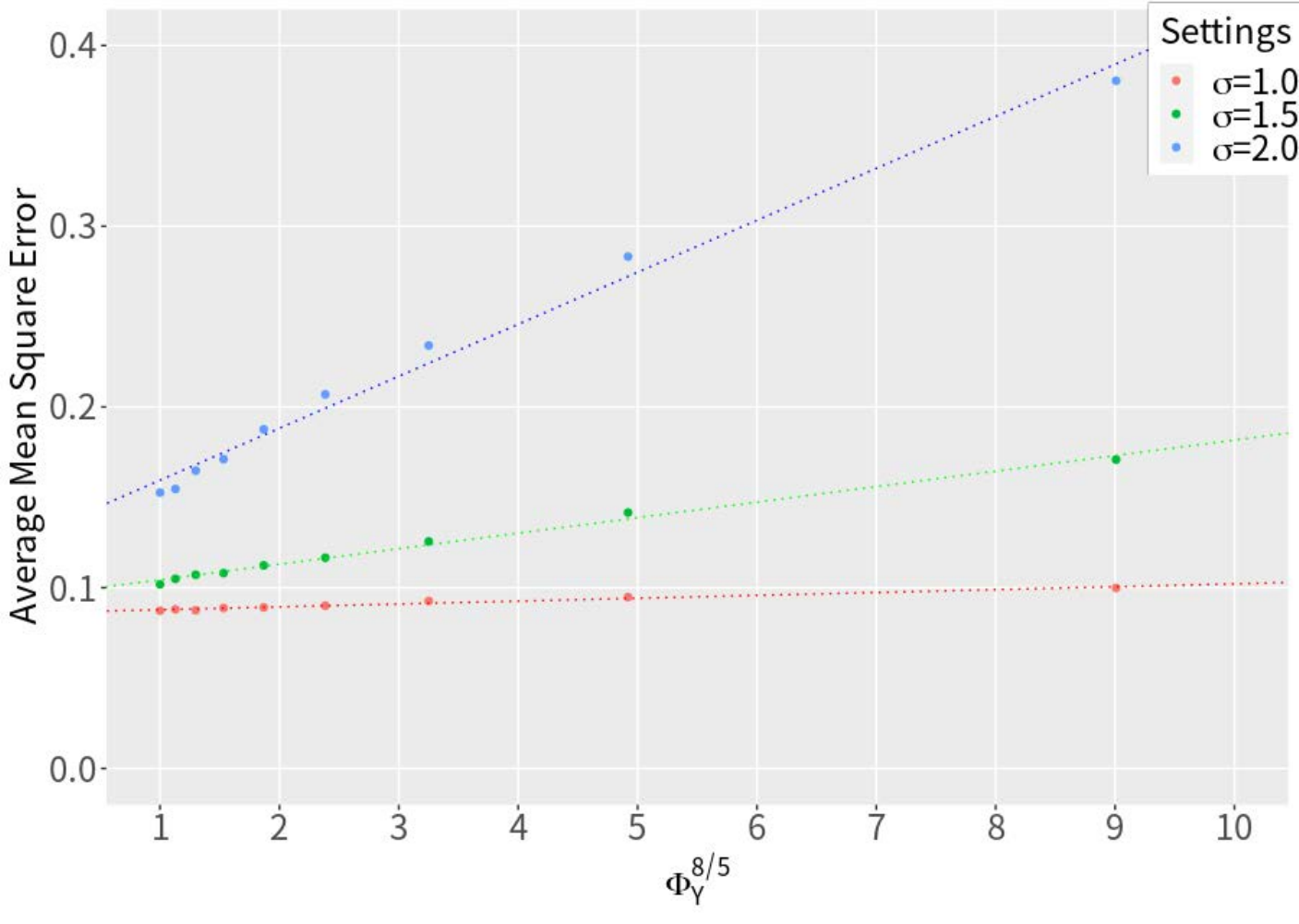}}
  \caption{The left panel illustrates the MSE$_t$ across time with different $\sigma_\xi$ and $\beta$ and the right panel shows that the average MSE versus $\Phi_{\mathcal{Y}}^{8/5}$ under different settings when $\Phi_{\mathcal{X}}=1$.  }
  \label{fig_2_1}
  \end{figure}

\newpage
  We also investigate the influence of the dependent noise on the proposed estimator.
The noises $\xi_{ji}$ are generated by
  \begin{align}\label{xi_phimixing_simu}
  E_0 &= U_0,\quad U_0 \sim \operatorname{WN}(0,\sigma_\xi^2\mathbf{1}_{m_1}\mathbf{1}^T_{m_2}),\nonumber\\
  E_{t}& = \beta E_{t-1} + \sqrt{1-\beta^2}U_{t},\quad U_{t}\sim \operatorname{WN}(0,\sigma_\xi^2\mathbf{1}_{m_1}\mathbf{1}^T_{m_2}),\\
  \xi_{ji} &= \langle E_j,X_{ji}\rangle,\quad j=1,2,\dots,T,\ i=1,2,\dots,n_t,\nonumber
  \end{align}
  where $\operatorname{WN}(0,\sigma_\xi^2\mathbf{1}_{m_1}\mathbf{1}^T_{m_2})$ generates the $m_1\times m_2$ random matrix whose elements are i.i.d. normal with mean zero and variance $\sigma_\xi^2$. The larger $\beta$ is, the stronger dependence exists among $\{\xi_{ji}\}_j$. The other parameters are set are the same as those in the independent case. Then $\Phi_{\mathcal{X}}=1$ and $\Phi_{\mathcal{Y}}$ can be calculated from \eqref{xi_phimixing_simu} which is larger than 1 when $\beta>0$.
  We apply the proposed method to recover $M_t^0$ for $1\le t\le T$ under the combinations of $\sigma_\xi=1,2$ and $\beta=0,0.3,0.6,0.9$.
  As illustrated in the left panel of Figure \ref{fig_2_1}, the MSE increases as $\sigma_\xi^2$ and/or $\beta$ enlarge(s). Note that according to (S.26) in Supplement, the MSE is proportional to $\Phi_{\mathcal{Y}}^{8/5}$ given $M_t,n,T$. We plot the average MSE versus $\Phi_{\mathcal{Y}}^{8/5}$ under different $\sigma_\xi^2$ in the right panel of Figure \ref{fig_2_1} which verifies the conclusion.
  
  \begin{figure}[!ht]
  \centering 
  \subfigbottomskip=2pt 
  \subfigcapskip=-5pt 
  \subfigure{
  \includegraphics[width=0.48\linewidth]{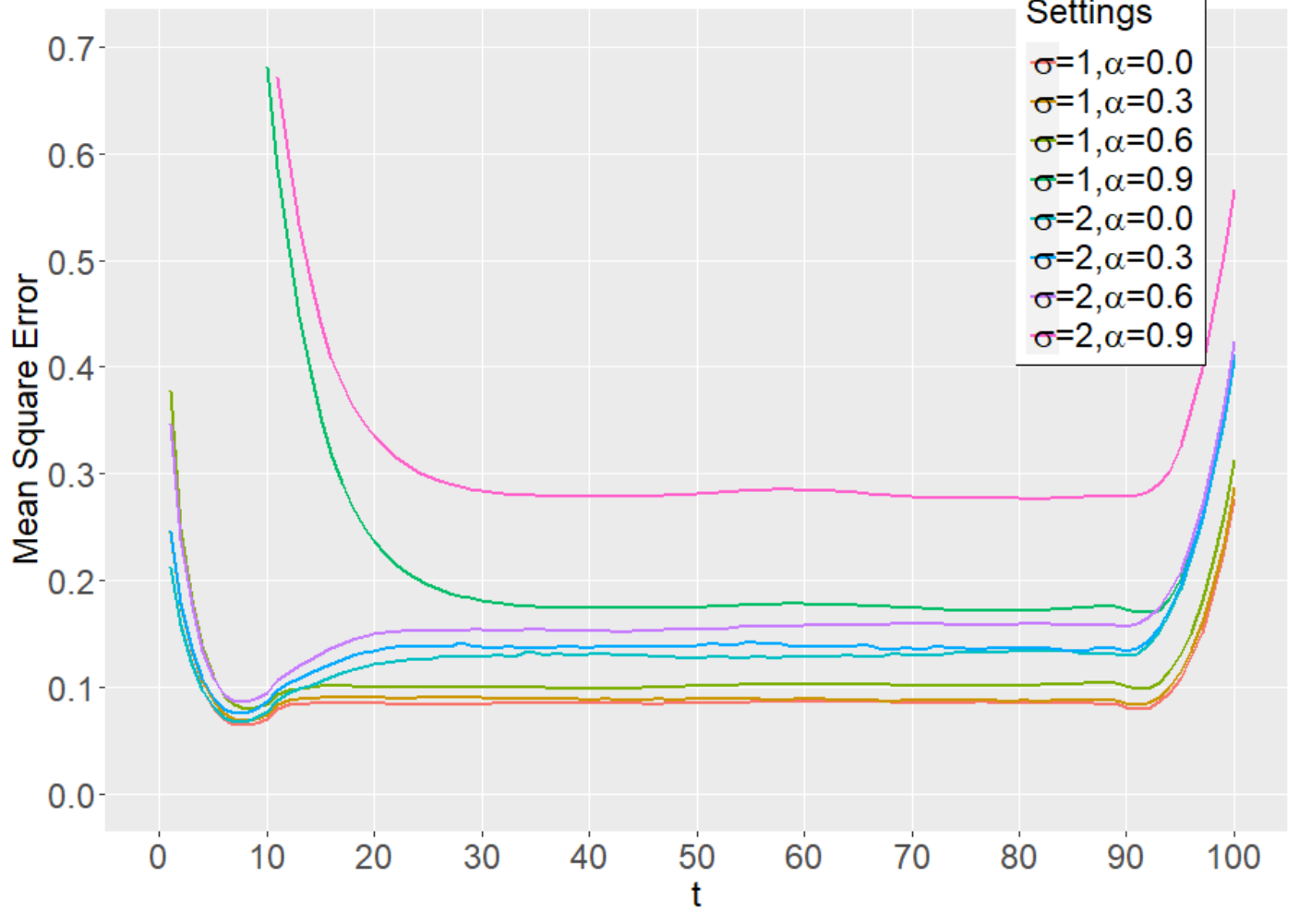}}
  \subfigure{
  \includegraphics[width=0.48\linewidth]{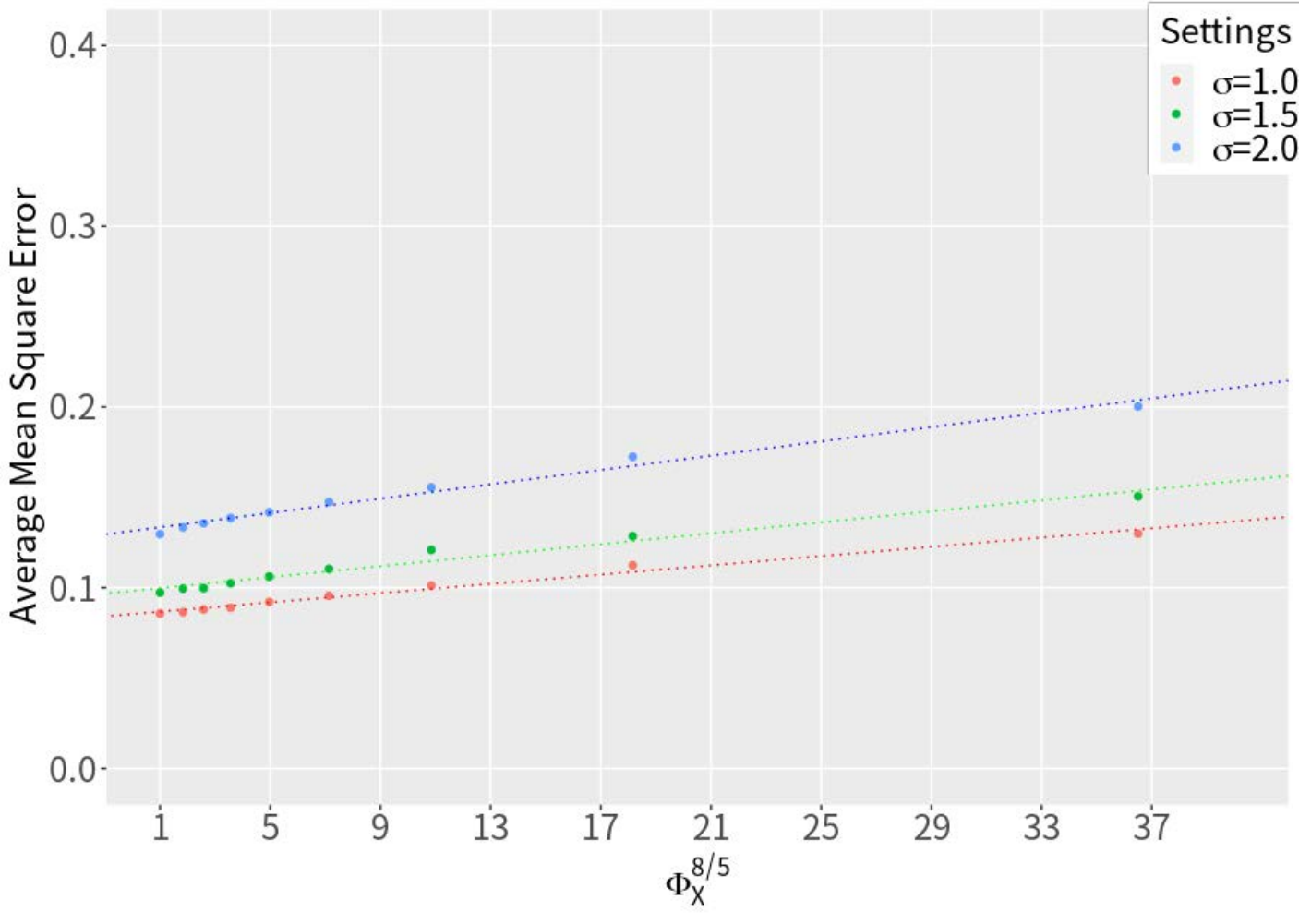}}
  \caption{The left panel illustrates the MSE$_t$ across time with different $\sigma_\xi$ and $\alpha$ and the right panel shows that the average MSE versus $\Phi_{\mathcal{X}}^{8/5}$ under different settings when $\Phi_{\mathcal{Y}}=1$.
  }
  \label{fig_2_2}
  \end{figure}
  
  Finally we study the influence of the dependence of design matrices on the estimation. We sample $X_{1i},i=1,\dots,n$ from $\mathcal{E}$ with uniform probabilities independently and generate $X_{ji},j\ge 2,i=1,2,\dots,n$ by preserving $\alpha n$ elements randomly in $X_{j-1, i},i=1,2,\dots,n$ and choosing $(1-\alpha)n$ new elements i.i.d. from $\mathcal{E}$. Then the dependence of design matrices across time becomes stronger as $\alpha$ increases. We set the remaining parameters the same as those in the independent case and have $\Phi_{\mathcal{Y}}=1$, $\Phi_{\mathcal{X}}\ge1$, where the equality holds if and only if $\alpha=0$. We plot the $\operatorname{MSE}_t$ across time under different $\sigma_\xi=1,2$ and $\alpha=0,0.3,0.6,0.9$ and verify the linear relationship between the average MSE and $\Phi_{\mathcal{X}}^{8/5}$ in Figure \ref{fig_2_2}. 
  We remark that the slopes of linear relationships in the right panels of Figure \ref{fig_2_1} and \ref{fig_2_2} are different. This is because the bound (S.26) in Supplement is obtained by the technique that $C_M\Phi_\mathcal{X}+\sigma_\xi\Phi_\mathcal{Y}\le(C_M\vee \sigma_\xi)(\Phi_\mathcal{X}\vee\Phi_\mathcal{Y})$. Hence the curves of the average MSE versus $\Phi_{\mathcal{Y}}^{8/5}$ become sharper as $\sigma_\xi$ increases, as shown in the right panel of Figure \ref{fig_2_1}, while the slopes of the average MSE versus $\Phi_{\mathcal{X}}^{8/5}$ under different $\sigma_\xi$ in the right panel of Figure \ref{fig_2_2} are similar.

\section{Real Data Examples}
\label{sec:data}
  In this section, we apply the proposed method to two real data examples for the dynamic matrix completion and compressed sensing problems, respectively. 
  
  The first example is the Netflix Prize Dataset \citep{netflix_dataset}, in which users ratings for movies received by Netflix were collected from
  October 1998 to December 2005 and can be downloaded at \href{https://www.kaggle.com/datasets/netflix-inc/netflix-prize-data}{https://www.kaggle.com/datasets/netflix-inc/netflix-prize-data}. The ratings are integer-valued from 1 to 5. We preprocess the dataset to remove those movies which are watched less than 25000 times. 
  
  \oldchange{For user selection, we consider two different filters. Filter 1 is to choose users whose rating times are more than 700 because users who watch a larger number of movies tend to offer more reliable ratings  and make better comparisons across different movies due to their broader exposure.} There remain $3036$ users and $1034$ movies with total 2337997 ratings. We utilize a highly efficient data merging approach that requires minimal computational resources. Specifically, we divide the observations into $T=100$ chronological time intervals. Within each interval, observations are aggregated as they occur at distinct time points. Then we randomly choose $4/5$ of them in each interval as training data and set the rest as test data. 
  \oldchange{At each time point $t$, there exists an underlying matrix $M_t\in \mathbb{R}^{m_1\times m_2}$ to be recovered, where the rows represent users and the columns represent movies. The observed  sparse matrix $Y_t\in \mathbb{R}^{m_1\times m_2}$ has elements at position $(j,k)$, representing the rating given by user $j$ to movie $k$ at time point $t$, or zero if no rating is available. Thus we can convert it to the trace regression model.}
  The average sample size at each time point is about $17803$ and thus the rate of observation samples in each time point is $\rho \approx 0.0057$. We use again the Epanechnikov kernel $K(x)$ and select the bandwidth $h$ in the same way as in the simulations. \oldchange{Filter 2 is to select users whose rating times are more than 30 because those users can represent the target groups of users who often using Netflix to watch movies. Then we conduct a random selection of 3000 individuals from the pool of eligible users for ease of computation.}
  
\begin{figure}[!ht]
  \centering 
  \subfigbottomskip=2pt 
  \subfigcapskip=-5pt 
  \subfigure{
  \includegraphics[width=0.48\linewidth]{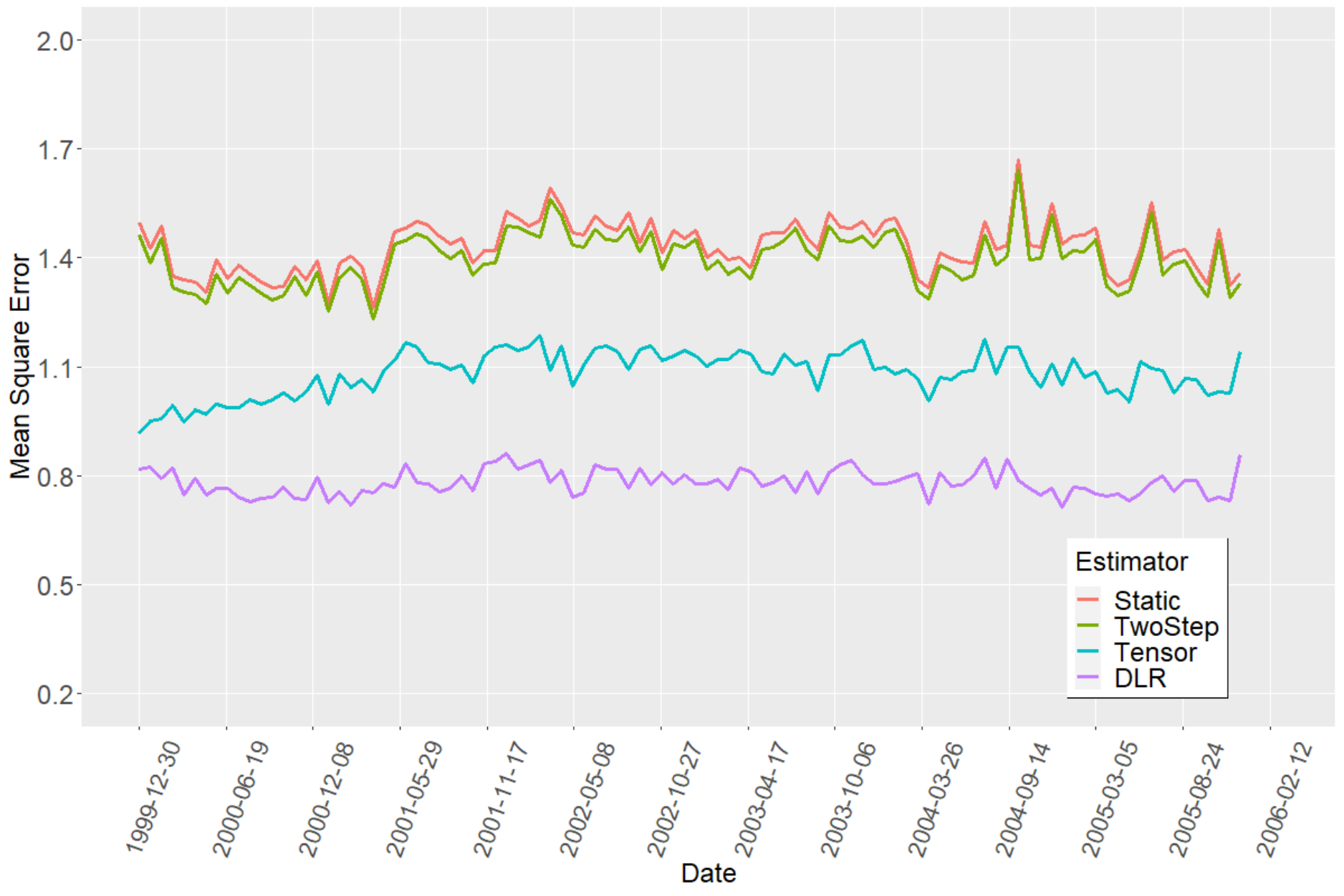}}
  \subfigure{
  \includegraphics[width=0.48\linewidth]{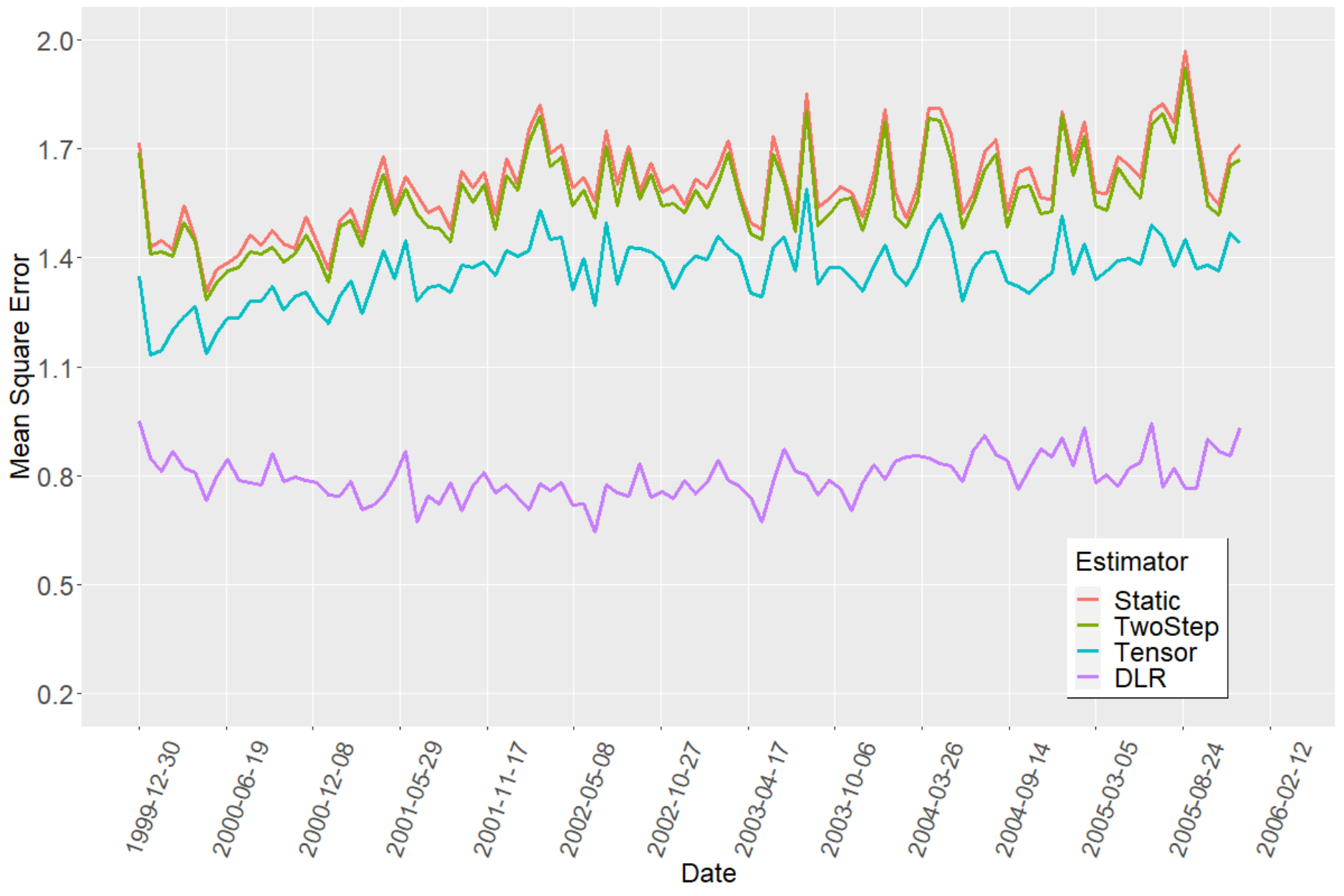}}
 \caption{The MSE$^*_t$ (\ref{eq:mse2}) at different time points for the proposed method and three benchmarks for the Netflix dataset. The left plot is the result for Filter 1 and the right plot is the result for Filter 2.}
  \label{fig_3}
  \end{figure}

  We use 5-fold cross validation to choose tuning parameter $\lambda$ and evaluate the performance of $\widetilde{M}^\lambda_t$ by calculating 
 \begin{equation} \label{eq:mse2}
  \mbox{MSE}^*_t=\frac{1}{n_{t}^\text{test}}\sum_{i=1}^{n_{t}^\text{test}} \left(\langle X_{t,i}^\text{test},\widetilde{M}^\lambda_t\rangle -Y_{t,i}^\text{test} \right)^2,
  \end{equation}
  using the test data and compare the proposed method with those three benchmarks mentioned in Section \ref{sec:simu}. Figure \ref{fig_3} shows that our dynamic method is more accurate than those benchmarks across the time domain. And the average MSE$^*$, i.e. $T^{-1}\sum_{t=1}^T \mbox{MSE}^*_t$, of our dynamic method and benchmarks Static, TwoStep and Tensor methods are compared in Table \ref{table:netflix}.
  
\begin{figure}[!ht]
  \centering 
  \subfigbottomskip=2pt 
  \subfigcapskip=-5pt 

  \subfigure{ 
  \includegraphics[width=0.18\linewidth]{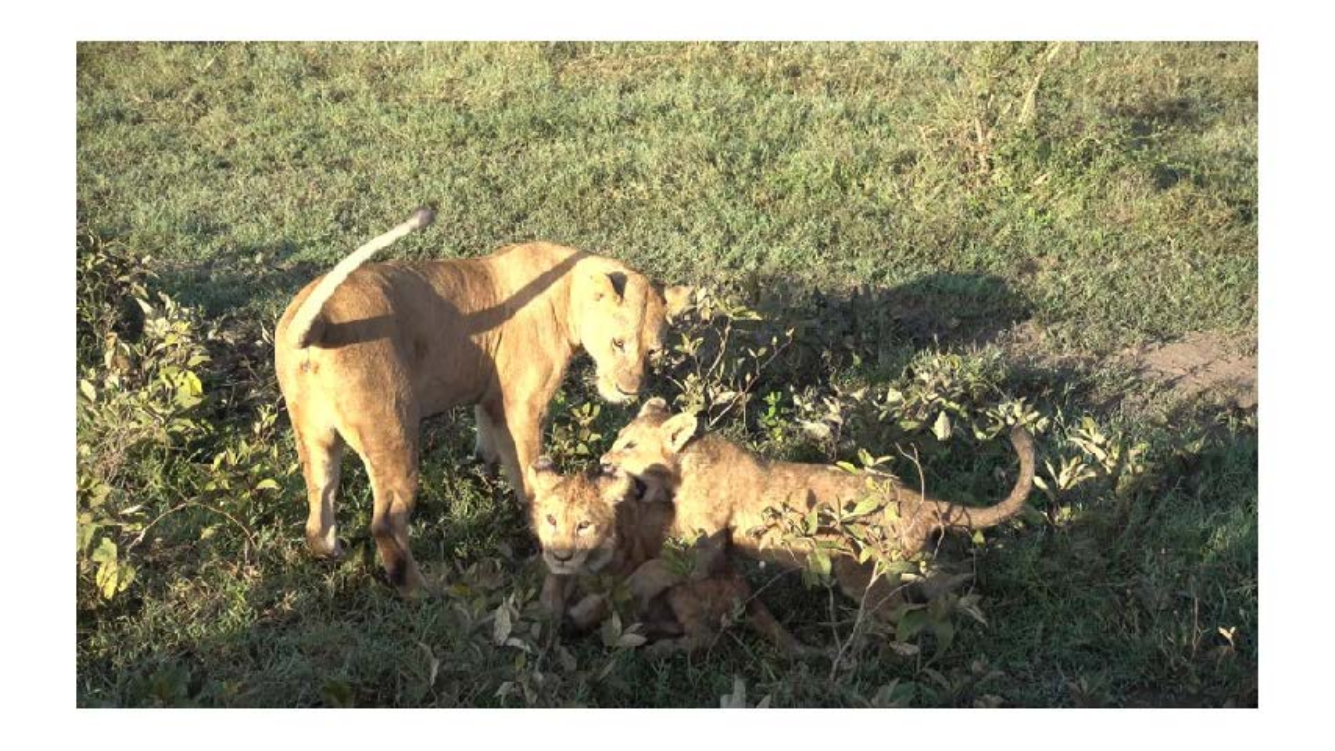}}
  \subfigure{
  \includegraphics[width=0.18\linewidth]{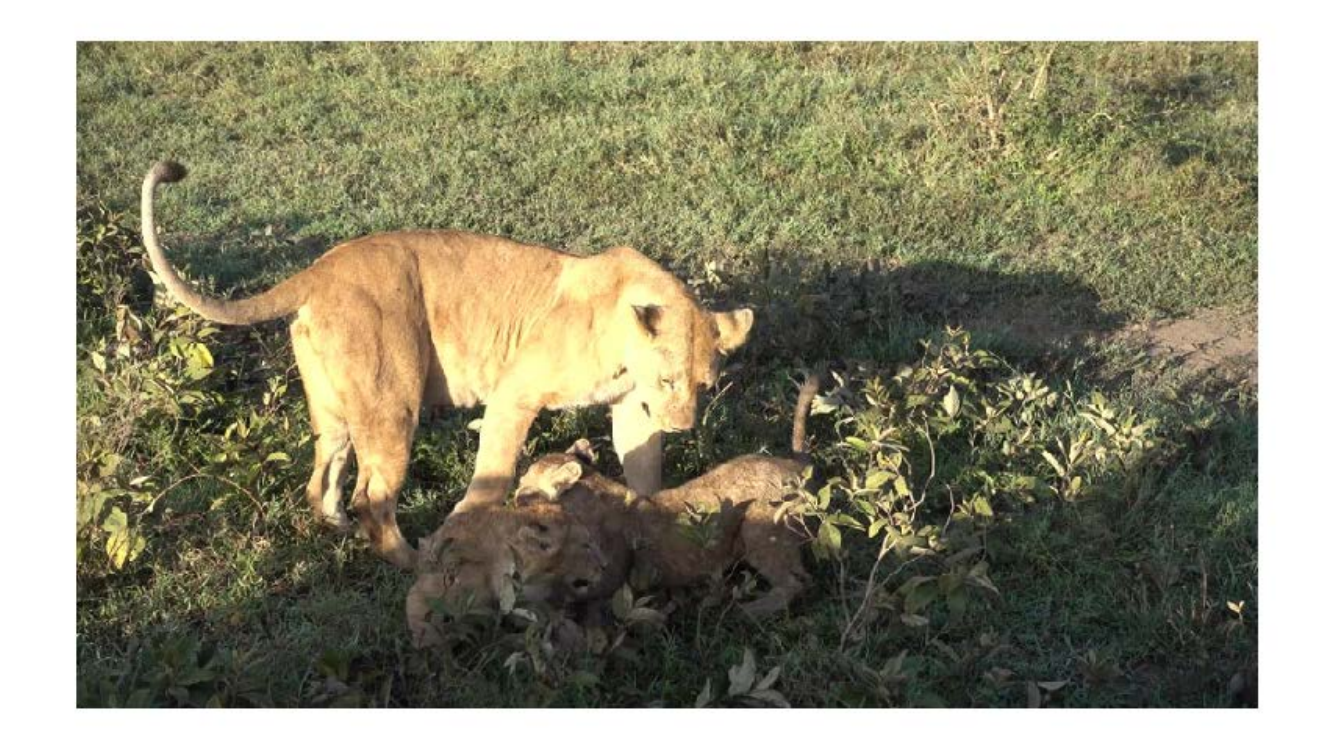}}
  \subfigure{
  \includegraphics[width=0.18\linewidth]{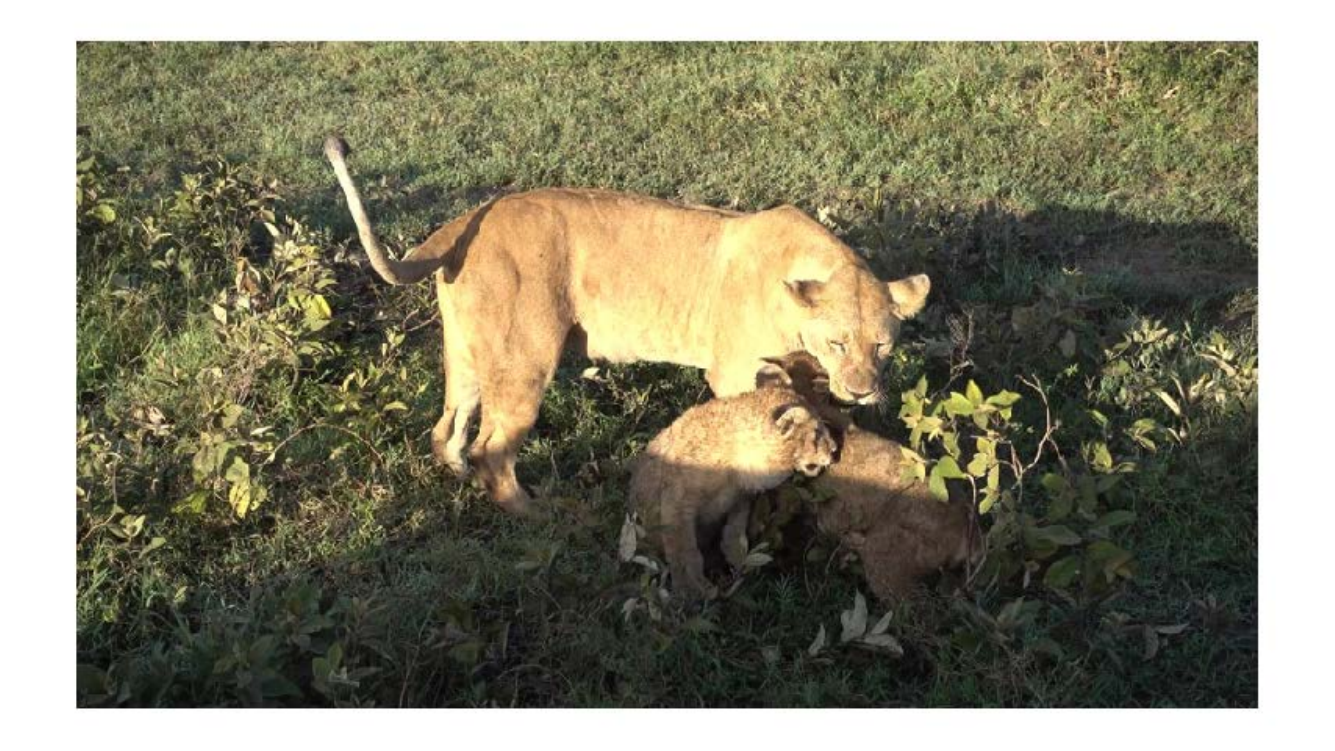}}
  \subfigure{
  \includegraphics[width=0.18\linewidth]{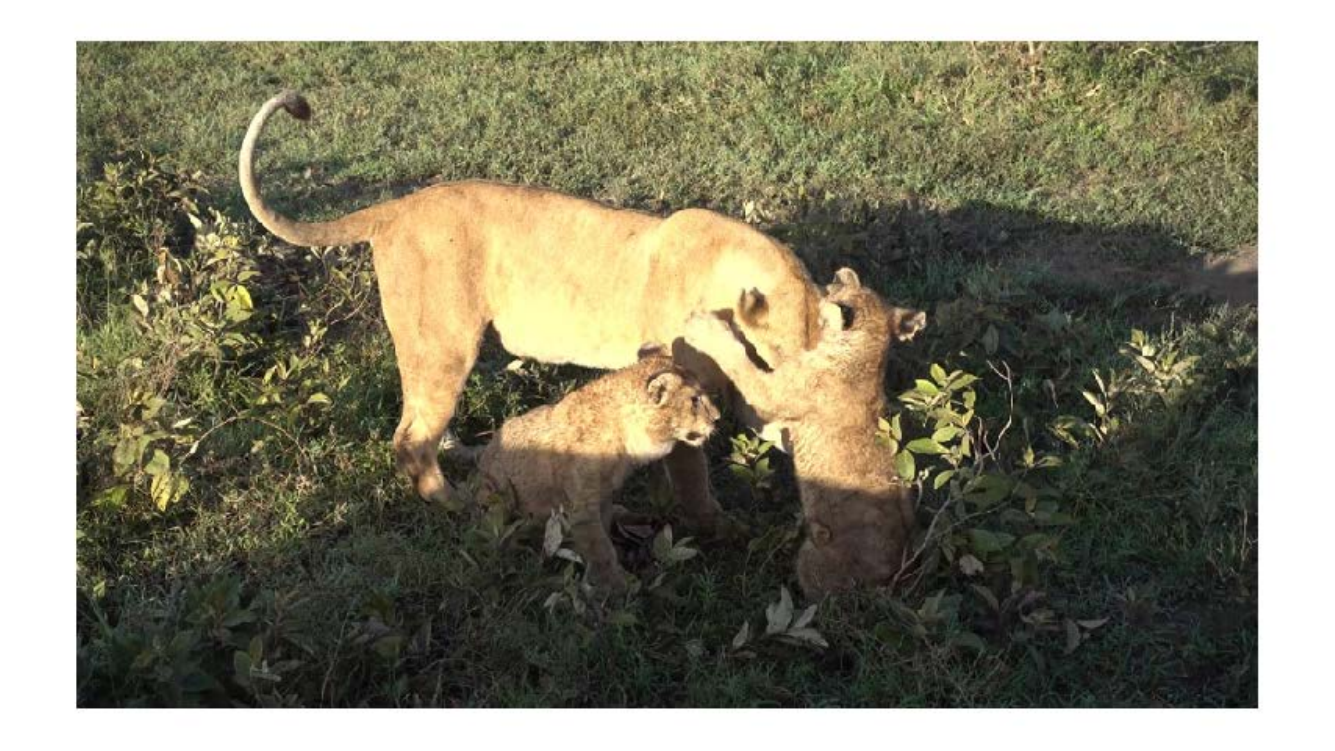}}
  \subfigure{
  \includegraphics[width=0.18\linewidth]{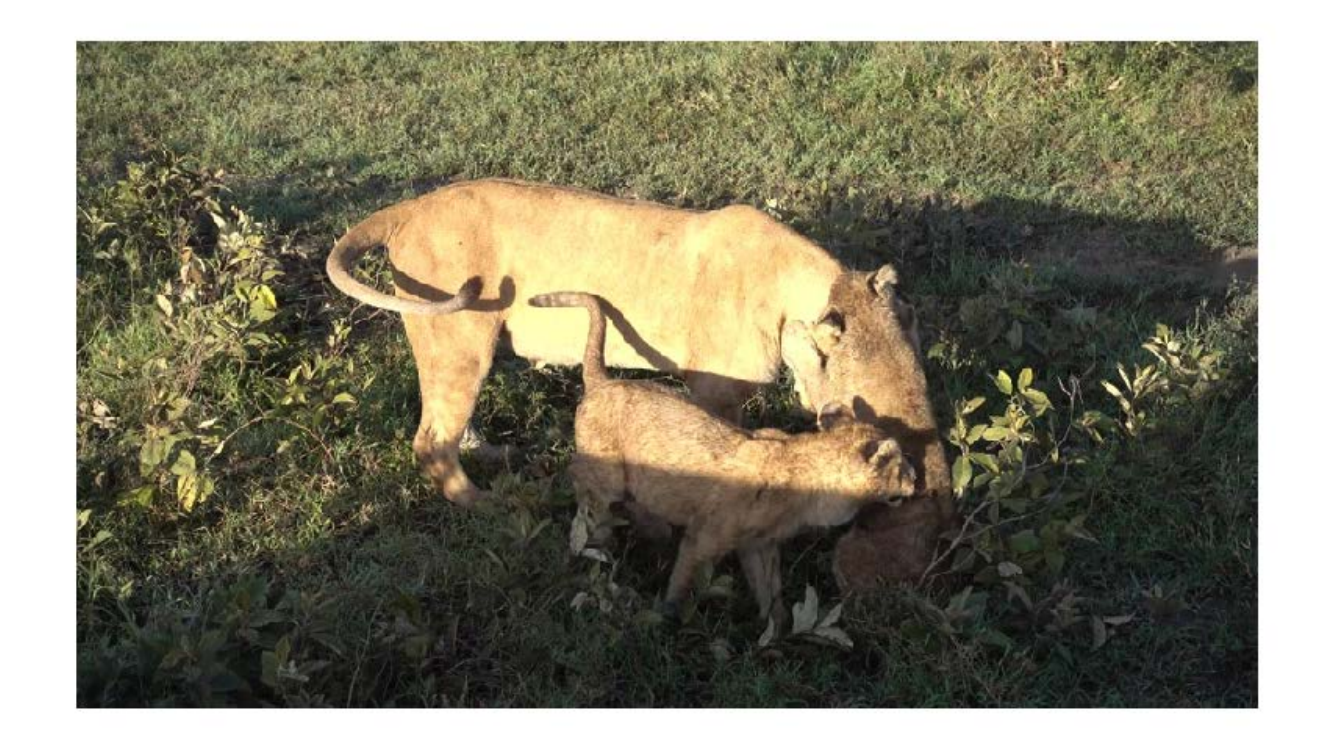}}

  \subfigure{
  \includegraphics[width=0.18\linewidth]{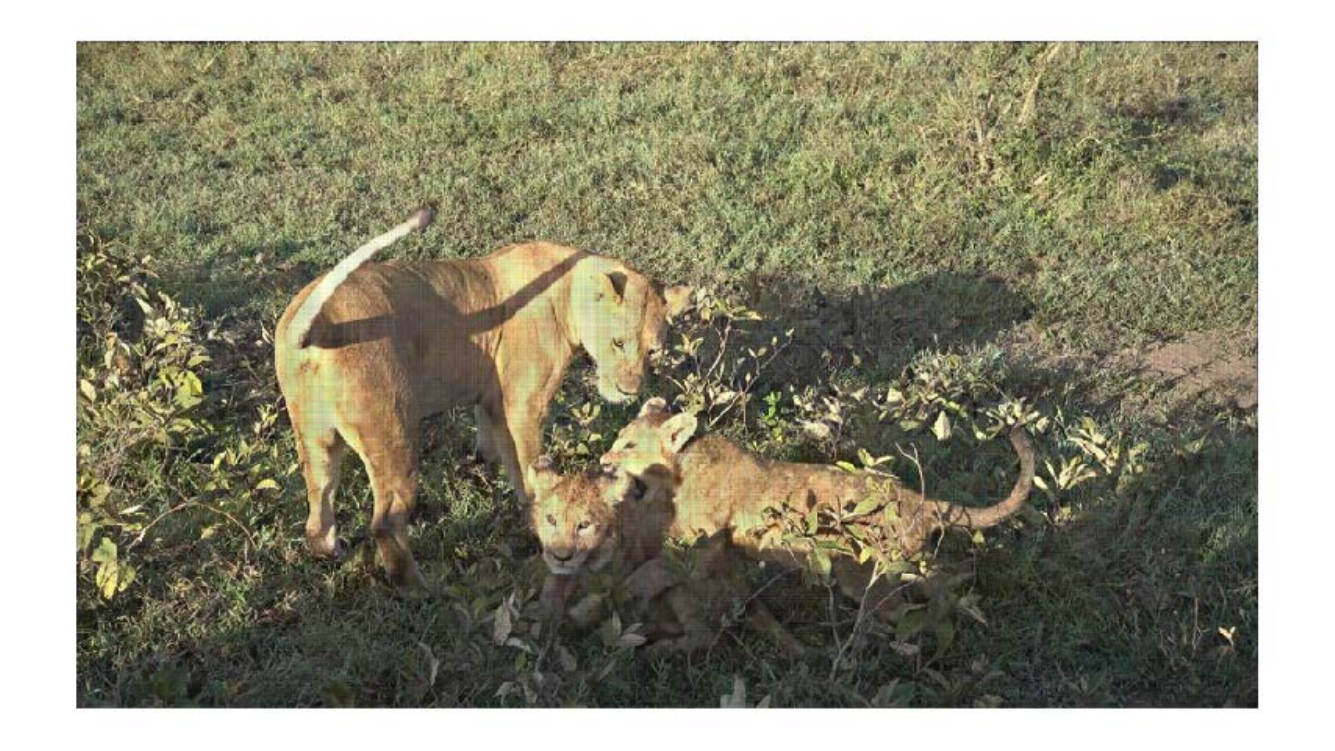}}
  \subfigure{
  \includegraphics[width=0.18\linewidth]{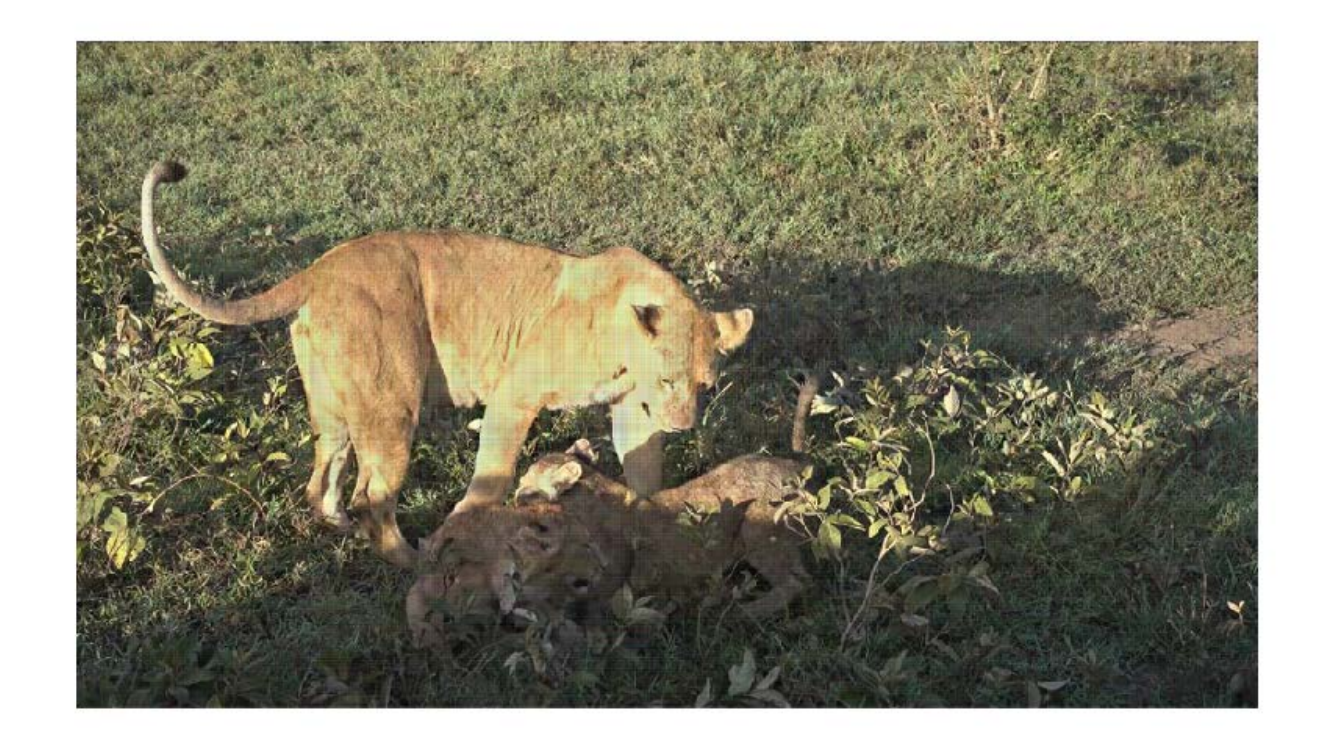}}
  \subfigure{
  \includegraphics[width=0.18\linewidth]{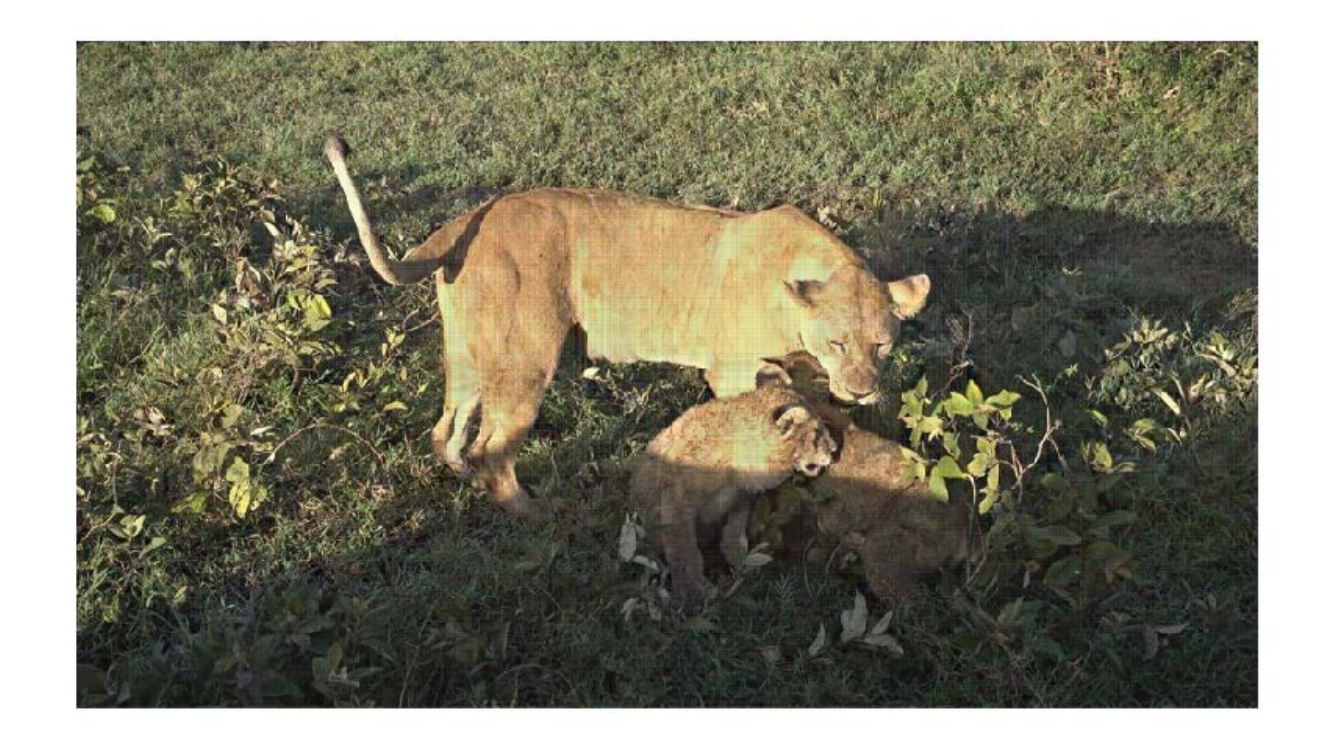}}
  \subfigure{
  \includegraphics[width=0.18\linewidth]{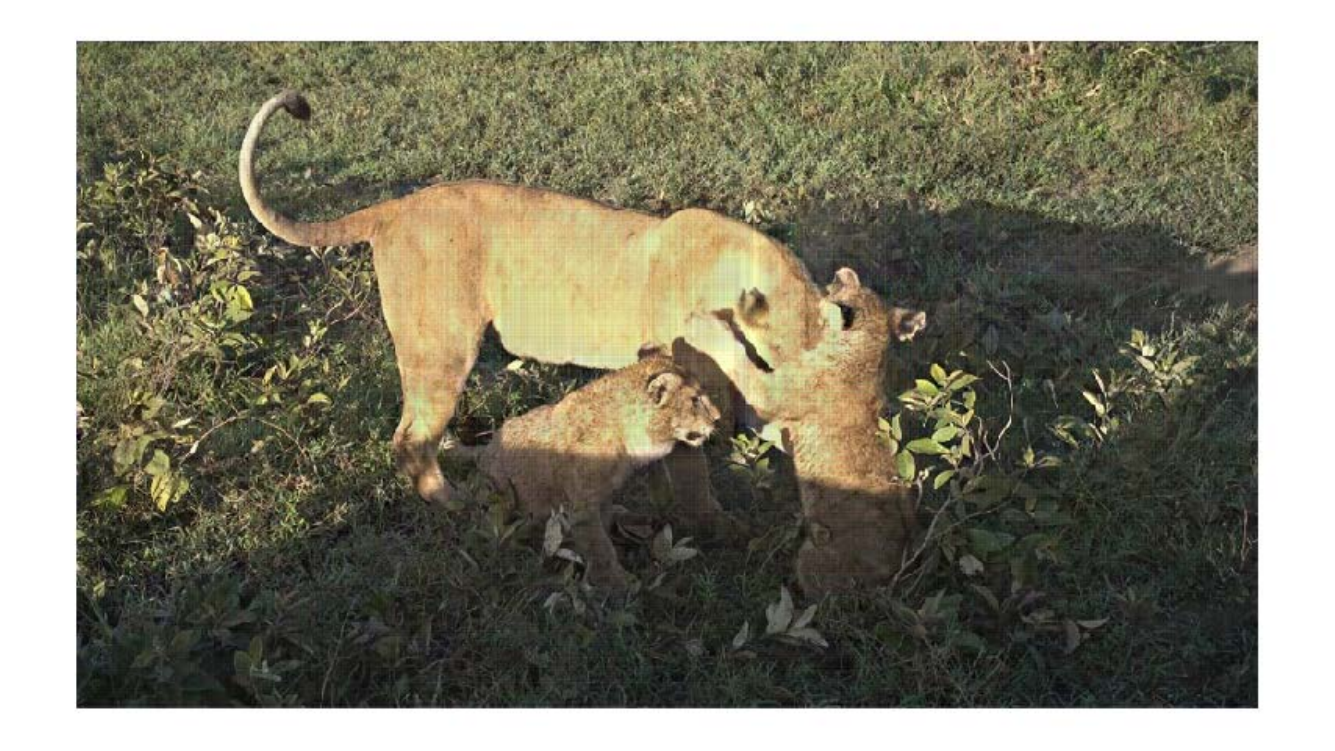}}
  \subfigure{
  \includegraphics[width=0.18\linewidth]{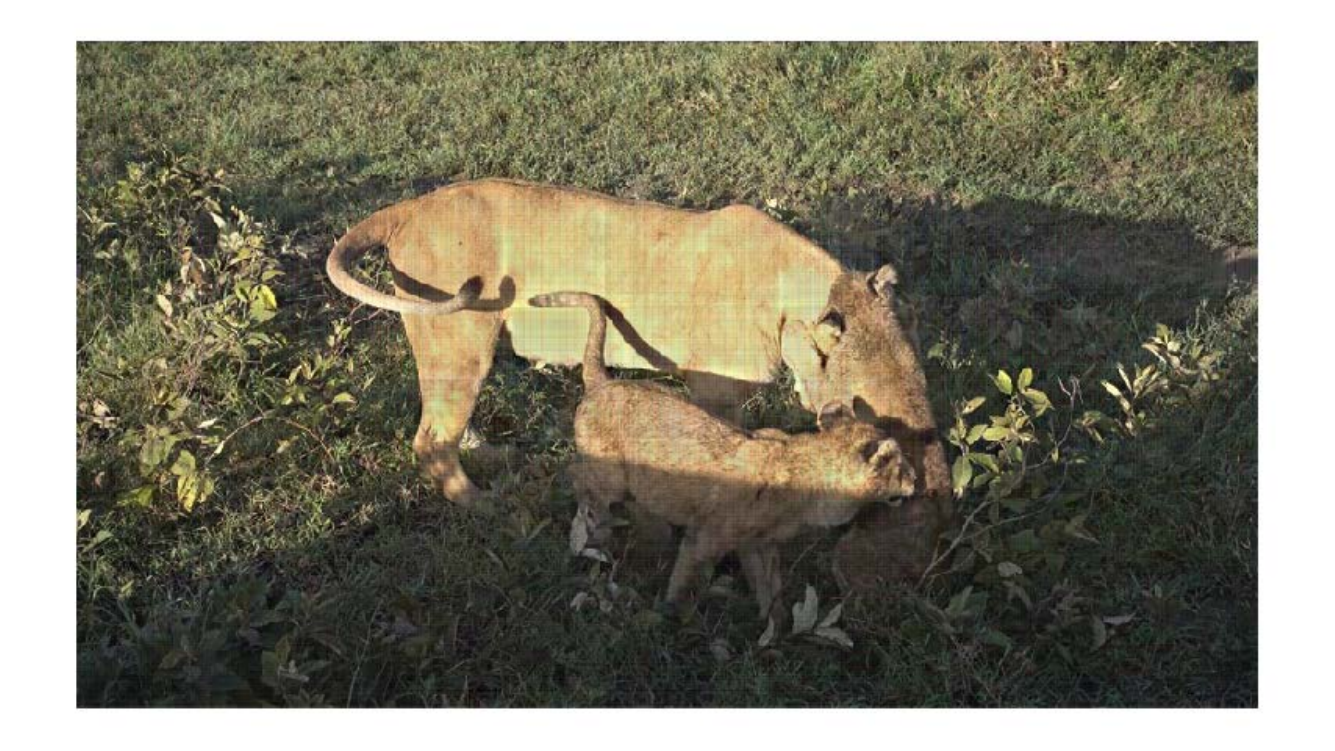}}

  \subfigure{ 
  \includegraphics[width=0.18\linewidth]{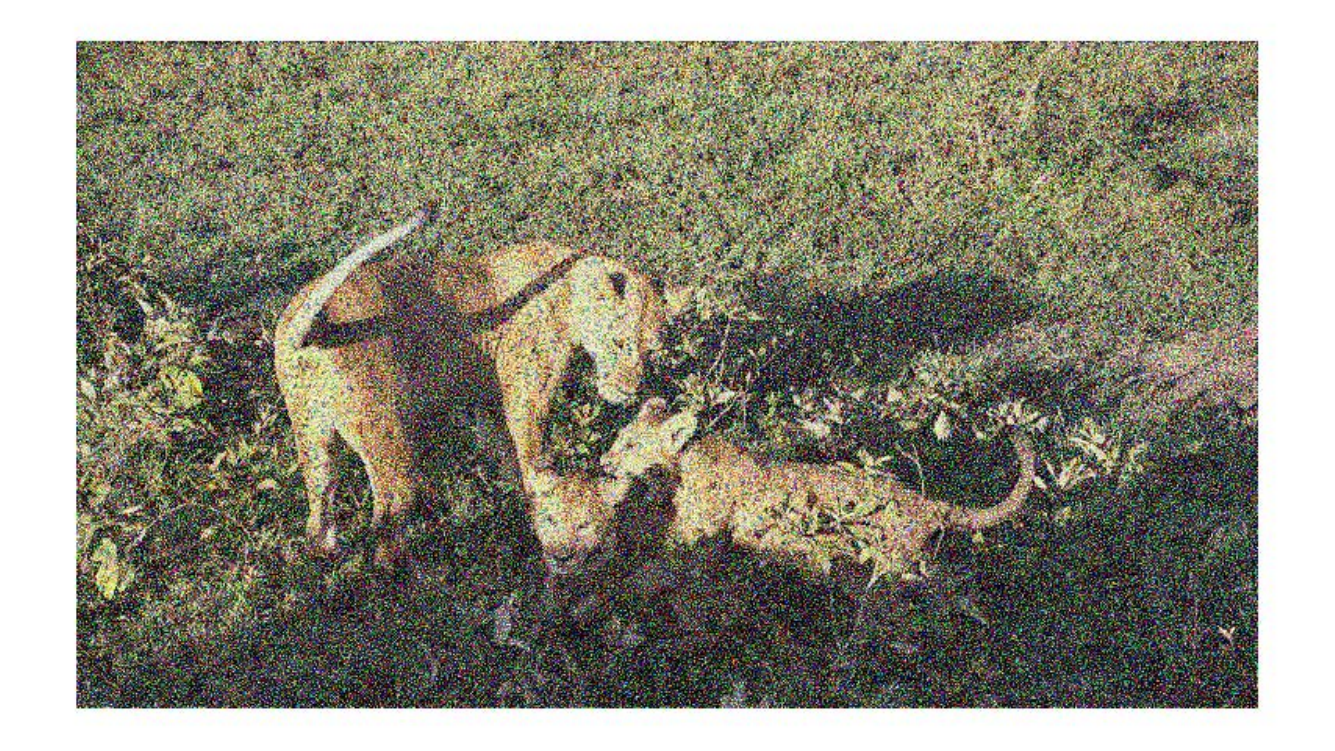}}
  \subfigure{
  \includegraphics[width=0.18\linewidth]{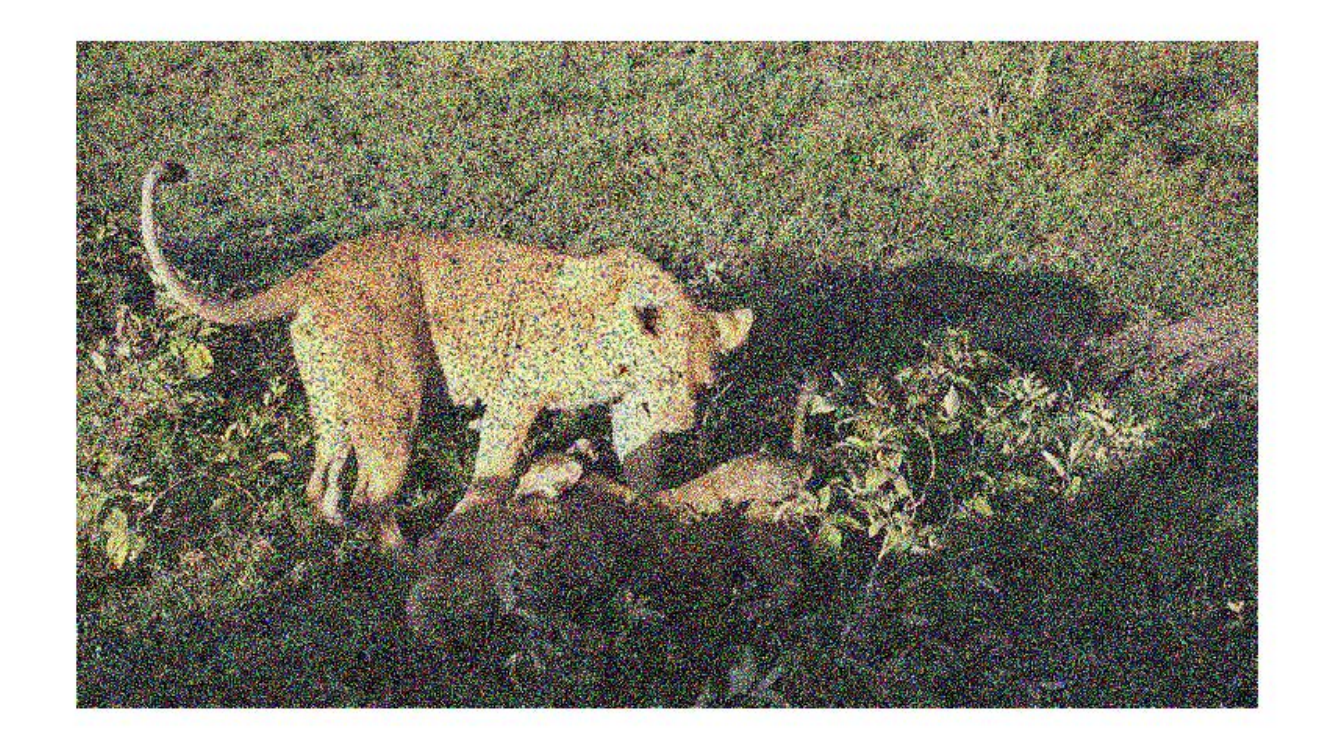}}
  \subfigure{
  \includegraphics[width=0.18\linewidth]{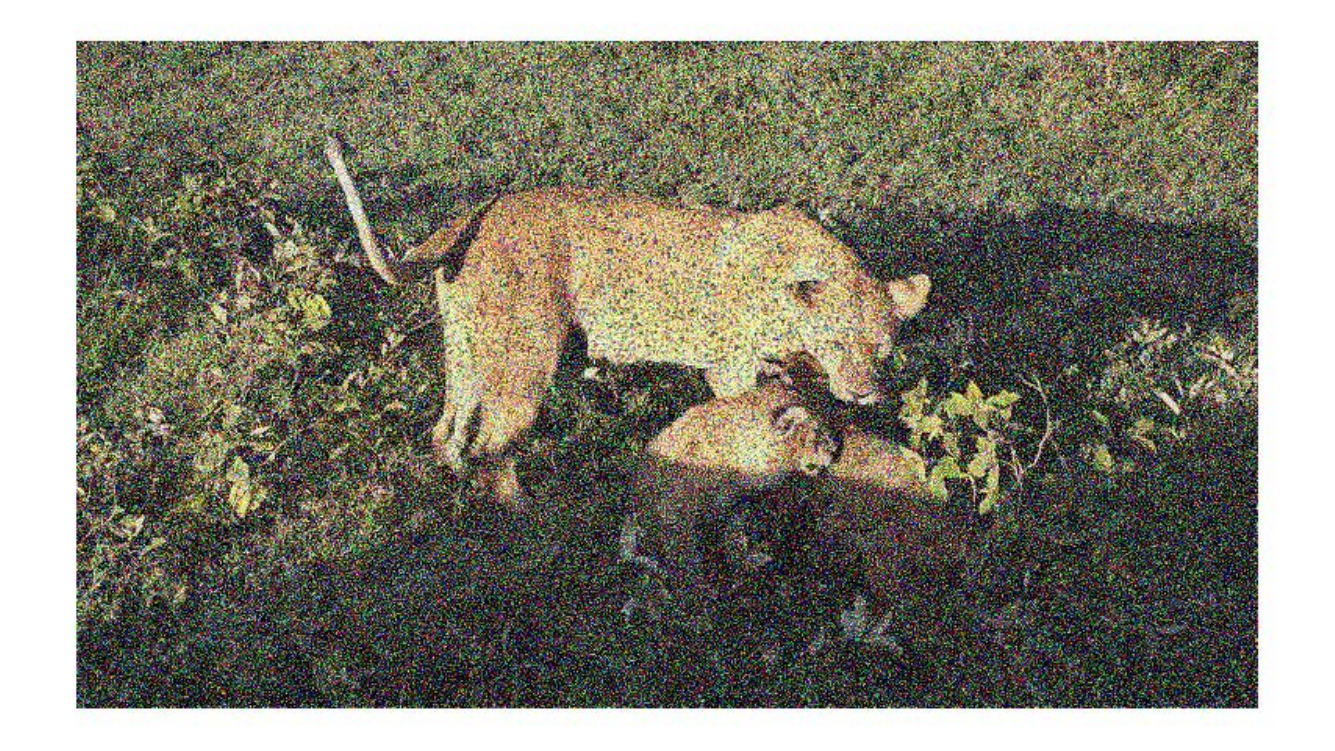}}
  \subfigure{
  \includegraphics[width=0.18\linewidth]{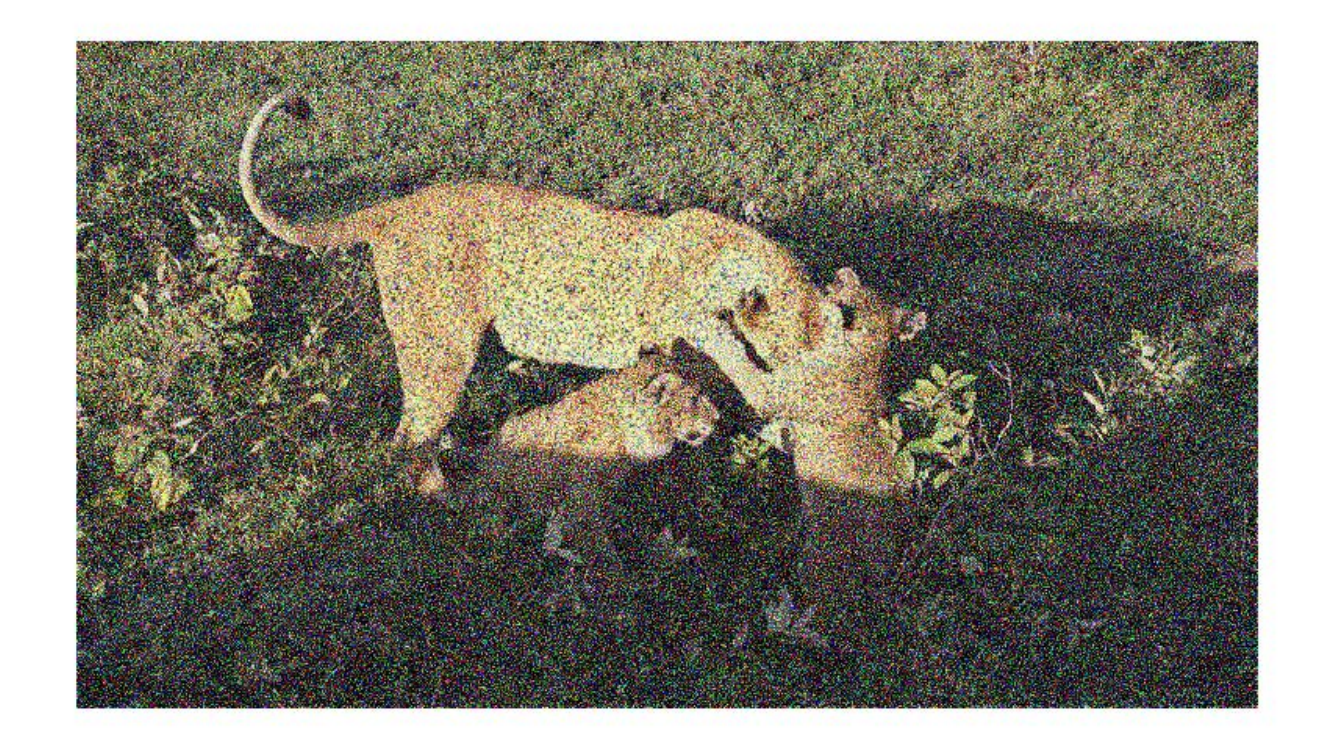}}
  \subfigure{
  \includegraphics[width=0.18\linewidth]{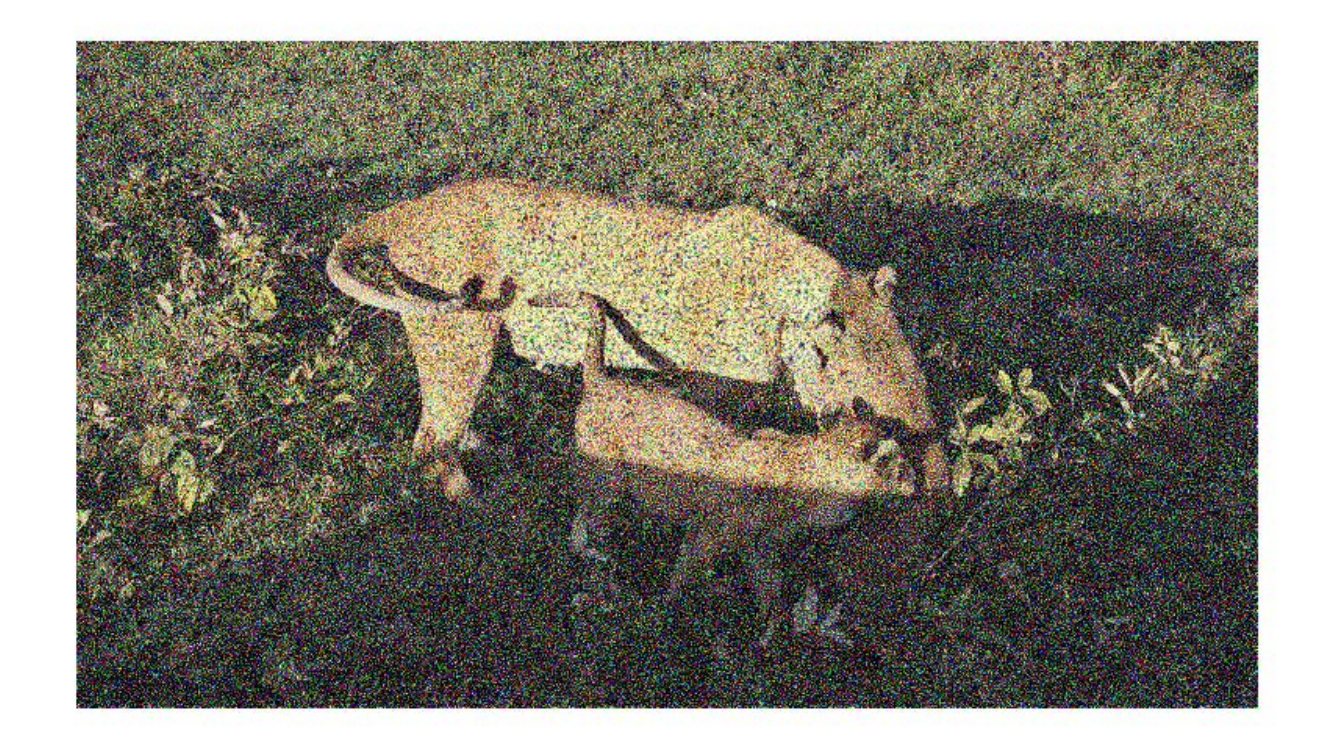}}

  \subfigure{
  \includegraphics[width=0.18\linewidth]{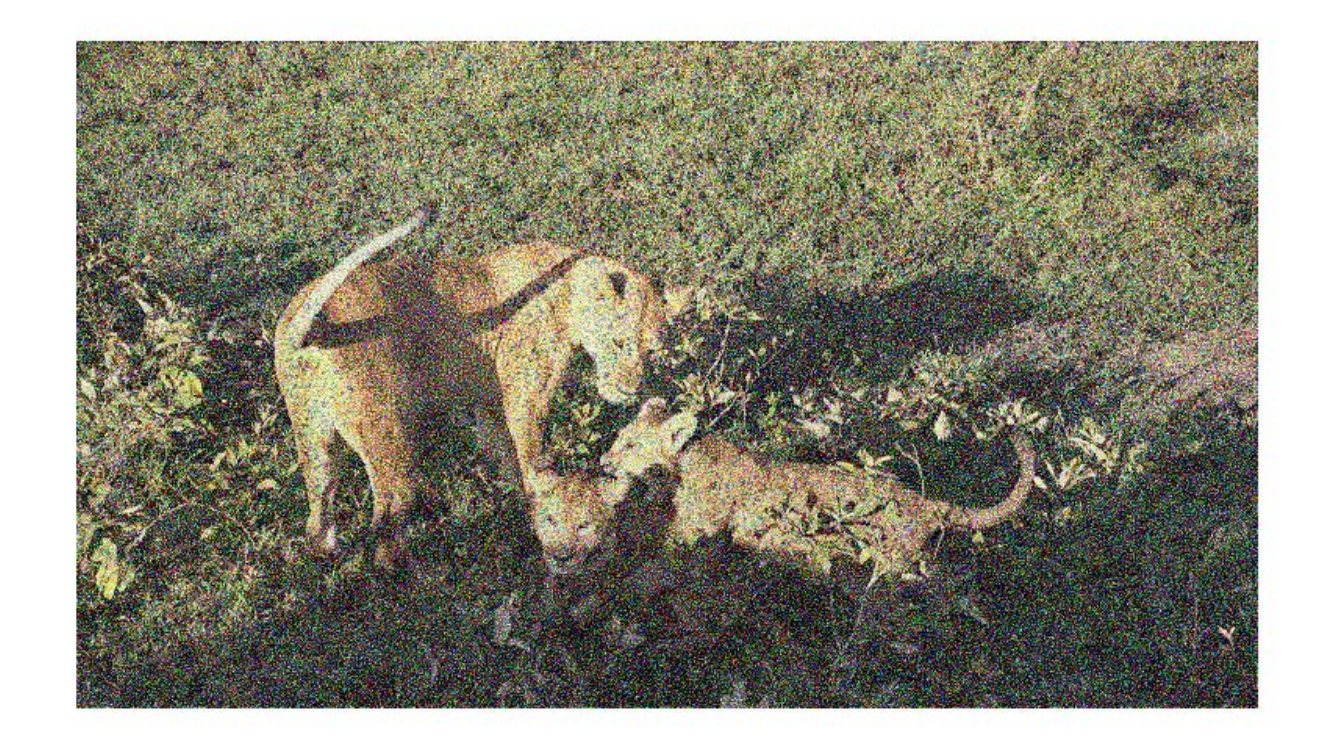}}
  \subfigure{
  \includegraphics[width=0.18\linewidth]{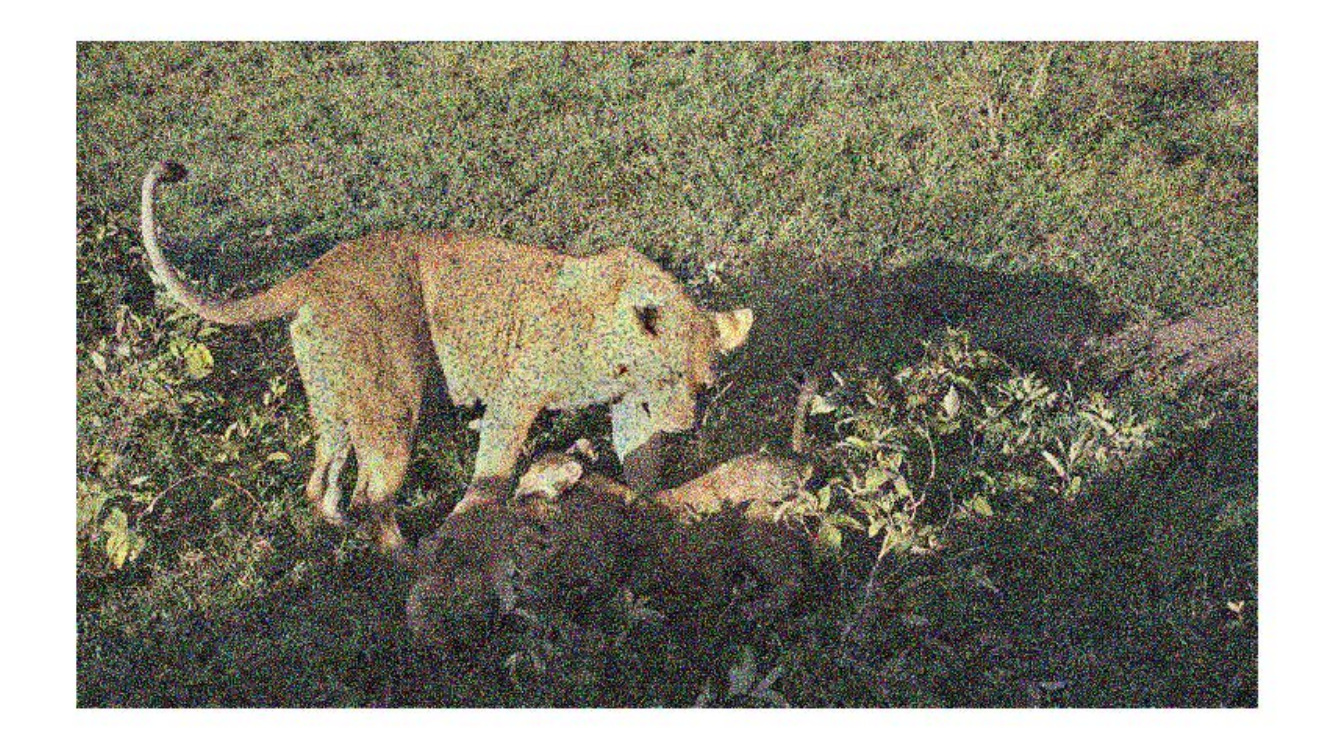}}
  \subfigure{
  \includegraphics[width=0.18\linewidth]{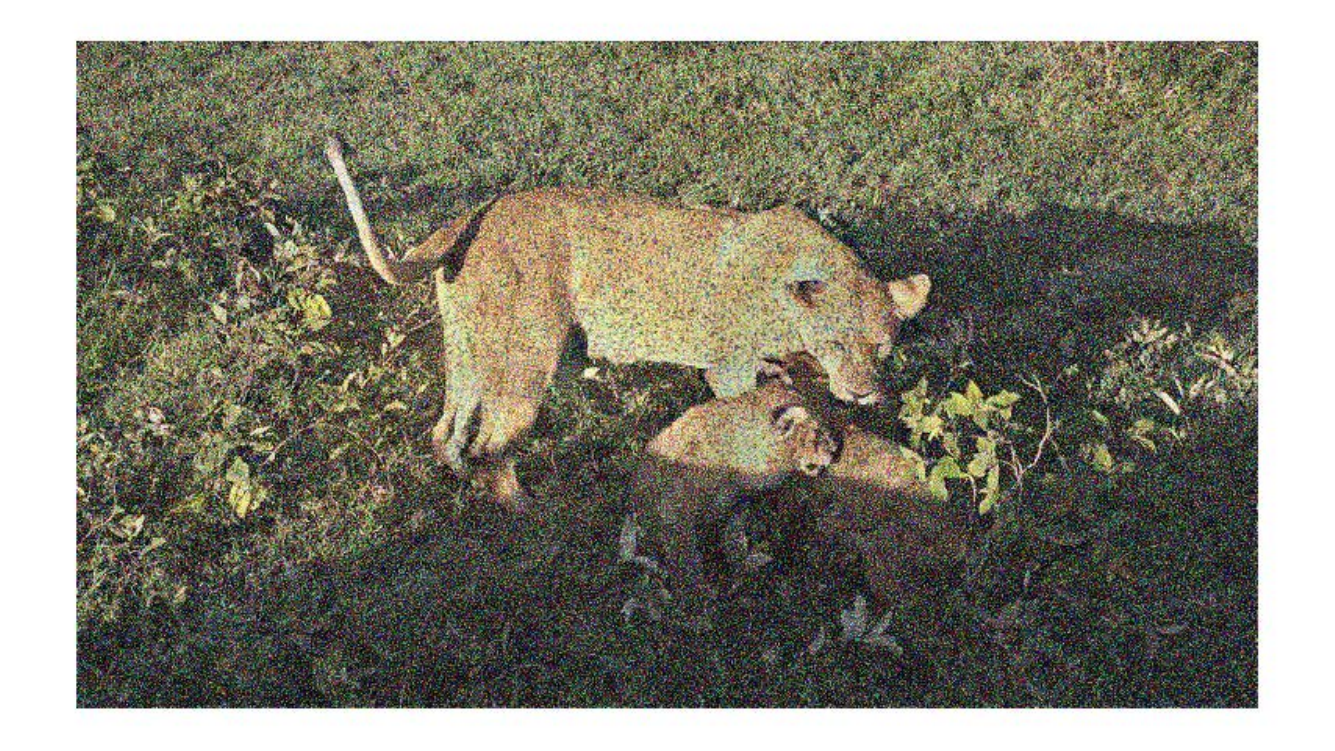}}
  \subfigure{
  \includegraphics[width=0.18\linewidth]{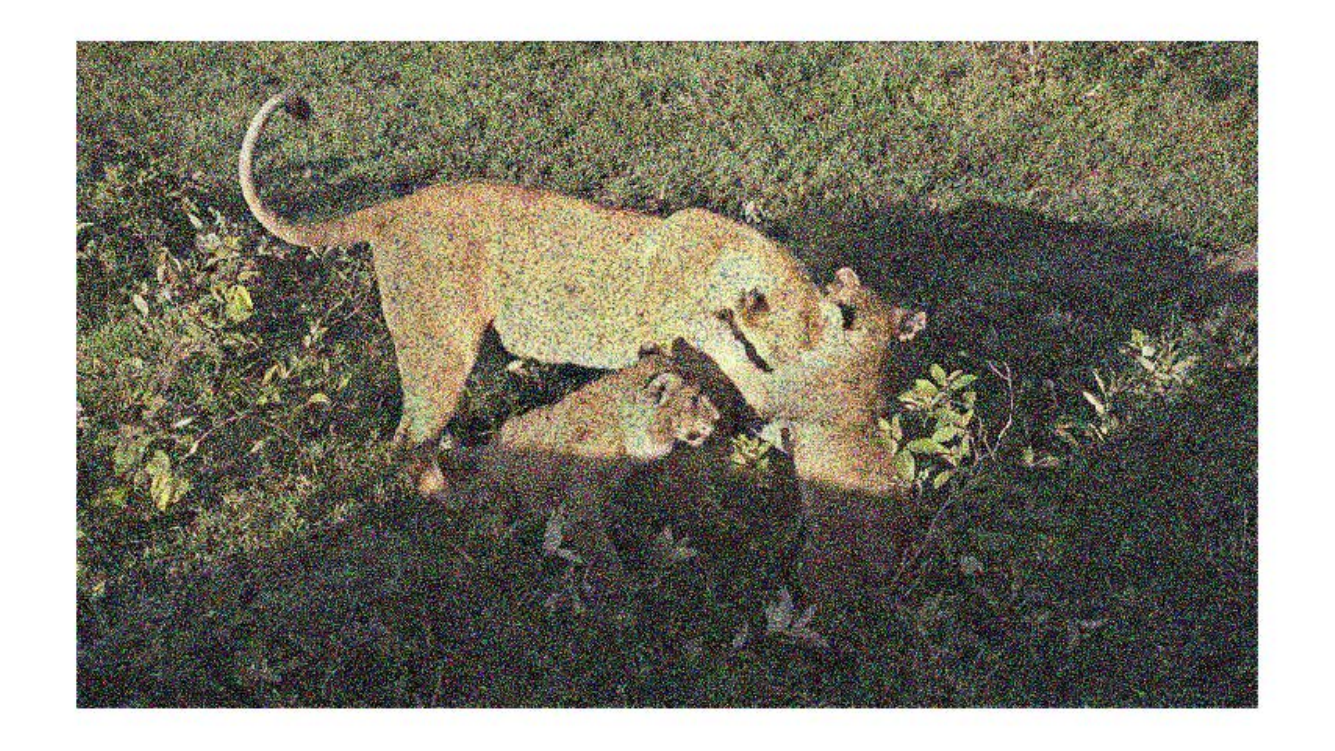}}
  \subfigure{
  \includegraphics[width=0.18\linewidth]{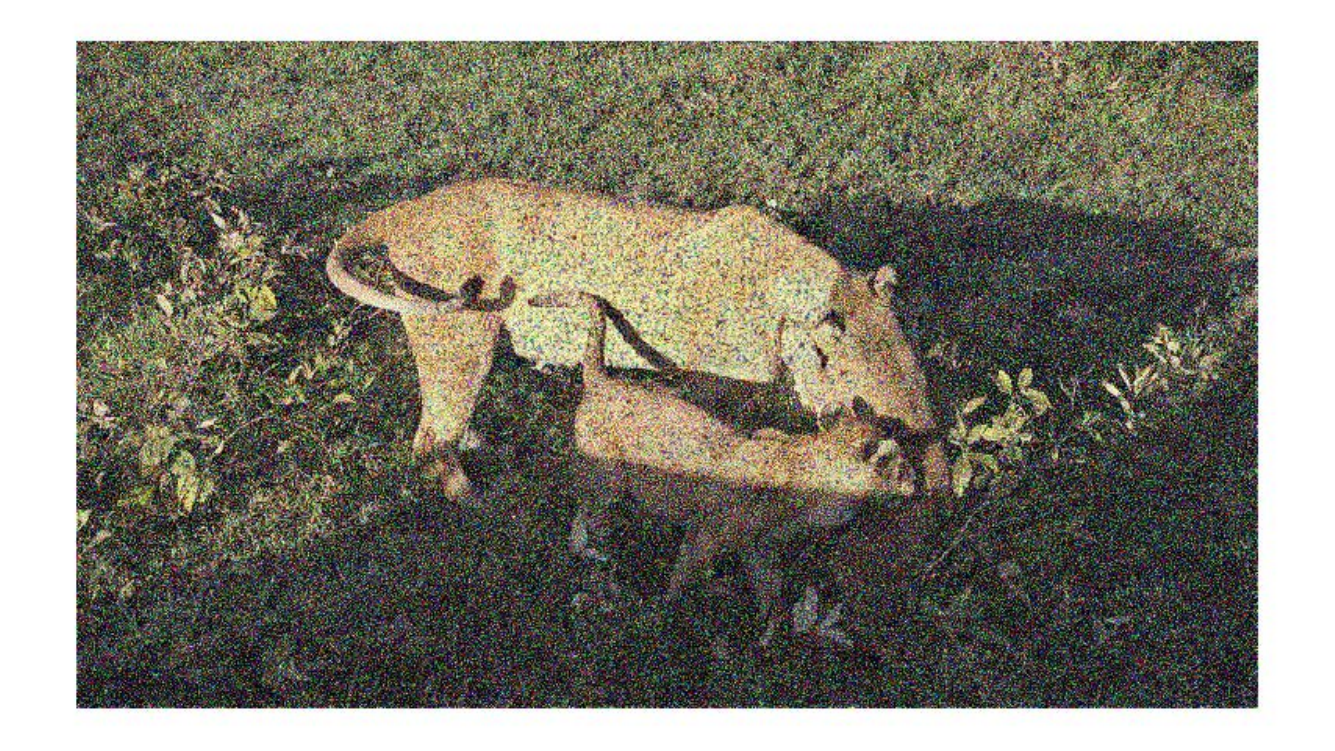}}
  \caption{The top five pictures are the $5$th, $25$th, $45$th, $65$th and $85$th original frames in the lions video, the second rows are the corresponding estimates of the proposed DLR method and the third and fourth rows are the benchmarks Static and TwoStep, respectively.}
  \label{fig_4}
  \end{figure}

\begin{table}[!ht]
     \centering
      \caption{The average MSE$^*$ for Netflix dataset.}
      \label{table:netflix}
     \begin{tabular}{c|c|c|c|c}
     \hline
         Setting & DLR & Static & TwoStep & Tensor\\
         \hline
         Filter 1 MSE$^*$& 0.781& 1.431&  1.397&1.082\\
         \hline
          Filter 2 MSE$^*$&0.796&  1.601&1.565 & 1.360\\
         \hline
     \end{tabular}
 \end{table}
  Another example is the compression and recovery of videos. We use the lion video from Davis 2017 dataset \citep{Pont-Tuset_arXiv_2017} and treat each frame as a matrix at each time point. The dataset at \href{https://davischallenge.org/davis2017/code.html\#unsupervised}{https://davischallenge.org/davis2017/code.html\#unsupervised} is publicly available. For compression, we first separate the matrices $M^0_t$ into the sparse part $S_t$ and the low-rank part $L_t$ using robust principal component analysis \citep{candes2011robust}. The design matrices $X_{ti}$ are the convolution kernel matrices with random centers $(a_{ti},b_{ti})$, i.e., $[X_{ti}]_{a_{ti},b_{ti}} = 4,[X_{ti}]_{a_{ti}\pm1,b_{ti}}=[X_{ti}]_{a_{ti},b_{ti}\pm1}=2,[X_{ti}]_{a_{ti}\pm1,b_{ti}\pm1}=1$. It is straightforward to verify that the distributions of $\{X_{ti}\}_{t,i}$ satisfy Assumption \ref{assump_distri}. Then we compress the low-rank part to obtain corresponding output $Y_{ti} = \langle X_{ti},L_t\rangle$. For recovery, we use the dynamic trace regression  to reconstruct the low-rank part, denoted as $\widetilde{L}_t^\lambda$. Then we recover each frame by $\widetilde{M}_t^\lambda = S_t +\widetilde{L}^\lambda_t$, while the sparse part $S_t$ is retained. We emphasize that we only need store the centers $(a_{ti},b_{ti})$, the output $Y_{ti}$ and the nonzero part of $S_t$ to recover the matrices $M_t^0$, which is considerably space-saving compared to storing the whole video. 
  In the lion video, each frame has $480\times 854$ pixels. We generate $Y_{ti}$ with the rate of observation samples $\rho=0.146$ and the file volume is reduced by 70\% after compression.   
  Here we use Static and TwoStep mentioned in Section \ref{sec:simu} to reconstruct $\widehat{L}_t^\lambda, \widehat{L}_t^{\text{t-s}}$ and recover $\widehat{M}^\lambda_t = S_t +\widehat{L}_t^\lambda$, $\widehat{M}^{\text{t-s}} = S_t +\widehat{L}_t^{\text{t-s}}$ as benchmarks.
  Figure \ref{fig_4} shows the original frames $M_t^0$ and the estimates $\widetilde{M}^\lambda_t, \widehat{M}_t^\lambda, \widehat{M}^{\text{t-s}}$ at different time points. The background and the smooth motions of all three lions are recovered well by the proposed DLR method, which can be seen visually better than two benchmark methods.  Recall the definition (\ref{eq:mse2}), the average MSE$^*=T^{-1}\sum_{t=1}^T \mbox{MSE}^*_t$ is $2.9\times 10^{-3}$ for our dynamic method while is $7.6\times 10^{-2}$ and $4.7\times 10^{-2}$ for the two benchmarks Static and TwoStep repsectively.

\bibliographystyle{agsm}
\bibliography{DMC}

\section{Proof of Theorem 1}\label{pf:thm_error_bound}
  
  We need the following lemma for proving Theorem 1.
 \begin{lemma}\label{lem1}
  \label{lemm_convex}
  Let $\Sigma\in \mathbb{S}_{+}^{m\times m}$ be a positive definite matrix with the smallest eigenvalue $\mu>0$. Then the following inequality
  \begin{equation}
  \label{eq_lemm_convex}
  \alpha^\top\Sigma(\alpha +\beta)\ge \mu \left(\|\alpha\|_2^2 - \kappa_{\Sigma}\|\alpha\|_2\|\beta\|_2\right)
  \end{equation}
  holds for any $\alpha,\beta \in \mathbb{R}^{m}$ 
  \end{lemma}
  \begin{pf}{\it of Lemma \ref{lem1}:}
This follows immediately from Assumption 1, since $\alpha^\top \Sigma \alpha \ge \mu \|\alpha\|_2^2$ and $\alpha^\top \Sigma \beta \ge - \kappa_{\Sigma} \mu \|\alpha\|_2 \|\beta\|_2$.
  \end{pf}

  \begin{pf}{\it of Theorem 1:}
  For brevity, we denote $$\omega_j = \omega_h(j-t),\quad \mathcal{C}_j(M,N) = \frac{1}{n_j}\sum_{i=1}^{n_j}\mathbb{E}\langle M,X_{j i}\rangle \langle N, X_{j i}\rangle$$
  for all $M,N \in \mathbb{R}_{m_1\times m_2}$.
  With the definition of $\widetilde{M}_t^{\lambda}$ ,
  \begin{equation*}
  \begin{aligned}
  & \sum_{j=1}^T \omega_j \mathcal{C}_j(\widetilde{M}_t^\lambda,\widetilde{M}_t^\lambda) -2 \sum_{j=1}^T \omega_j \left\langle \frac{1}{n_j}\sum_{i=1}^{n_j} Y_{j i}X_{j i}, \widetilde{M}_t^\lambda \right\rangle + \lambda \|\widetilde{M}_t^\lambda \|_1\\
  \le &
  \sum_{j=1}^T \omega_j \mathcal{C}_j(M,M) -2 \sum_{j=1}^T \omega_j \left\langle \frac{1}{n_j}\sum_{i=1}^{n_j} Y_{ji}X_{ji}, M \right\rangle + \lambda \|M\|_1
  \end{aligned}
  \end{equation*}
  holds for any $M \in \mathbb{R}^{m_1\times m_2}$.
  With the identity
  \begin{equation*}
  \left\langle \frac{1}{n_j}\sum_{i=1}^{n_j} Y_{ji}X_{ji}, M \right\rangle  =
  \langle \Delta_j ,M \rangle + \mathcal{C}_j(M_j^0,M),
  \end{equation*}
  we have 
  \begin{equation*}
  \begin{aligned}
  & \sum_{j= 1}^T \omega_j  \mathcal{C}_j(\widetilde{M}_t^\lambda,\widetilde{M}_t^\lambda) - 2\sum_{j=1}^T \omega_j \mathcal{C}_j(M_j^0,\widetilde{M}_t^\lambda) \\
  \le &\sum_{j= 1}^T \omega_j \mathcal{C}_j(M,M) - 2\sum_{j=1}^T \omega_j \mathcal{C}_j( M_j^0,M)   +2\sum_{j=1}^T \omega_j \langle \Delta_j,\widetilde{M}_t^\lambda - M\rangle +\lambda(\|M\|_1-\|\widetilde{M}_t^\lambda\|_1).
  \end{aligned}
  \end{equation*}
  Using $\langle \cdot,*\rangle \le \|\cdot\|_{\infty}\|*\|_{1}$, one can obtain
  \begin{equation*}
  \begin{aligned}
  &\sum_{j=1}^T \omega_j \mathcal{C}_j(\widetilde{M}_t^\lambda - M_j^0,\widetilde{M}_t^\lambda - M_j^0)\\ \le & \sum_{j = 1}^T \omega_j \mathcal{C}_j(M-M_j^0,M-M_j^0) + 2\|\sum_{j= 1}^T \omega_j \Delta_j\|_{\infty}\|\widetilde{M}_t^\lambda - M\|_1 +\lambda (\|M\|_1-\|\widetilde{M}_t^\lambda\|_1).
  \end{aligned} 
  \end{equation*}
  Under the assumption that $2\|\sum_{j=1}^T \omega_j \Delta_j\|_{\infty} \le \lambda$,  we have
  \begin{equation*}
  \sum_{j=1}^T \omega_j \mathcal{C}_j(\widetilde{M}_t^\lambda - M_j^0,\widetilde{M}_t^\lambda - M_j^0)\le \sum_{j = 1}^T \omega_j \mathcal{C}_j(M-M_j^0,M-M_j^0) + 2\lambda\|M\|_1.
  \end{equation*}
  Set $M=M_t^0$, then
  \begin{equation*}
  \sum_{j=1}^T \omega_j \mathcal{C}_j(\widetilde{M}_t^\lambda - M_j^0,\widetilde{M}_t^\lambda - M_j^0)\le \sum_{j = 1}^T \omega_j \mathcal{C}_j(M_t^0-M_j^0,M_t^0-M_j^0) + 2\lambda\|M_t^0\|_1.
  \end{equation*}
  With the identity
  \begin{align*}
  &\mathcal{C}_j (\widetilde{M}_t^\lambda - M_j^0,\widetilde{M}_t^\lambda - M_j^0) \\=&  \mathcal{C}_j(\widetilde{M}_t^\lambda - M_t^0,\widetilde{M}_t^\lambda - M_t^0) + \mathcal{C}_j(M_t^0-M_j^0,M_t^0-M_j^0) + 2\mathcal{C}_j(\widetilde{M}_t^\lambda-M_t^0,M_t^0-M_j^0),
  \end{align*}
  we have
  \begin{equation}\label{ineq:general}
  \sum_{j=1}^T \omega_j \mathcal{C}_j(\widetilde{M}_t^\lambda - M_t^0,\widetilde{M}_t^\lambda - M_t^0) +2\sum_{j=1}^T \omega_j \mathcal{C}_j(\widetilde{M}_t^\lambda-M_t^0,M_t^0-M_j^0)\le 2\lambda\|M_t^0\|_1.
  \end{equation}
  If $\left\|\widetilde{M}_t^\lambda -M_t^0\right\|_2 < 2 \left\|M_t^0-\sum_{j=1}^T \omega_j M_j^0\right\|_2$,
  we can maintain the bound 
  \begin{equation}\label{ineq:general_firstterm_left}
  \left\|\widetilde{M}_t^\lambda -M_t^0\right\|_2 < 2\delta_hM_t.
  \end{equation}
  If $\left\|\widetilde{M}_t^\lambda -M_t^0\right\|_2 \ge 2 \left\|M_t^0-\sum_{j=1}^T \omega_j M_j^0\right\|_2$,
  with Assumption 1, we know that 
  \begin{equation}
  \label{eq_proof_C}
  \mathcal{C}_j(M,N)=\myvec({M})^\top \Sigma \myvec({N}).
  \end{equation}
  So \eqref{ineq:general} can be expressed as
  \begin{align*}
 & \left(\myvec({\widehat{M}}_t^\lambda) - \myvec({M}_t^0)\right)^\top \Sigma \left[ \left(\myvec({\widehat{M}}_t^\lambda) - \myvec({M}_t^0)\right) + \left(2 \myvec({M}_t^0) - 2\sum_{j=1}^T \omega_j \myvec({M}_j^0)\right)\right]\\ \le& 2\lambda \|M_t^0\|_1.
  \end{align*}
  From Lemma \ref{lemm_convex}, 
  \begin{equation}
    \label{eq:homo1}
  \begin{aligned}
  & \left(\myvec({\widehat{M}}_t^\lambda) - \myvec({M}_t^0)\right)^\top \Sigma \left[ \left(\myvec({\widehat{M}}_t^\lambda)- \myvec({M}_t^0)\right) + \left(2 \myvec({M}_t^0) - 2\sum_{j=1}^T \omega_j \myvec({M}_j^0)\right)\right]\\ \ge& 
  \mu\left\|\widetilde{M}_t^\lambda - M_t^0\right\|_2^2 - 2\mu\kappa_{\Sigma}\left\|\widetilde{M}_t^\lambda - M_t^0\right\|_2\left\|M_t^0 - \sum_{j=1}^T \omega_j M_j^0\right\|_2.
  \end{aligned}
  \end{equation}
  Thus we have
  \begin{equation*}
  \mu\left\|\widetilde{M}_t^\lambda - M_t^0\right\|_2^2 - 2\mu\kappa_{\Sigma} \left\|\widetilde{M}_t^\lambda - M_t^0\right\|_2\left\|M_t^0 - \sum_{j=1}^T \omega_j M_j^0\right\|_2 - 2\lambda \|M_t^0\|_1 \le 0.
  \end{equation*}
  By solving this inequality, we have 
  \begin{equation}\label{ineq:general_firstterm_right}
  \left\|\widetilde{M}_t^\lambda -M_t^0\right\|_2 \le \frac{4\mu\kappa_{\Sigma}\delta_hM_t+\sqrt{\mu^2(\delta_h M_t)^2+8\mu\lambda \|M_t^0\|_1}}{2\mu}\lesssim \delta_hM_t+\sqrt{\frac{2\lambda \|M_t^0\|_1}{\mu}}.
  \end{equation}
  Combining \eqref{ineq:general_firstterm_left} and \eqref{ineq:general_firstterm_right},
  \begin{equation}\label{ineq:general_firstterm}
  \left\|\widetilde{M}_t^\lambda -M_t^0\right\|_2 \lesssim \delta_h M_t+\max\left\{\delta_h M_t,\sqrt{\frac{2\lambda \|M_t^0\|_1}{\mu}}\right\}\le \delta_h M_t + \left((\delta_h M_t)^2+ 2\mu^{-1}\lambda \|M_t^0\|_1 \right)^{1/2}.
  \end{equation}
  Because $\widetilde{M}_t^\lambda$ is the minimizer of (6), there exists a sub-gradient matrix $B$ satisfying that for all $M\in \mathbb{R}^{m_1\times m_2}$,
  $$\left\langle B, \widetilde{M}_t^\lambda- M\right\rangle \le 0,$$
  which equals to that there exists $\widehat{V}_t\in\partial \|\widetilde{M}_t^\lambda \|_1$ satisfying that
  \begin{equation*}
  2\sum_{j=1}^T \omega_j\left( \mathcal{C}_j ( \widetilde{M}_t^\lambda, \widetilde{M}_t^\lambda-M) - \left\langle\frac{1}{n_j}\sum_{i=1}^{n_j}Y_{ji}X_{ji},\widetilde{M}_t^\lambda-M\right\rangle\right) +\lambda \langle \widehat{V}_t,\widetilde{M}_t^\lambda -M\rangle \le 0.
  \end{equation*}
  It is equivalent to 
  \begin{equation*}
  \begin{aligned}
  & 2 \sum_{j=1}^T \omega_j \mathcal{C}_j(\widetilde{M}_t^\lambda - M_j^0, \widetilde{M}_t^\lambda - M) +\lambda \langle \widehat{V}_t - V,\widetilde{M}_t^\lambda - M\rangle\\ \le &
  -\lambda \langle V,\widetilde{M}_t^\lambda-M\rangle  + 2\sum_{j=1}^T\omega_j \langle \Delta_j, \widetilde{M}_t^\lambda-M\rangle ,
  \end{aligned}
  \end{equation*}
  where $V \in\partial\|M\|_1$.\\
  Denote that $M = \sum_{j=1}^r \sigma_j u_jv_j^\top$ with support $S_1,S_2$. $V$ have representation from \cite{waston_1992}:
  $$V = \sum_{j =1}^r u_jv_j^\top + P_{S_1^\perp}WP_{S_2^\perp},\ \|W\|_{\infty}\le 1.$$
  We choose a $W$ subject to 
  $$\langle P_{S_1^\perp} WP_{S_2^\perp}, \widetilde{M}_t^\lambda-M\rangle =  \langle W, P_{S_1^\perp} \widehat{M}_{t}^\lambda P_{S_2^\perp}\rangle = \|P_{S_1^\perp}\widehat{M}_{t}^\lambda P_{S_2^\perp}\|_1.$$
  With the monotonicity of sub-gradients, we have $\langle \widehat{V}_t -V,\widetilde{M}_t^\lambda-M\rangle \ge 0 $. So
  \begin{equation*}
  \begin{aligned}
  & 2 \sum_{j=1}^T \omega_j \mathcal{C}_j( \widetilde{M}_t^\lambda - M_j^0, \widetilde{M}_t^\lambda - M) +\lambda \|P_{S_1^\perp}\widetilde{M}_t^\lambda P_{S_2^\perp}\|_1\\ \le &
  -\lambda \langle \sum_{j=1}^r u_jv_j^\top,\widetilde{M}_t^\lambda-M\rangle  + 2\sum_{j=1}^T \omega_j \langle \Delta_j, \widetilde{M}_t^\lambda-M\rangle.
  \end{aligned}
  \end{equation*}
  Using the identity
  \begin{equation*}
  \begin{aligned}
  & 2 \sum_{j=1}^T \omega_j\mathcal{C}_j(\widetilde{M}_t^\lambda - M_j^0, \widetilde{M}_t^\lambda - M) \\= &\sum_{j=1}^T \omega_j\left(\mathcal{C}_j(\widetilde{M}_t^\lambda - M_j^0,\widetilde{M}_t^\lambda - M_j^0) +\mathcal{C}_j(\widetilde{M}_t^\lambda - M,\widetilde{M}_t^\lambda - M)  - \mathcal{C}_j(M-M_{j}^0,M-M_{j}^0)\right)
  \end{aligned}
  \end{equation*}
  and the inequality 
  $$-\langle \sum_{j=1}^r u_jv_j^\top,\widetilde{M}_t^\lambda - M\rangle =- \langle \sum_{j=1}^r u_jv_j^\top,P_{S_1}(\widetilde{M}_t^\lambda - M)P_{S_2}\rangle \le \|P_{S_1}(\widetilde{M}_t^\lambda-M)P_{S_2}\|_1,$$
  we have
  \begin{equation}
  \begin{aligned}
  \label{eq_proof_error_bound_1}
  & \sum_{j=1}^T \omega_j\mathcal{C}_j(\widetilde{M}_t^\lambda - M_j^0,\widetilde{M}_t^\lambda - M_j^0) +  \sum_{j=1}^T \omega_j\mathcal{C}_j(\widetilde{M}_t^\lambda - M,\widetilde{M}_t^\lambda - M)  + \lambda\|P_{S_1^\perp}\widetilde{M}_t^\lambda P_{S_2^\perp}\|_1 \\\le 
  & 
  \sum_{j=1}^T \omega_j \mathcal{C}_j(M-M_{j}^0,M-M_{j}^0) + \lambda \|P_{S_1}(\widetilde{M}_t^\lambda-M)P_{S_2}\|_1 +2\sum_{j=1}^T\omega_j \langle \Delta_j, \widetilde{M}_t^\lambda-M\rangle .
  \end{aligned}
  \end{equation}
  With identity
  \begin{equation*}
  \begin{aligned}
  \langle \Delta_j,\widetilde{M}_t^\lambda-M\rangle& = \langle P_M(\Delta_j),\widehat{M}^\lambda_t-M\rangle + \langle P_{S_1^\perp}\Delta_j P_{S_2^\perp}, \widehat{M}^\lambda_t - M\rangle \\
  & = \langle P_M(\Delta_j),\widehat{M}^\lambda_t-M\rangle + \langle P_{S_1^\perp}\Delta_j P_{S_2^\perp}, \widehat{M}^\lambda_t\rangle,
  \end{aligned}
  \end{equation*}
  we have that 
  \begin{align*}
  & \sum_{j=1}^T \omega_j \langle \Delta_j,\widetilde{M}_t^\lambda-M\rangle\\=&\left\langle P_M\left(\sum_{j=1}^T \omega_j\Delta_j\right),\widehat{M}^\lambda_t-M\right\rangle + \left\langle P_{S_1^\perp}\left(\sum_{j=1}^T \omega_j\Delta_j\right) P_{S_2^\perp}, \widehat{M}^\lambda_t\right\rangle\\\le&
  \left\|P_M\left(\sum_{j=1}^T\omega_j\Delta_j\right)\right\|_2\left\|\widetilde{M}_t^\lambda-M\right\|_2 + \left\|P_{S_1^\perp}\left(\sum_{j=1}^T\omega_j\Delta_j\right) P_{S_2^\perp}\right\|_\infty \left\|P_{S_1^\perp}\widetilde{M}_t^\lambda P_{S_2^\perp}\right\|_1.
  \end{align*}
  With
  \begin{equation*}
  \begin{aligned}
  \left\|P_M\left(\sum_{j=1}^T\omega_j\Delta_j\right)\right\|_2 & \le \sqrt{\rank\left[P_A\left(\sum_{j=1}^T\omega_j\Delta_j\right)\right]}\left\|P_M\left(\sum_{j=1}^T\omega_j\Delta_j\right)\right\|_\infty \\&\le
  \sqrt{2\rank(M)} \|\sum_{j=1}^T\omega_j\Delta_j\|_{\infty}\ 
  \end{aligned}
  \end{equation*}
  and 
  \begin{equation*}
  \left\|P_{S_1^\perp}\left(\sum_{j=1}^T \omega_j\Delta_j\right) P_{S_2^\perp}\right\|_\infty\le \|\sum_{j=1}^T\omega_j\Delta_j\|_{\infty},
  \end{equation*}
  we have
  \begin{equation}
  \label{eq_proof_error_bound_2}
  \sum_{j=1}^T \omega_j\langle \Delta_j,\widetilde{M}_t^\lambda - M\rangle \le  \sqrt{2\rank(M)}\|\sum_{j=1}^T \omega_j\Delta_j\|_{\infty}\|\widetilde{M}_t^\lambda - M\|_2 + \|\sum_{j=1}^T \omega_j\Delta_j\|_{\infty}\|P_{S_1^\perp}\widetilde{M}_t^\lambda P_{S_2^\perp}\|_1.
  \end{equation}
  Also we have 
  \begin{equation}
  \label{eq_proof_error_bound_3}
  \begin{aligned}
  \|P_{S_1}(\widetilde{M}_t^\lambda-M)P_{S_2}\|_1 & \le \sqrt{\rank(M)}\|P_{S_1}(\widetilde{M}_t^\lambda-M)P_{S_2}\|_2
  \\ & \le \sqrt{\rank(M)}\|\widetilde{M}_t^\lambda- M\|_2.
  \end{aligned}
  \end{equation}
  With (\ref{eq_proof_error_bound_1}), (\ref{eq_proof_error_bound_2}) and (\ref{eq_proof_error_bound_3}),
  \begin{equation*}
  \begin{aligned}
  & \sum_{j=1}^T \omega_j\left[\mathcal{C}_j (\widetilde{M}_t^\lambda - M_j^0,\widetilde{M}_t^\lambda - M_j^0) +  \mathcal{C}_j(\widetilde{M}_t^\lambda - M,\widetilde{M}_t^\lambda - M)\right] \\+& \left(\lambda-2 \|\sum_{j=1}^T \omega_j\Delta_j\|_{\infty}\right)\|P_{S_1^\perp}\widetilde{M}_t^\lambda P_{S_2^\perp}\|_1 \\\le 
  & 
  \sum_{j=1}^T \omega_j \mathcal{C}_j(M-M_{j}^0,M-M_{j}^0) + \sqrt{\rank(M)}\left(\lambda + 2\sqrt{2}\|\sum_{j=1}^T\omega_j\Delta_j\|_{\infty}\right)\|\widetilde{M}_t^\lambda -M\|_2.
  \end{aligned}
  \end{equation*}
  With $\lambda \ge 2\|\sum_{j=1}^T \omega_j\Delta_j\|_{\infty}$, 
  \begin{equation*}
  \begin{aligned}
  & \sum_{j=1}^T \omega_j\mathcal{C}_j(\widetilde{M}_t^\lambda - M_j^0,\widetilde{M}_t^\lambda - M_j^0) +  \sum_{j=1}^T \omega_j\mathcal{C}_j(\widetilde{M}_t^\lambda - M,\widetilde{M}_t^\lambda - M) \\\le 
  & 
  \sum_{j=1}^T \omega_j \mathcal{C}_j(M-M_{j}^0,M-M_{j}^0) + \lambda(1+\sqrt{2})\sqrt{\rank(M)}\|\widetilde{M}_t^\lambda -M\|_2.
  \end{aligned}
  \end{equation*}
  Set $M=M_t^0$, so 
  \begin{equation*}
  \begin{aligned}
  & \sum_{j=1}^T \omega_j\mathcal{C}_j(\widetilde{M}_t^\lambda - M_j^0,\widetilde{M}_t^\lambda - M_j^0) +  \sum_{j=1}^T \omega_j\mathcal{C}_j(\widetilde{M}_t^\lambda - M_t^0,\widetilde{M}_t^\lambda - M_t^0) \\\le 
  & 
  \sum_{j=1}^T \omega_j \mathcal{C}_j(M_t^0-M_{j}^0,M_t^0-M_{j}^0) + \lambda(1+\sqrt{2})\sqrt{\rank(M)}\|\widetilde{M}_t^\lambda -M_t^0\|_2.
  \end{aligned}
  \end{equation*}
  With the identity
  \begin{equation*}
  \mathcal{C}_j(\widetilde{M}_t^\lambda - M_j^0,\widetilde{M}_t^\lambda - M_j^0) = \mathcal{C}_j(\widetilde{M}_t^\lambda - M_t^0) +2\mathcal{C}_j(\widetilde{M}_t^\lambda-M^0_t,M_t^0 - M_j^0) +\mathcal{C}_j(M_t^0-M_j^0,M_t^0-M_j^0),
  \end{equation*}
  we have 
  \begin{align*}
 & \sum_{j=1}^T \omega_j\left[\mathcal{C}_j(\widetilde{M}_t^\lambda - M_t^0,\widetilde{M}_t^\lambda - M_t^0) +\mathcal{C}_j(\widetilde{M}_t^\lambda-M^0_t,M_t^0 - M_j^0)\right]\\\le& 
  \frac{1+\sqrt{2}}{2} \lambda\sqrt{\rank(M_t^0)}\|\widetilde{M}_t^\lambda -M_t^0\|_2.
  \end{align*}
  If $\|\widetilde{M}_t^\lambda -M_t^0\|_2 < \left\|M_t^0-\sum_{j=1}^T \omega_j M_j^0\right\|_2$, we maintain the bound
  \begin{equation}
  \label{ineq:general_secondterm_left}
  \|\widetilde{M}_t^\lambda -M_t^0\|_2 < \delta_hM_t.
  \end{equation}
  If $\|\widetilde{M}_t^\lambda -M_t^0\|_2 \ge  \left\|M_t^0-\sum_{j=1}^T \omega_j M_j^0\right\|_2$,
  using (\ref{eq_proof_C}) and Lemma \ref{lemm_convex}, we know that
  \begin{equation}
  \begin{aligned}
  \label{eq:homo2}
  &\sum_{j=1}^T \omega_j\left[\mathcal{C}_j(\widetilde{M}_t^\lambda - M_t^0,\widetilde{M}_t^\lambda - M_t^0) +\mathcal{C}_j(\widetilde{M}_t^\lambda-M^0_t,M_t^0 - M_j^0)\right]
  \\=&
  \left(\myvec({\widehat{M}}_t^\lambda) - \myvec({M}_t^0)\right)^\top \Sigma \left(  \left(\myvec({\widehat{M}}_t^\lambda) - \myvec({M}_t^0)\right) + \left(M_t^0 - \sum_{j=1}^T\omega_jM_j^0\right)\right)\\\ge &
  \mu\left\|\widetilde{M}_t^\lambda - M_t^0\right\|_2^2 -  \mu\kappa_{\Sigma} \left\|\widetilde{M}_t^\lambda - M_t^0\right\|_2\left\|M_t^0 - \sum_{j=1}^T \omega_j M_j^0\right\|_2.
  \end{aligned}
  \end{equation}
  So we have
  \begin{equation}
  \label{ineq:general_secondterm_right}
  \mu\left\|\widetilde{M}_t^\lambda - M_t^0\right\|_2^2 -  \mu\kappa_{\Sigma} \left\|\widetilde{M}_t^\lambda - M_t^0\right\|_2\left\|M_t^0 - \sum_{j=1}^T \omega_j M_j^0\right\|_2 \le \frac{1+\sqrt{2}}{2}\lambda \sqrt{\rank(M_t^0)}\left\|\widetilde{M}_t^\lambda - M_t^0\right\|_2.
  \end{equation}
  Combining \eqref{ineq:general_secondterm_left} and \eqref{ineq:general_secondterm_right}, we have 
  \begin{equation}
  \label{ineq:general_secondterm}
  \left\|\widetilde{M}_t^\lambda - M_t^0\right\|_2\lesssim \delta_hM_t+\frac{1+\sqrt{2}}{2} \frac{\lambda}{\mu}\sqrt{r_t},
  \end{equation}
  where $r_t=\rank(M_t^0)$.
  With \eqref{ineq:general_firstterm} and \eqref{ineq:general_secondterm}, the proof is completed.
  
\end{pf}

\section{Proofs of Theorem 2 and Corollary 1, 2}
 \begin{prop}\label{prop:bias}
  Under Assumption 2 and 3, if $h\rightarrow0$ and $(m_1m_2)^{1/2}Th^2\rightarrow\infty$ as $m_1,m_2,T\rightarrow\infty$, then
  \begin{equation*}
  \delta_h M_t\le\frac{1}{2} \alpha(K)(m_1m_2)^{1/2}D_2 h^2+o\left((m_1m_2)^{1/2}h^2\right).
  \end{equation*}
  \end{prop}
  \begin{pf}{\it of Proposition \ref{prop:bias}:} 
  Note that $T^{-1}\ll (m_1m_2)^{-1/2}h^2$.
  The conclusion follows from the classical nonparametric results (of fixed design) that
  \begin{align*}
  \left\|M_t^0 - \sum_{j=1}^T \omega_j M_j^0\right\|_2 
  &= \left\|\sum_{j=1}^T \frac{1}{Th}K\left(\frac{j-t}{Th}\right) \left\{M\left(\frac{j}{T}\right)-  M\left(\frac{t}{T}\right)\right\}\right\|_2\\ 
  &= \left\|\int\frac{1}{h}K\left(\frac{u-t/T}{h}\right)\left\{M(u)-M\left(\frac{t}{T}\right)\right\}du \right\|_2+O\left(T^{-1}\right)\\ 
  &=\left\|\int K(z)\left\{M\left(zh+\frac{t}{T}\right)-M\left(\frac{t}{T}\right)\right\}dz\right\|_2+O\left(T^{-1}\right)\\ 
  &=\frac{1}{2}\alpha(K)\left\|\nabla^2M(\frac{t}{T})\right\|_2  h^2+o\left((m_1m_2)^{1/2}h^2\right)\\ 
  &\le \frac{1}{2} \alpha(K)(m_1m_2)^{1/2}D_2 h^2+o\left((m_1m_2)^{1/2}h^2\right).
  \end{align*}
  \end{pf}  

  Now we prove two lemmas  which are needed to finish the proofs Lemma \ref{lem_lambda_indpdt} and Theorem 2. For brevity, define that $\|\cdot\|\overset{\triangle}{=}\|\cdot\|_\infty$. 
  We extend proposition 2 in \cite{koltchinskii2011neumann} to obtain the following Lemma \ref{bernstein_lemma}.
  \begin{lemma}
  \label{bernstein_lemma}
  Let $Z_1,Z_2,\dots,Z_n\in (\mathcal{B}(\mathbb{R}^{m_1\times m_2}), \mathbb{R}^{m_1\times m_2}, \mu^{\otimes m_1m_2})$ be independent, mean-zero 
  random variables satisfying that $$\|Z_i\|_{L^2} \le \sigma,\quad \|Z_i\|_{\psi(\alpha)} \le K, \quad \|\cdot\|_{\psi(\alpha)}:= \inf\left\{t>0,\mathbb{E}\left[\exp\left(\frac{\|\cdot\|^\alpha}{t^\alpha}\right)\right]\le 2
  \right\}.$$
  Let $\alpha >1$ and $a=(a_1,a_2,\dots,a_n)^\top\in \mathbb{R}^n$. There exists a constant $C>0$ such that, for all $t>0$, with probability at least $1-e^{-t}$
  \begin{equation}
  \label{bernstein_ineq}
  \left\|\sum_{i=1}^n a_i Z_i\right\|\le C\max \left(\sqrt{\|a\|_2^2 \sigma^2(t+\log(2m))}, \|a\| K\left(\log \frac{K}{\sigma}\right)^{1/\alpha}(t + \log(2m))\right).
  \end{equation}
  \end{lemma}

\begin{pf}{\it of Lemma \ref{bernstein_lemma}:}
  We follow the proof of proposition 2 in \cite{koltchinskii2011neumann} here.\\
  Let 
  $$S_n := \sum_{i=1}^n a_i\widetilde{Z}_i,\quad \widetilde{Z} = \left(\begin{matrix}0&Z\\Z^\top &0\\\end{matrix}\right)\in \mathbb{R}^{m\times m}.$$
  We know that
  $\|S_n\|<s $ if and only if $-s I_{m}\prec S_n \prec s I_{m}$. Therefore, 
  \begin{equation*}
  P(\|S_n\|\ge s ) = P(S_n \not \preceq s I_{m}) +P(S_n\not  \succeq -s I_{m})
  \end{equation*}
  and
  \begin{equation*}
  P(S_n\not \preceq s I_{m}) = P\left(e^{\lambda S_n}\not \preceq e^{\lambda s I_{m}}\right)\le P\left(\Tr(e^{\lambda S_n})\ge  e^{\lambda s}\right) \le e^{-\lambda s} \mathbb{E}\Tr(e^{\lambda S_n}).
  \end{equation*}
  To bound $\mathbb{E}\Tr(e^{\lambda S_n})$, we use independence and Golden-Thompson inequality 
  \begin{equation*}
  \begin{aligned}
  \mathbb{E}\Tr(e^{\lambda S_n}) & = \mathbb{E}\Tr\left(e^{\lambda S_{n-1} + \lambda a_n Z_n}\right)\\&\le 
  \mathbb{E}\Tr\left(e^{\lambda S_{n_1}}e^{\lambda a_n Z_n}\right)\\&=
  \Tr\left(\mathbb{E}e^{\lambda S_{n-1}}\mathbb{E}e^{\lambda a_n Z_{n}}\right)\\& \le 
  \mathbb{E}\Tr\left(e^{\lambda S_{n-1}}\right)\left\|\mathbb{E}e^{\lambda a_n Z_n}\right\|.
  \end{aligned}
  \end{equation*}
  By induction, we have 
  \begin{equation*}
  \mathbb{E}\Tr(e^{\lambda S_n}) \le m\prod_{i=1}^n \left\|\mathbb{E}e^{\lambda a_i Z_i}\right\|.
  \end{equation*}
  It remains to bound the norm $\|\mathbb{E}e^{\lambda a_i Z_i}\|$. Using Taylor expansion and $\mathbb{E}(Z_i)=0$, we have
  \begin{align*}
  \mathbb{E} e^{\lambda a_i Z_i}&=I_{m}+\mathbb{E} \lambda^{2}a_i^2 Z_i^{2}\left[\frac{1}{2 !}+\frac{\lambda a_i Z_i}{3 !}+\frac{\lambda^{2}a_i^2 Z_i^{2}}{4 !}+\ldots\right] \\&\leq 
  I_{m}+\lambda^{2} a_i^2\mathbb{E} Z_i^{2}\left[\frac{1}{2 !}+\frac{\lambda a_i\|Z_i\|}{3 !}+\frac{\lambda^{2}a_i^2\|Z_i\|^{2}}{4 !}+\ldots\right]\\&=I_{m}+\lambda^{2}a_i^2 \mathbb{E} Z_i^{2}\left[\frac{e^{\lambda a_i\|Z_i\|}-1-\lambda a_i\|Z_i\|}{\lambda^{2}a_i^2\|Z_i\|^{2}}\right]. 
  \end{align*}
  Therefore, for all $\tau >0$ we have
  \begin{align*}
  &\left\| \mathbb{E} e^{\lambda a_i Z_i}\right\|\\&\le 1 + \lambda^{2}a_i^2 \left\|\mathbb{E} Z_i^{2}\left[\frac{e^{\lambda a_i\|Z_i\|}-1-\lambda a_i\|Z_i\|}{\lambda^{2}a_i^2\|Z_i\|^{2}}\right]\right\| 
  \\& \le
  1 + \lambda^{2} \left\|\mathbb{E} a_i^2 Z_i^{2}\right\|\left[\frac{e^{\lambda \tau}-1-\lambda \tau}{\lambda^2\tau^2}\right] + \lambda^2 \mathbb{E}\|a_iZ_i\|^2\left[\frac{e^{\lambda a_i\|Z_i\|}-1-\lambda a_i\|Z_i\|}{\lambda^{2}a_i^2\|Z_i\|^{2}}\right]I(\|a_iZ_i\|\ge \tau).
  \end{align*}
  Let $$\tau = C_2 \|a_i\|_\infty K (\log \frac{K}{\sigma})^{1/\alpha}, \quad \lambda\tau \le 1,$$
  we have 
  \begin{equation*}
  \left\|\mathbb{E} e^{\lambda a_i Z_i} \right\|\le 1 + C_1 \lambda^2 a_i^2 \sigma^2 \le \exp\{C_1\lambda^2 a_i^2 \sigma^2\}.
  \end{equation*}
  So when
  \begin{equation*}
  P(\|S_n\|\ge s) \le 2m \exp\{-\lambda s + C_1\lambda^2 \|a\|_2^2 \sigma^2\}
  \end{equation*}
  and $\lambda$ is chosen to be $$\min\left\{\frac{1}{C_2\|a\|_{\infty}K\left(\log K/\sigma \right)^{1/\alpha}}, \frac{s}{2C_1 \|a\|_2^2 \sigma^2}\right\},$$
  which can directly deduce (\ref{bernstein_ineq}).
  
\end{pf}

  \begin{lemma}
  \label{lemma_psi_rela}
  If X,Y are random variables with $\|X\|_{\psi(\alpha)}\le K_1$, $\|Y\|_{\psi(\beta)}\le K_2$, and there exists $\gamma \ge 1$ that $\frac{1}{\alpha} + \frac{1}{\beta} = \frac{1}{\gamma}\le 1$, then we have
  \begin{equation}
  \label{psi_rela}
  \|X Y\|_{\psi(\gamma)} \le K_1K_2.
  \end{equation}
  \end{lemma}
\begin{pf}{\it of Lemma \ref{lemma_psi_rela}:}
  Using Young's
  inequality we know that for all $\theta \in (0,1)$
  \begin{equation*}
  \frac{\left(\|X\|^\gamma\right)^{\alpha/\gamma}}{ t^{\alpha \theta} \alpha /\gamma} +  \frac{\left(\|Y\|^\gamma\right)^{\beta/\gamma}}{t^{\beta(1-\theta)}\beta /\gamma} \ge
  \frac{\left(\|X Y\|\right)^{\gamma}}{t^\gamma}.
  \end{equation*}
  Thus we have
  \begin{equation*}
  \begin{aligned}
  \mathbb{E}\left[\exp\left(\frac{\|X Y\|^{\gamma}}{t^\gamma}\right)\right]& \le \mathbb{E}\left[\exp\left( \frac{\left(\|X\|^\gamma\right)^{\alpha/\gamma}}{t^{\alpha\theta}\alpha /\gamma } +  \frac{\left(\|Y\|^\gamma\right)^{\beta/\gamma}}{t^{\beta(1-\theta)}\beta /\gamma } \right)\right]
  \\& \le 
  \frac{\gamma}{\alpha}\mathbb{E}\left[\exp\left(\frac{(\|X\|^\alpha}{t^{\alpha\theta}}\right)\right] +
  \frac{\gamma}{\beta}\mathbb{E}\left[\exp\left(\frac{(\|Y\|^\beta}{t^{\beta (1-\theta)}}\right)\right] &\le 2,
  \end{aligned}
  \end{equation*}
  where we choose $t, \theta$ by $t^{\theta}=K_1,t^{1-\theta}=K_2$.\\
  So we have 
  \begin{equation*}
  \mathbb{E}\left[\exp\left(\frac{\|X Y\|^{\gamma}}{(K_1K_2)^\gamma}\right)\right] \le 2.
  \end{equation*}
  
\end{pf}

  We construct the following theorem to bound $\|\sum_{j=1}^T \omega_h(j-t)\Delta_j\|$ and Lemma \ref{lem_lambda_indpdt} is an immediate result of the following theorem. For brevity, we denote $\omega_j = \omega_h(j-t)$.

  \begin{thm}
  \label{lem_lambda_indpdt_full}
    Under Assumption 4-5, for all $t>0$, with probability at least $1-3e^{-t}$, 
    \begin{equation*}
    \begin{aligned}
    \left\|\sum_{j=1}^T \omega_j\Delta_j \right\| & \le C_1\max\left(\sigma \sqrt{\frac{(t+\log(2m))}{N}},\frac{K\left(\log \frac{K}{\sigma}\right)^{1/\gamma}(t + \log(2m))}{W}\right) \\&+ C_2\mu_X \max\left(\sigma_\xi \sqrt{\frac{t}{N}},\frac{K_1 (\log \frac{K_1}{\sigma_\xi})^{1/\alpha}t}{W}
    \right) \\&+ C_3\max\left(\varsigma \sqrt{\frac{(t+\log(2m))}{N}},\frac{\mathcal{K}\left(\log \frac{\mathcal{K}}{\varsigma}\right)^{2/\beta}(t + \log(2m))}{W}\right),
    \end{aligned}
    \end{equation*}
    where $$N=\frac{1}{\sum_{j=1}^T \omega_j^2/n_j},\quad W  =\frac{1}{\max_{j} \omega_j/n_j}$$
    and constants with respect to $\omega_j, n_j$ such that
    \begin{equation*}
    \begin{aligned}
    & K=  \max \| \xi_{ji} (X_{ji} - \mathbb{E}X_{ji})\|_{\psi(\gamma)} \lesssim K_1K_2,\\
    & \sigma= \max \| \xi_{ji} (X_{ji} - \mathbb{E}X_{ji})\|_{L^{2}}\lesssim K_1K_2 2^{1/\gamma},\\
    & \mu_X = \max \|\mathbb{E} X_{ji}\| \lesssim K_2,\\
    &\sigma_\xi = \max\|\xi_{ji}\|_{L^2} \lesssim K_1 2^{1/\alpha},\\
    & \mathcal{K}=\max \| \langle M_j^0, X_{ji}\rangle X_{ji} - \mathbb{E}(\langle M^0_j, X_{ji}\rangle X_{ji})\|_{\psi(\beta/2)} \lesssim C_* K_2^2,\\
    & \varsigma = \max\|\langle M_j^0, X_{ji}\rangle X_{ji} - \mathbb{E}(\langle M^0_j, X_{ji}\rangle X_{ji})\|_{L^2} \lesssim C_* K_2^2 2^{2/\beta}.
    \end{aligned}
    \end{equation*}
  \end{thm}

\begin{pf}{\it of Theorem \ref{lem_lambda_indpdt_full}:}
  Because that 
  \begin{equation}
  \begin{aligned}
  \left\|\sum_{j=1}^T \omega_j \Delta_j\right\|
  &=\left\|\sum_{j=1}^{T} \sum_{i=1}^{n_{j}} \frac{1}{n_{j}} \omega_j\left(Y_{j i} X_{j i}-E Y_{j i} X_{j i}\right)\right\| \\ 
  & \le\left\|\sum_{j=1}^{T} \sum_{i=1}^{n_{j}} \frac{1}{n_{j}} \omega_j\xi_{ji}X_{ij}\right\|\\
  &+\left\|\sum_{j=1}^{T} \sum_{i=1}^{n_{j}} \frac{1}{n_{j}} \omega_j\left\{\left\langle M_0(j),X_{ji}\right\rangle X_{ji}-E\left(\left\langle M_0(j),X_{ji}\right\rangle X_{ji}\right)\right\}\right\|\\
  &\le\left\|\sum_{j=1}^{T} \sum_{i=1}^{n_{j}} \frac{1}{n_{j}} \omega_j\xi_{ji}(X_{ij}-EX_{ij})\right\|+\left\|\sum_{j=1}^{T} \sum_{i=1}^{n_{j}} \frac{1}{n_{j}} \omega_j\xi_{ji}\mathbb{E}X_{ij}\right\|\\
  &+\left\|\sum_{j=1}^{T} \sum_{i=1}^{n_{j}} \frac{1}{n_{j}} \omega_j\left\{\left\langle M_0(j),X_{ji}\right\rangle X_{ji}-E\left(\left\langle M_0(j),X_{ji}\right\rangle X_{ji}\right)\right\}\right\|\\&
  \label{eq:variance}
  :=
  \mathcal{D}_1 +\mathcal{D}_2 +\mathcal{D}_3. 
  \end{aligned}
  \end{equation}
  \begin{itemize}
  \item[a.] For $\mathcal{D}_1$.

  We denote that $Z_{ji} := \xi_{ji} (X_{ji} - \mathbb{E}X_{ji})$. Using Lemma \ref{lemma_psi_rela}, $Z_{ji}$ are independent, mean-zero random variables satisfying that 
  \begin{equation*}
   \|Z_{ji}\|_{\psi(\gamma)} \le K,\quad \|Z_{ji}\|_{L^{2}}\le\sigma (2)^{1/\gamma}.
  \end{equation*}

  Set $a_{ji} = {\omega_j}/{n_j}$. Using Lemma \ref{bernstein_lemma}, we have that for all $t>0$, with probability at least $1-e^{-t}$
  \begin{equation}
  \label{d1_bound}
  \mathcal{D}_1 = \left\|\sum_{j=1}^{T}\sum_{i=1}^{n_j}a_{ji}Z_{ji}\right\| \le  C\max \left(\sigma \sqrt{\frac{(t+\log(2m))}{N}},\frac{K\left(\log \frac{K}{\sigma}\right)^{1/\gamma}(t + \log(2m))}{W}\right).
  \end{equation}
  \item[b.] For $\mathcal{D}_2$.

  Directly using Lemma \ref{bernstein_lemma} with $\|\mathbb{E} X_{ji}\| \le  \mu_X,  \|\xi_{ji}\|_{L^2} \le \sigma_\xi$ and $a_{ji} = \frac{\omega_j}{n_j}$ , we know that for all $t\ge 0$, with probability at least $1-e^{-t}$
  \begin{equation}
  \label{d2_bound}
  \mathcal{D}_2 = \left\|\sum_{j=1}^T\sum_{i=1}^{n_j}a_{ji}\mu_X \xi_{ji}\right\| \le C \mu \max\left(\sigma_\xi \sqrt{\frac{t}{N}},\frac{K_1 (\log \frac{K_1}{\sigma_\xi})^{1/\alpha}t}{W}
  \right).
  \end{equation}
  \item[c.] For $\mathcal{D}_3$.
  Denote that $Z_{ji} = \langle M_j^0, X_{ji}\rangle X_{ji} - \mathbb{E}(\langle M^0_j, X_{ji}\rangle X_{ji})$ and $a_{ji} = \frac{\omega_j}{n_j}$ with 
  \begin{equation*}
 \|Z_{ji}\|_{\psi(\beta/2)} \le   \mathcal{K}, \quad \|Z_{ji}\|_{L^2} \le\varsigma .
  \end{equation*}
  Using Lemma \ref{bernstein_lemma}, we know that for all $t\ge0$, with probability at least $1-e^{-t}$
  \begin{equation}
  \label{d3_bound}
  \mathcal{D}_3 = \left\|\sum_{j=1}^T \sum_{i=1}^{n_j} a_{ji} Z_{ji}\right\| \le C\max\left(\varsigma \sqrt{\frac{(t+\log(2m))}{N}},\frac{\mathcal{K}\left(\log \frac{\mathcal{K}}{\varsigma}\right)^{2/\beta}(t + \log(2m))}{W}\right).
  \end{equation}
  \end{itemize}

  From (\ref{d1_bound}), (\ref{d2_bound}), (\ref{d3_bound}), we know that
  for all $t>0$, with probability at least $1-3e^{-t}$, 
  \begin{equation*}
  \begin{aligned}
  \left\|\sum_{j=1}^T \omega_j\Delta_j \right\| & \le C_1\max\left(\sigma \sqrt{\frac{(t+\log(2m))}{N}},\frac{K\left(\log \frac{K}{\sigma}\right)^{1/\gamma}(t + \log(2m))}{W}\right) \\&+ C_2\mu_X\max\left(\sigma_\xi \sqrt{\frac{t}{N}},\frac{K_1 (\log \frac{K_1}{\sigma_\xi})^{1/\alpha}t}{W}
  \right) \\&+ C_3\max\left(\varsigma \sqrt{\frac{(t+\log(2m))}{N}},\frac{\mathcal{K}\left(\log \frac{\mathcal{K}}{\varsigma}\right)^{2/\beta}(t + \log(2m))}{W}\right).
  \end{aligned}
  \end{equation*}
  
\end{pf}

 \begin{lemma}
  \label{lem_lambda_indpdt}
  Under Assumption 4 and 5, if $h\rightarrow0$ and $(m_1m_2)^{1/2}Th^2\rightarrow\infty$ as $m_1,m_2,T\rightarrow\infty$, $nTh\gg (K_*/\sigma_*)^2\log(m_1+m_2)$, then with probability at least $1-3/(m_1+m_2)$, 
  \begin{equation}\label{lambda_bound_indpdt}
  \mathcal{W}_h\Delta_{t}  \le
  C_1\sigma_*\sqrt{\frac{\log(m_1+m_2)}{nTh}},
  \end{equation}
  where  $C_1>0$ is a constant  independent to $\sigma_*,K_*,T,h,m_1,m_2$ such that
  \begin{align*}
     \sigma_*& =\max \left\{\sigma ,\mu_X\sigma_\xi,\varsigma \right\},\\
  K_*&=\max\left\{\mathcal{K}\left(\log \frac{\mathcal{K}}{\varsigma}\right)^{2/\beta},\mu_X K_1 \left(\log \frac{K_1}{\sigma_\xi}\right)^{1/\alpha}, K\left(\log \frac{K}{\sigma}\right)^{1/\gamma}\right\},
  \end{align*}
  in which $\mu_X  =\max_{i,j} \|\mathbb{E} X_{ji}\|_{\infty},\ \sigma_\xi = \max_{i,j}\|\xi_{ji}\|_{L^2}$ and
  \begin{equation*}
    \begin{aligned}
     & K=  \max_{i,j} \left \| \xi_{ji} (X_{ji} - \mathbb{E}X_{ji})\right\|_{\psi(\gamma)},\quad \mathcal{K}=\max_{i,j} \| \langle M_j^0, X_{ji}\rangle X_{ji} - \mathbb{E}(\langle M^0_j, X_{ji}\rangle X_{ji})\|_{\psi(\beta/2)},\\
  & \sigma= \max_{i,j} \| \xi_{ji} (X_{ji} - \mathbb{E}X_{ji})\|_{L^{2}}, \quad \varsigma =\max_{i,j}\|\langle M_j^0, X_{ji}\rangle X_{ji} - \mathbb{E}(\langle M^0_j, X_{ji}\rangle X_{ji})\|_{L^2},
    \end{aligned}
  \end{equation*}
  with $\|X\|_{L^2}=\|\mathbb{E}XX^T\|_{\infty}^{1/2}\vee\|\mathbb{E}X^TX\|_{\infty}^{1/2}$.
  \end{lemma}
  \begin{pf}{\it of Lemma \ref{lem_lambda_indpdt}:}

  Define that $$\mathcal{N}_h(t)=\{j:t-Th<j<t+Th\},\quad \bar{n}_t=\frac{1}{|\mathcal{N}_h(t)|}\sum_{j\in\mathcal{N}_h(t)}n_j,\quad \check{n}_t=\min_{j\in\mathcal{N}_h(t)}n_j.$$
  Recall the definition of $\omega_j$ that
  $\omega_j=\omega_h(j-t)=\frac{K\left(\frac{j-t}{Th}\right)}{\sum_{k=1}^T K\left(\frac{k-t}{Th}\right)},
  $ we have that
  \begin{align*}
  N&= \frac{1}{\sum_{j=1}^T \omega_h(j-t)^2/n_j}
  = \frac{\left\{\sum_{k=1}^T\frac{1}{Th}K\left(\frac{k-t}{Th}\right)\right\}^2}{\sum_{j=1}^T  \frac{1}{T^2h^2}K\left(\frac{j-t}{Th}\right)^2/n_j}\\ 
  &\asymp\frac{1}{\sum_{j=1}^T  \frac{1}{n_jT^2h^2}K\left(\frac{j-t}{Th}\right)^2}
  \asymp\left\{\frac{1}{\bar{n}_tTh}\sum_{j=1}^T\frac{1}{Th}K\left(\frac{j-t}{Th}\right)^2\right\}^{-1}\\
  &=\bar{n}_tThR(K)^{-1},
  \end{align*}
  and
  \begin{align*}
  W&=\frac{1}{\max_j \omega_h(j-t)/n_j}=\left\{\max_j\frac{\frac{1}{Th}K\left(\frac{j-t}{Th}\right)}{n_j\frac{1}{Th}\sum_{k=1}^T K\left(\frac{k-t}{Th}\right)}\right\}^{-1}\\
  &\asymp\left\{\max_j{\frac{1}{n_jTh}K\left(\frac{j-t}{Th}\right)}\right\}^{-1}
  \asymp \check{n}_tThK(0)^{-1}.
  \end{align*}
  
  With the assumption $\bar{n}_t,\check{n}_t \asymp n$, for $t\le\log(m_1+m_2)$, when $nTh\gg\{K_*/\sigma_*\}^2 \log(m_1+m_2)$, the first terms in \eqref{d1_bound}, \eqref{d2_bound} and \eqref{d3_bound} dominate those bounds which gives
  $$\mathcal{W}_h\Delta_{t}  \le {C_1}\sigma_*\sqrt{\frac{s+\log(m_1+m_2)}{nTh}}\ll1.$$

  By setting $t=\log (m_1+m_2)$, we can deduce Lemma \ref{lem_lambda_indpdt} directly.
  
\end{pf}

\begin{pf}{\it of Theorem 2:}
With Lemma \ref{lem_lambda_indpdt} and Theorem 1, we have $$\left(\delta_h M_{t}^2 + 2\mu^{-1}\mathcal{W}_h\Delta_{t}\left\|M_t^0\right\|_1\right)^{1/2}\gg{(1+\sqrt{2})}{\mu^{-1}}{\mathcal{W}_h\Delta_{t}}\sqrt{r_t},$$ which leads to the following conclusion. 
  From Lemma \ref{lem_lambda_indpdt}, we know that for all $t>0$, with probability at least $1-3e^{-t}$,
  \begin{equation*}
  \left\|\sum_{j=1}^T \omega_j\Delta_j \right\|_\infty  \le {C_1}\max\left\{\sigma^*\sqrt{\frac{t+\log(m_1+m_2)}{\bar{n}_t Th}},K^*\frac{t + \log(m_1+m_2)}{\check{n}_t Th}\right\}.
  \end{equation*}
  Choose that $$t=\log(m_1+m_2),\quad 
  \lambda = 2\sqrt{2}C_1\sigma^*\sqrt{\frac{\log(m_1+m_2)}{\bar{n}_t Th}}.$$
  If $(\check{n}_t^2/\bar{n}_t) T h \gg\{K^*/\sigma^*\}^2 \log(m_1+m_2)$, than $$\sigma^*\sqrt{\frac{t+\log(m_1+m_2)}{\bar{n}_t Th}}\gg K^*\frac{t + \log(m_1+m_2)}{\check{n}_t Th}.$$
   From Proposition \ref{prop:bias}
  \begin{equation*}
       (m_1m_2)^{-1/2}\left\|M_t^0 - \sum_{j=1}^T \omega_j M_j^0\right\|_2 \le \frac{1}{2} \alpha(K)D_2 h^2.
  \end{equation*}
  With Theorem 1, so
  \begin{align*}
  (m_1m_2)^{-1/2}\left\|\widetilde{M}_t^\lambda - M_t^0\right\|_2 \le \frac{1}{2}C_2\alpha(K)D_2h^2 + \left(2+\sqrt{2}\right)C_1\sigma^*\left(\frac{r_t\log(m_1+m_2)}{\mu^2m_1m_2 \bar{n}_t Th}\right)^{1/2}
  \end{align*}
  holds with probability at least $1-3/(m_1+m_2)$.

\end{pf}



\begin{pf}{\it of Corollary 1:} Plug in the upper bound of $\mathcal{W}_h\Delta_t$ in Lemma \ref{lem_lambda_indpdt}, Corollary 1 can be obtained immediately.
\end{pf}

\begin{pf}{\it of Corollary 2:}

  From the distribution of $X_{ji}$ and Assumption 4, we have that
  $$\gamma=\alpha,\quad\beta=\infty,\quad\mu=(m_1m_2)^{-1},$$
  and 
  \begin{align*}
  & \sigma\le \sigma_\xi\sqrt{\frac{1}{m_1\wedge m_2}},\quad \mu_X\sigma_\xi= \sigma_\xi\sqrt{\frac{1}{m_1\wedge m_2}},\quad\varsigma\le M\sqrt{\frac{1}{m_1\wedge m_2}},\\
  &   K\asymp\sigma_\xi,\quad K_1\asymp \sigma_\xi,\quad \mathcal{K}\asymp M.
  \end{align*}
  Then it can be obtained that
  \begin{align*}
  &\sigma^* = (C_M\vee \sigma_\xi)\sqrt{\frac{1}{m_1\wedge m_2}},\\ 
  &K^*/\sigma^* \asymp (m_1\wedge m_2)^{1/2}\log^{1/\alpha}(m_1\wedge m_2).
  \end{align*}
  Then $(K^*/\sigma^*)^2\log(m_1+m_2)\asymp(m_1\wedge m_2)\log^{2/\alpha}(m_1\wedge m_2)\log(m_1+m_2)$.
  Note that $\log(m_1+m_2)>\log(m_1\wedge m_2)$, when $(\check{n}_t^2/\bar{n}_t) T h\gg (m_1\wedge m_2)\log^{1+2/\alpha}(m_1+m_2)$, the first term in \eqref{lambda_bound_indpdt} dominates the bound. Apply Theorem 2 and Corollary 1 and the proof is finished.
  
\end{pf}

\section{Heterogeneous Scenario}
\label{supp:hetero}
Consider the heterogeneous assumption first.
\begin{assump}
\label{assump_hete}
(Heterogeneous Assumption)
The second moment matrix $\Sigma_t$ of $X_t$ is positive definite with smallest eigenvalue $\mu_t > 0$ for $t=1,2,\cdots,T$.
\end{assump}

Here we provide the corresponding results of Theorem 1, Theorem 2 and Corollary 1 under the heterogeneous scenario. The key difference between homogeneous and heterogeneous scenarios is that the covariance matrix used for time points $t$ is $\Sigma$ in homogeneous case and $\sum_{j=1}^T \omega_h(j-t) \Sigma_j$ in heterogeneous case. Define $\nu_t = \sum_{j=1}^T\omega_h (j-t) \mu_t$, and the difference is using $\nu_t$ to replace $\mu$ in each theorems and corollaries for the upper bound of $\widetilde{M}_t^\lambda$.

\begin{thm}
(Heterogeneous Theorem 1) 
  Under Assumption \ref{assump_hete}, if $\lambda \ge 2\mathcal{W}_h\Delta_{t}$, then
  \begin{equation}
   \label{eq_thm_error_bound_hete}
  \left\|\widetilde{M}_t^\lambda - M_t^0\right\|_2 \le\delta_h M_t +\min\left\{ \left((\delta_h M_t)^2 + \frac{2\lambda}{\nu_t}\left\|M_t^0\right\|_1\right)^{1/2}, \frac{1+\sqrt{2}}{2} \frac{\lambda}{\nu_t}\sqrt{r_t}\right\}.
  \end{equation}
  When selecting $\lambda=2\mathcal{W}_h\Delta_{t}$, we have
  \begin{equation}
  \label{proposed_bound_hete}
  \left\|\widetilde{M}_t^\lambda - M_t^0\right\|_2 \le\delta_h M_{t} + \min\left\{\left((\delta_h M_{t})^2 + 2\nu_t^{-1}\mathcal{W}_h\Delta_{t}\left\|M_t^0\right\|_1\right)^{1/2},{(1+\sqrt{2})}{\nu_t^{-1}}{\mathcal{W}_h\Delta_{t}}\sqrt{r_t}\right\}.
  \end{equation}
\end{thm}
\begin{pf}
    In the Proof of Theorem 1, the whole proof is not changed but changing \eqref{eq:homo1} as 
     \begin{equation*}
  \begin{aligned}
  & \left(\myvec({\widehat{M}}_t^\lambda) - \myvec({M}_t^0)\right)^\top \Sigma \left[ \left(\myvec({\widehat{M}}_t^\lambda)- \myvec({M}_t^0)\right) + \left(2 \myvec({M}_t^0) - 2\sum_{j=1}^T \omega_j \myvec({M}_j^0)\right)\right]\\ \ge& 
  \nu_t\left\|\widetilde{M}_t^\lambda - M_t^0\right\|_2^2 - \nu_t\left\|\widetilde{M}_t^\lambda - M_t^0\right\|_2\left\|M_t^0 - \sum_{j=1}^T \omega_j M_j^0\right\|_2.
  \end{aligned}
  \end{equation*}
    and changing \eqref{eq:homo2} as
    \begin{equation*}
  \begin{aligned}
  &\sum_{j=1}^T \omega_j\left[\mathcal{C}_j(\widetilde{M}_t^\lambda - M_t^0,\widetilde{M}_t^\lambda - M_t^0) +\mathcal{C}_j(\widetilde{M}_t^\lambda-M^0_t,M_t^0 - M_j^0)\right]
  \\=&
  \left(\myvec({\widehat{M}}_t^\lambda) - \myvec({M}_t^0)\right)^\top \Sigma \left(  \left(\myvec({\widehat{M}}_t^\lambda) - \myvec({M}_t^0)\right) + \left(M_t^0 - \sum_{j=1}^T\omega_jM_j^0\right)\right)\\\ge &
  \nu_t\left\|\widetilde{M}_t^\lambda - M_t^0\right\|_2^2 -  \nu_t\left\|\widetilde{M}_t^\lambda - M_t^0\right\|_2\left\|M_t^0 - \sum_{j=1}^T \omega_j M_j^0\right\|_2.
  \end{aligned}
  \end{equation*}
\end{pf}

  \begin{thm}\label{thm_total_hete}
  (Heterogeneous Theorem 2) 
  Under Assumption 1-5, let $n T h \gg(K_*/\sigma_*)^2 \log(m_1+m_2)$, $h\rightarrow0$ and $(m_1m_2)^{1/2}Th^2\rightarrow\infty$ as $n,m_1,m_2,T\rightarrow\infty$, when
  $$\lambda = 2C_1\sigma_*\sqrt{\frac{\log(m_1+m_2)}{n Th}}, $$
  where $C_1$ and $\sigma_*$ are defined in Lemma \ref{lem_lambda_indpdt} of Supplementary Material, then 
  with probability at least $1-3/(m_1+m_2)$,
  \begin{equation} \label{err_bound_indpdt_hete}   
  \begin{aligned}
  (m_1m_2)^{-1/2}\left\|\widetilde{M}_t^\lambda - M_t^0\right\|_2 \le \frac{1}{2}\alpha(K)D_2h^2 +\left(1+\sqrt{2}\right)C_1\sigma_*\left(\frac{r_t\log(m_1+m_2)}{\nu_t^2m_1m_2 n Th}\right)^{1/2}+o(h^2),
  \end{aligned}
  \end{equation}
  where $\alpha(K)$ is defined in (10) and $D_2$ in Assumption 2.
  \end{thm}
  
\begin{cor}
  \label{cor_general_result_hete}
  (Heterogeneous Corollary 1)
  Under assumptions of Theorem \ref{thm_total_hete}, when 
  $$ nT\gg\max\{K_*^{5/2}\sigma_*^{-3}(\nu_t^{2}m_1m_2)^{1/4}\log(m_1+m_2),\ \sigma_*^2(\nu_t^{2}m_1m_2)^{-1}\log(m_1+m_2)\},$$
  with probability at least $1-3/(m_1+m_2)$,
  \begin{equation}\label{eq_general_bound_hete}
  (m_1m_2)^{-1/2}\left\|\widetilde{M}_t^\lambda - M_t^0\right\|_2 \le C_2\left(\frac{\sigma_*^2r_t\log(m_1+m_2)}{\nu_t^2m_1m_2 n T}\right)^{2/5},
  \end{equation}
  where $C_2 =1/2\left[(2+2\sqrt{2})C_1\right]^{4/5}\left[\alpha(K)D_2\right]^{1/5}$.
  \end{cor}

\section{Definition of $\phi-$mixing}
\label{def:phi}
The dependence of two $\sigma$-field $\mathcal{A}$ and $\mathcal{B}$ is measured by
  \begin{equation*}
    \phi(\mathcal{A},\mathcal{B}) = \sup \left\{\left|P(A) - \frac{P(A\cap B)}{P(B)}\right|;A\in \mathcal{A}, B\in \mathcal{B},P(B)\not =0 \right\}.
  \end{equation*}
  And for a sequence of $\sigma$-field $\mathcal{A}_j,j\in \mathbb{N}^*$, the $\phi$-coefficients $\phi_\mathcal{A}(k),k\in \mathbb{N}$ is defined as 
  \begin{equation*}
    \phi_{\mathcal{A}}(k) = \sup_{|j-s|\le k }\phi(\mathcal{A}_j,\mathcal{A}_s).
  \end{equation*}
  If $\lim_{k\rightarrow \infty} \phi_\mathcal{A}(k)=0$, the sequence of $\sigma$-field $\mathcal{A}_j,j\in \mathbb{N}^*$ is said to be $\phi$-mixing. 
  
\section{Proofs of 
Theorem 3 and Corollary 3
, \ref{cor_matrix_completion_mixing}}
\begin{lemma}
  \label{lem_lambda_dependent}
  Under Assumption 4 and 6, if $h\rightarrow0$ as $n,T\rightarrow\infty$, $Th\gg\log(m_1+m_2)$ and $nTh \gg \left(K_*/\sigma_*\right)^2\log^3 (m_1+m_2)$, with probability at least $1-3/(m_1+m_2)$,
   \begin{equation*}
  \mathcal{W}_h\Delta_{t}  \le \mathcal{C}_1 (\Phi_{\mathcal{X}}\vee \Phi_{\mathcal{Y}})\sigma_* \sqrt{\frac{\log (m_1+m_2)}{nTh}},
  \end{equation*}
  where $\mathcal{C}_1>0$ is a constant independent to $\sigma_*,K_*,T,h,m_1,m_2$.
  \end{lemma}
\begin{pf}{\it of Lemma \ref{lem_lambda_dependent} and Theorem 3:}
  Denote $\mathbf{X}_j = (X_{j1},X_{j2},\dots,X_{jn_j})^\top, \mathbf{\xi}_j= (\xi_{j1},\xi_{j2},\dots,\xi_{jn_j})^\top $. We know that $\sigma(\mathbf{X}_{j}-\mathbb{E} \mathbf{X}_{j}) \subseteq \mathcal{X}_j$. Define $ \mathcal{Z}_j$ such that $\mathcal{Z}_j = \sigma((\mathbf{X}^\top_j,\mathbf{\xi}^\top_j)^\top)$, then
  \begin{align*}
  &\left|P(A_x\otimes A_y)-\frac{P(A_x\otimes A_y\cap B_x\otimes B_y)}{P(B_x\otimes B_y)}\right|\\=&\left|P(A_x)\left(P(A_y) - \frac{P(A_y\cap B_y)}{P(B_y)}\right) + \frac{P(A_y\cap B_y)}{P(B_y)}\left(P(A_x) - \frac{P(A_x\cap B_x)}{P(B_x)}\right)\right|
  \\\le & \phi_{\mathcal{X}}(|j-l|) +\phi_{\mathcal{Y}}(|j-l|) - \phi_{\mathcal{X}}(|j-l|) \phi_{\mathcal{Y}}(|j-l|),
  \end{align*}
  where $A_x\otimes A_y \in \mathcal{Z}_j,\ B_x\otimes B_y \in \mathcal{Z}_l$.
  So we know that $\phi_{\mathcal{Z}} \le \phi_{\mathcal{X}} +\phi_{\mathcal{Y}} - \phi_{\mathcal{X}}\phi_{\mathcal{Y}}$, which means that
  \begin{equation*}
  \Phi_\mathcal{Z} = \sum_{k=1}^\infty \sqrt{\phi_{\mathcal{Z}}(k)}\le \Phi_\mathcal{X}+ \Phi_{\mathcal{Y}}<\infty.
  \end{equation*}\\
  Let $Z_{ji} = \frac{\omega_j}{n_j}\xi_{ji}\left(X_{ji} - \mathbb{E}X_{ji}\right)$ and $Z_{j} = \sum_{i=1}^{n_j} Z_{ji}$. Using Lemma \ref{bernstein_lemma}, we know that
  \begin{equation*}
  P\left(\left\|Z_{j}\right\|\ge s\right)  \le 2m \exp\left\{-C\min\left\{\frac{n_j s^2}{w^2
  _j \sigma^2}, \frac{n_j s}{\omega_j K(\log\frac{K}{\sigma})^{1/\gamma}}\right\}\right\}.
  \end{equation*}
  Denote the bounded function 
  $$g((\mathbf{X}_j^\top,\mathbf{\xi}_j^\top)^\top) = \frac{\omega_j}{n_j}(\mathbf{X}_j-\mathbb{E}\mathbf{X}_j)^\top\mathbf{\xi}_j\mathbf{1}_{\left\| \frac{\omega_j}{n_j}(\mathbf{X}_j-\mathbb{E}\mathbf{X}_j)^\top\mathbf{\xi}_j\right\|\le s},$$
  we know that
  \begin{align*}
  P\left(\left\|\sum_{j=1}^T Z_{j}\right\|\ge t \right) & \le 
  \sum_{j=1}^T P(\|Z_j\|\ge s) + P\left(\left\|\sum_{j=1}^T Z_j\mathbf{1}_{\|Z_{j}\|\le s}\right\|\ge t\right)\\&=
  \sum_{j=1}^T P(\|Z_j\|\ge s) + P\left(\left\|\sum_{j=1}^\top g((\mathbf{X}_j^\top,\mathbf{\xi}_j^\top)^\top)\right\|\ge t\right).
  \end{align*}
  With the tail distribution assumptions for $X_j,\xi_j$, when $s \le \frac{\omega_j\sigma^2}{K(\log\frac{K}{\sigma})^{1/\gamma}}$, we have the inequalities
  \begin{align*}
  \left\| \mathbb{E}\sum_{j=1}^T g((\mathbf{X}_j^\top,\mathbf{\xi}_j^\top)^\top)\right\| &= \left\|\sum_{j=1}^T Z_j \mathbf{1}_{\|Z_j\|\ge s} \right\|\\& \le \sum_{j=1}^T\left|
  \int_{s}^\infty P(\|Z_j\|\ge s) d x + s P(\|Z_j\|\ge s) \right| \\& \le \left(s+\frac{c} {s n T^2h^2}\right) \sum_{j=1}^T P(\|Z_j\|\ge s).
  \end{align*}
  And when $s \ge \frac{\omega_j\sigma^2}{K(\log\frac{K}{\sigma})^{1/\gamma}}$, we have that 
  \begin{align*}
  \left\| \mathbb{E}\sum_{j=1}^T g((\mathbf{X}_j^\top,\mathbf{\xi}_j^\top)^\top)\right\| \le \left(s + \frac{c K(\log \frac{K}{\sigma})^{1/\gamma}}{nTh}\right)\sum_{j=1}^T P(\|Z_j\|\ge s).
  \end{align*}
  Meanwhile we have the bound
  \begin{equation*}
  \left\|\mathbb{E}\sum_{j=1}^T g^2((\mathbf{X}_j^\top,\mathbf{\xi}_j^\top)^\top)\right\| \le \sum_{j=1}^T \frac{\omega_j^2}{n_j} \sigma_\xi^2\sigma_X^2 = \frac{\sigma^2}{N}.
  \end{equation*}
  Similar to Theorem 3 in \cite{samson2000concentration}, with the matrix value function $g(\cdot)$ and norm $\|\cdot\|_{\infty}$
  \begin{align*}
  & P\left(\left\|\sum_{j=1}^T g((\mathbf{X}_j^\top,\mathbf{\xi}_j^\top)^\top)\right\|\ge t\right) \\\le &P\left(\left\|\sum_{j=1}^T g((\mathbf{X}_j^\top,\mathbf{\xi}_j^\top)^\top) - \mathbb{E}\sum_{j=1}^T g((\mathbf{X}_j^\top,\mathbf{\xi}_j^\top)^\top)\right\|\ge t- \left\| \mathbb{E}\sum_{j=1}^T g((\mathbf{X}_j^\top,\mathbf{\xi}_j^\top)^\top)\right\|\right)\\ \le &
  2m \exp\left\{-\frac{1}{8\Phi_Z^2}\min\left\{\frac{t}{2s},\frac{N t^2}{16\sigma^2}\right\}\right\},
  \end{align*}
  when $t \ge  \left\| \mathbb{E}\sum_{j=1}^T g((\mathbf{X}_j^\top,\mathbf{\xi}_j^\top)^\top)\right\|$.\\
  Thus we have 
  \begin{equation*}
  \begin{aligned}
  &P\left(\left\|\sum_{j=1}^T Z_{j}\right\|\ge t \right) \\ \le&
  \sum_{j=(t-h/2)\vee 0}^{(t+h/2)\wedge T} 2m \exp\left\{-C\min\left\{\frac{n_j s^2}{w^2
  _j \sigma^2}, \frac{n_j s}{\omega_j K(\log\frac{K}{\sigma})^{1/\gamma}}\right\}\right\} \\+ &2m \exp\left\{-\frac{1}{8\Phi_\mathcal{Z}^2}\min\left\{\frac{t}{2s},\frac{N t^2}{16\sigma^2}\right\}\right\}\\\le&
  \sum_{j=(t-h/2)\vee 0}^{(t+h/2)\wedge T} 2m\exp\left\{-\frac{C n_j s^2}{\omega_j^2\sigma^2}\right\} +2m \exp\left\{-\frac{1}{8\Phi_\mathcal{Z}^2}\min\left\{\frac{t}{2s},\frac{N t^2}{16\sigma^2}\right\}\right\} \\\lesssim&
  2m\exp\left\{-\frac{C n T^2 h^2 s^2}{\sigma^2} + \log T h\right\} +2m \exp\left\{-\frac{1}{8\Phi_\mathcal{Z}^2}\min\left\{\frac{t}{2s},\frac{N t^2}{16\sigma^2}\right\}\right\}
  \end{aligned}
  \end{equation*}
  when $t\ge  \left\| \mathbb{E}\sum_{j=1}^T g((\mathbf{X}_j^\top,\mathbf{\xi}_j^\top)^\top)\right\| $ and $s \le \frac{\omega_j\sigma^2}{K(\log\frac{K}{\sigma})^{1/\gamma}}$.\\
  And similarly 
  \begin{align*}
      & P\left(\left\|\sum_{j=1}^T Z_{j}\right\|\ge t \right) \\&\lesssim
  2m\exp\left\{-C\frac{n T h s}{K (\log \frac{K}{\sigma})^{1/\gamma}} +\log Th\right\}+2m \exp\left\{-\frac{1}{8\Phi_\mathcal{Z}^2}\min\left\{\frac{t}{2s},\frac{N t^2}{16\sigma^2}\right\}\right\}
  \end{align*}
  when $t \ge  \left\| \mathbb{E}\sum_{j=1}^T g((\mathbf{X}_j^\top,\mathbf{\xi}_j^\top)^\top)\right\| $ and $s \ge \frac{\omega_j\sigma^2}{K(\log\frac{K}{\sigma})^{1/\gamma}}$.\\
  When $$n\gtrsim \left(K(\log K/\sigma)^{1/\gamma}/\sigma\right)^2 \log (Thm),$$ choose 
  \begin{equation*}
      s\asymp n^{-1/2}(Th)^{-1}\sigma \log^{1/2} (Thm),\quad t\asymp n^{-1/2}(Th)^{-1/2}\sigma \Phi_\mathcal{Z} \log^{1/2} m.
  \end{equation*}
  If $$Th \gg \Phi_\mathcal{Z}^{2}\log (Thm)\log m,$$ we have that $t\gg s\Phi_\mathcal{Z}^2\log m.$\\
  Meanwhile, when $$n \lesssim \left(K(\log K/\sigma)^{1/\gamma}/\sigma\right)^2 \log (Thm),$$ choose 
  \begin{equation*}
      s \asymp n^{-1}(Th)^{-1}K(\log K/\sigma)^{1/\gamma}\log (Thm),\quad t \asymp n^{-1/2}(Th)^{-1/2}\sigma \Phi_\mathcal{Z} \log^{1/2} m.
  \end{equation*}
  If $$nTh \gg (K(\log K/\sigma)^{1/\gamma}/\sigma)^2\Phi_\mathcal{Z}^2\log^2 (Thm) \log m,$$ we have that $t \gg s\Phi_\mathcal{Z}^2 \log m$.\\
  Thus when $$Th \gg \Phi_\mathcal{Z}^{2}\log (Thm)\log m$$ and $$nTh \gg (K(\log K/\sigma)^{1/\gamma}/\sigma)^2\Phi_\mathcal{Z}^2\log^2 (Thm) \log m,$$ the inequality
   \begin{equation*}
  P\left(\left\|\sum_{j=1}^T \sum_{i=1}^{n_j} \frac{\omega_j}{n_j}\xi_{ji}(X_{ji}-\mathbb{E}(X_{ji}))\right\|\ge t \right)  \le \frac{1}{m}.
  \end{equation*}
  holds where $t \asymp n^{-1/2}(Th)^{-1/2}\sigma \Phi_\mathcal{Z} \log^{1/2} m.$\\
  Similarly, we choose $t\asymp n^{-1/2}(Th)^{-1/2}\sigma_\xi \mu_X \Phi_\mathcal{Y}\log^{1/2} m$ and
  \begin{equation*}
  P\left(\left\|\sum_{j=1}^T\sum_{i=1}^{n_j} \frac{\omega_j}{n_j}\xi_{j i} \mathbb{E}(X_{ji})\right\|\ge t\right)\le \frac{1}{m}
  \end{equation*}
  holds when $Th \gg \Phi_{\mathcal{Y}}^2 \log (Thm)\log m$ and $$nTh \gg (K_1(\log K_1/\sigma_\xi)^{1/\alpha}/\sigma_\xi)^2\Phi_\mathcal{Y}^2\log^2 (Thm) \log m.$$
  And choose $t\asymp n^{-1/2}(Th)^{-1/2} \varsigma \Phi_\mathcal{X}\log^{1/2} m$, we have
  \begin{equation*}
  P\left(\left\|\sum_{j=1}^{T} \sum_{i=1}^{n_{j}} \frac{\omega_j}{n_{j}} \left\{\left\langle M_0(j),X_{j i}\right\rangle X_{ji}-E\left(\left\langle M_0(j),X_{j i}\right\rangle X_{j i}\right)\right\}\right\|\ge t\right)\le \frac{1}{m}
  \end{equation*}
  holds when $Th \gg \Phi_{\mathcal{X}}^2 \log (Thm)\log m$ and $nTh \gg (\mathcal{K}(\log \mathcal{K}/\varsigma)^{2/\beta}/\varsigma)^2\Phi_\mathcal{X}^2\log^2 (Thm) \log m$.

  Summarizing the above results with $\Phi_\mathcal{Z} \le \Phi_\mathcal{X}+\Phi_\mathcal{Y}$, we know that when $Th \gg (\Phi^2_{\mathcal{X}}\vee \Phi_\mathcal{Y}^2 )\log (Thm)$ and $nTh \gg \left(K^*/\sigma^*\right)^2(\Phi_\mathcal{X}^2\vee \Phi_{\mathcal{Y}}^2)\log^2(Thm)\log m$, there exists a constant $\mathcal{C}_1>0$, with probability at least $1-3/m$,
  \begin{equation*}
  \left\|\sum_{j=1}^T \omega_j \Delta_j\right\| \le \mathcal{C}_1 (\Phi_{\mathcal{X}}\vee \Phi_{\mathcal{Y}})\sigma^* \sqrt{\frac{\log (m_1+m_2)}{nTh}}.
  \end{equation*}
  Choose $$\lambda = 2\mathcal{C}_1(\Phi_{\mathcal{X}}\vee \Phi_{\mathcal{Y}})\sigma^* \sqrt{\frac{\log (m_1+m_2)}{nTh}}$$
  and using Theorem 1 and proof of Theorem 2, we have that
  \begin{equation*}
      (m_1m_2)^{-1/2}\left\|\widetilde{M}_t^\lambda - M_t^0\right\|_2 \le \frac{1}{2}\alpha(K) D_2 h^2 + (1+\sqrt{2})\mathcal{C}_1(\Phi_{\mathcal{X}}\vee \Phi_{\mathcal{Y}}) \sigma^*\left(\frac{r_t\log(m_1+m_2)}{\mu^2m_1m_2nTh}\right)^{1/2}.
  \end{equation*}
  Note that $Th \gg \log(Th)$ when $Th\rightarrow\infty$, we can represent the condition $Th \gg (\Phi^2_{\mathcal{X}}\vee \Phi_\mathcal{Y}^2 )\log (Thm)$ and $nTh \gg \left(K^*/\sigma^*\right)^2(\Phi_\mathcal{X}^2\vee \Phi_{\mathcal{Y}}^2)\log^2(Thm)\log m$ as $Th \gg (\Phi^2_{\mathcal{X}}\vee \Phi_\mathcal{Y}^2 )\log m$ and $nTh \gg \left(K^*/\sigma^*\right)^2(\Phi_\mathcal{X}^2\vee \Phi_{\mathcal{Y}}^2)\log^3 m$  respectively.
  
\end{pf}

\begin{cor}
  \label{cor_matrix_completion_mixing}
  Under assumptions of Theorem 3,  when $X_{ji}$ are i.i.d. uniformly distributed on $\mathcal{E}$, $\xi_{ji}$ are independently follow sub-exponential mean-zero distributions, $Th \gg\log (m_1+m_2)$ and $n T h \gg (m_1\wedge m_2)\log^{3+2/\alpha} (m_1+m_2)$,
  then with probability at least $1-3/(m_1+m_2)$,
  \begin{align*}
  (m_1m_2&)^{-1/2}\left\|\widetilde{M}_t^\lambda - M_t^0\right\|_2 \\&\le
  \frac{1}{2}\alpha(K)D_2h^2+
  (1+\sqrt{2})\mathcal{C}_1(\Phi_{\mathcal{X}}\vee \Phi_{\mathcal{Y}}) (C_M\vee \sigma_\xi)\left(\frac{r_t(m_1\vee m_2)\log (m_1+m_2)}{nTh}\right)^{1/2}.
  \end{align*}
  When $nT\gg (m_1\vee m_2)\log^{(1+5/2\alpha)}(m_1+m_2)$ and $n^{-1}T^4\gg (m_1\vee m_2)^{-1}\log^4(m_1+m_2)$,  we select
  $$h=\mathcal{C}_{h}\left(\frac{(C_M\vee \sigma_\xi)^2(\Phi_{\mathcal{X}}\vee \Phi_{\mathcal{Y}})^2r_t(m_1\vee m_2)\log(m_1+m_2)}{nT}\right)^{1/5},$$
  then
  \begin{equation}\label{eq_upper_bound_mc_mixing}
  (m_1m_2)^{-1/2}\left\|\widetilde{M}_t^\lambda - M_t^0\right\|_2 \le \mathcal{C}_2(\Phi_{\mathcal{X}}\vee \Phi_{\mathcal{Y}})^{4/5}(C_M\vee \sigma_\xi)^{4/5}\left(\frac{r_t(m_1\vee m_2)\log(m_1+m_2)}{nT}\right)^{2/5},
  \end{equation}
  where $\mathcal{C}_1,\mathcal{C}_h,\mathcal{C}_2$ are the same constants as above.
  \end{cor}
 
The proofs of Corollary 3 
and \ref{cor_matrix_completion_mixing} are similar to those of Corollary 1 
 and 2, which we do not repeat.

\section{Application to Compressed Sensing}
  \begin{cor}
  \label{cor_compress_sensing}(Independent case)
  Under Assumption 1-5, when $X_{ji}$ are random matrices with independent mean-zero sub-gaussian elements with variance $\sigma_X^2$, $\xi_{ji}$ are independently follow sub-gaussian distributions, $nTh\gg m_1\vee m_2$, $h\rightarrow0$ and $(m_1m_2)^{1/2}Th^2\rightarrow\infty$ as $n,m_1,m_2,T\rightarrow\infty$, with probability at least $1-2/(m_1+m_2)$,
  {\small\begin{align*}     
 & (m_1m_2)^{-1/2}\left\|\widetilde{M}_t^\lambda - M_t^0\right\|_2 \\&\le \frac{1}{2}\alpha(K)D_2h^2 +\left(1+\sqrt{2}\right)C_1\left[\eta\vee \{M (m_1 m_2)^{1/2}\}\right]\left(\frac{r_t\log(m_1+m_2)}{(m_1\wedge m_2)n Th}\right)^{1/2}+o(h^2),
  \end{align*}}
  When $nT\gg(m_1\vee m_2)\log(m_1+m_2)$, let 
  \begin{equation*}
    h = C_h\left(\frac{\left\{\eta^2 \vee(M^2 m_1 m_2)\right\}r_t\log(m_1+m_2)}{(m_1\wedge m_2)n T}\right)^{1/5},
  \end{equation*}
  then 
  \begin{equation}\label{eq_upper_bound_cs}
     (m_1m_2)^{-1/2}\left\|\widetilde{M}_t^\lambda - M_t^0\right\|_2 \le C_2 \left(\frac{\left\{\eta^2 \vee(M^2 m_1 m_2)\right\}r_t\log(m_1+m_2)}{(m_1\wedge m_2)n T}\right)^{2/5},
  \end{equation}
  where $\eta = \sigma_\xi/\sigma_X$ and $C_1,C_h,C_2$ are the same constants as above.
  \end{cor}
  
  Theorem 6 in \cite{koltchinskii2011neumann} presented the error bound for static matrix compressed sensing which is
  \begin{equation*}
      (m_1m_2)^{-1/2}\left\|\widehat{M}_t^\lambda - M_t^0\right\|_2\le C_4\eta\left(\frac{ r_t}{(m_1\wedge m_2)n}\right)^{1/2}.
  \end{equation*}
  Similarly to the matrix completion, when $T\gg\eta^{-1}\{\eta^2 \vee(M^2 m_1 m_2)\}\{n(m_1\wedge m_2)\}^{1/4}\log(m_1+m_2)$, our dynamic method gives a sharper bound than the static method. 
  
  From Corollary 2 and Corollary \ref{cor_compress_sensing}, when the variances of observation errors $\xi_{ji}$ are small enough comparing to the variances of $X_{ji}$, i.e., $\sigma_\xi\le M$ in matrix completion setting and $\eta \le M(m_1m_2)^{1/2}$ in compressed sensing setting, the upper bound (15) and \eqref{eq_upper_bound_cs} have the same order such that
  \begin{equation*}
     (m_1m_2)^{1/2}\left\|\widetilde{M}_t^\lambda - M_t^0\right\|_2 \le C_2 M^{4/5}\left(\frac{r_t(m_1\vee m_2)\log(m_1+m_2)}{n T}\right)^{2/5}.
  \end{equation*}

\begin{pf}{\it of Corollary \ref{cor_compress_sensing}:}
  Because $\mathbb{E}X_{ji}=0$, we know that $\mu_X=0$ and (b) in \eqref{eq:variance} becomes zero. From the distribution of $X_{ji}$ and Assumption 4, we know that $\beta=2$ and $K_2 \asymp (m_1\vee m_2)^{1/2}\sigma_X$ by Theorem 4.4.5 in \cite{vershynin2018high}. Also, it is easy to check that $K\asymp\sigma_\xi \sigma_X(m_1\vee m_2)^{1/2},\ \sigma = \sigma_\xi\sigma_X (m_1\vee m_2)^{1/2}$.
  With that
  \begin{align*}
  &\left\|\mathbb{E}(\langle M_j^0,X_{j i}\rangle X_{j i}- \mathbb{E}(\langle M_j^0,X_{j i}\rangle X_{j i})(\langle M_j^0,X_{j i}\rangle X_{j i}- \mathbb{E}(\langle M_j^0,X_{j i}\rangle X_{j i})^T\right\|_{\infty}
  \\\le &M^2 m_1m_2(m_1\vee m_2) \sigma^4_X
  \end{align*}
  and with Lemma \ref{lemma_psi_rela}
  \begin{align*}
      \left\|\langle M_j^0, X_{ji}\rangle X_{ji} - \mathbb{E}\langle M_j^0, X_{ji}\rangle X_{ji} \right\|_{\psi(1)} & \asymp \left\|\langle M_j^0,X_{ji}\rangle X_{ji} \right\|_{\psi(1)}
      \\& \le \left\|\langle M_j^0,X_{ji}\rangle\right\|_{\psi(2)}\left\|X_{ji}\right\|_{\psi(2)}
      \\&\le M (m_1m_2)^{1/2} \sigma_X^2,
  \end{align*}
  we have that $\varsigma\le M (m_1m_2)^{1/2}(m_1\vee m_2)^{1/2}\sigma^2_X,\mathcal{K}\lesssim M(m_1m_2)^{1/2}\sigma_X^2$.

  Then it can be obtained that 
  \begin{align*}
      &\sigma^* = \left[\sigma_{\xi}\vee M \sigma_X(m_1 m_2)^{1/2}\right](m_1\vee m_2)^{1/2}\sigma_X\\
      & K^*/\sigma^* \asymp (m_1\vee m_2)^{-1/2} \log(1/(m_1\vee m_2)).
  \end{align*}
  Apply Theorem 2 and Corollary 1, the proof is completed.
  \end{pf}
  
  \begin{cor}
  \label{cor_compress_sensing_mixing}(Dependent case.)
  Under assumptions of Theorem 3,  when $X_{ji}$ are random matrices with independent mean-zero sub-gaussian elements with variance $\sigma_X^2$ and $\xi_{ji}$ are independently follow sub-gaussian distributions, $Th \gg \log (m_1+m_2)$ and $nTh \gg(m_1\vee m_2)$, with probability at least $1-2/(m_1+m_2)$,
  \begin{align*}     
  &(m_1m_2)^{-1/2}\left\|\widetilde{M}_t^\lambda - M_t^0\right\|_2\\& \le \frac{1}{2}\alpha(K)D_2h^2 +\left(1+\sqrt{2}\right)\mathcal{C}_1(\Phi_{\mathcal{X}}\vee \Phi_{\mathcal{Y}})\left[\eta\vee M (m_1 m_2)^{1/2}\right]\left(\frac{r_t\log(m_1+m_2)}{(m_1\wedge m_2)nTh}\right)^{1/2}.
  \end{align*}
  When $nT\gg (m_1\vee m_2)\log (m_1+m_2)$ and $n^{-1}T^4\gg (m_1\vee m_2)^{-1}\log^4(m_1+m_2)$, we select 
  \begin{equation*}
    h = \mathcal{C}_h\left(\frac{\left(\eta^2 \vee M^2 m_1 m_2\right)(\Phi_{\mathcal{X}}\vee \Phi_{\mathcal{Y}})^2r_t\log(m_1+m_2)}{(m_1\wedge m_2)nT}\right)^{1/5},
  \end{equation*}
  then 
  \begin{equation*}
     (m_1m_2)^{1/2}\left\|\widetilde{M}_t^\lambda - M_t^0\right\|_2 \le \mathcal{C}_2(\Phi_{\mathcal{X}}\vee \Phi_{\mathcal{Y}})^{4/5} \left(\frac{\left(\eta^2 \vee M^2 m_1 m_2\right)r_t\log(m_1+m_2)}{(m_1\wedge m_2)nT}\right)^{2/5},
  \end{equation*}
  where $\eta = \sigma_\xi/\sigma_X$ and $\mathcal{C}_1,\mathcal{C}_h,\mathcal{C}_2$ are the same constants as above.
  \end{cor}
The proof of Corollary \ref{cor_compress_sensing_mixing} 
is similar to that of Corollary \ref{cor_compress_sensing}, which we do not repeat.

\section{Proof of Theorem 4}
\begin{pf}{\it of Theorem 4:}
  First we know that \begin{align*}
      \nabla f_t(M) = 2\sum_{j=1}^T \omega_j \left[\frac{1}{n_j}\sum_{i=1}^{n_j}\mathbb{E} \langle M,X_{ji}\rangle X_{ji}  - \frac{1}{n_j}\sum_{i=1}^{n_j}Y_{ji}X_{ji}\right].
  \end{align*}
  This gives
  \begin{align*}
      \left\|\nabla f_t(M) - \nabla f_t(N)\right\|_2 &= \left\|2\sum_{j=1}^T \frac{\omega_j}{n_j}\sum_{i=1}^{n_j} \mathbb{E}\langle M - N,X_{ji}\rangle X_{ji}\right\|_2
      \\&\le 2\left\|\sum_{j=1}^T \frac{\omega_j}{n_j}\sum_{i=1}^{n_t} \left(\|X_{ji}\|_2X_{ji}\right)\right\|_2\|M-N\|_2.
  \end{align*}
  Define that $2L_f = 4\left\|\sum_{j=1}^T \omega_h(j-t)/n_j\sum_{i=1}^{n_j} \left(\|X_{ji}\|_2X_{ji}\right)\right\|_2$,
  \cite{toh2010accelerated} proved that for any $k>1$,
  \begin{equation}
  \label{eq_algorithm}
  F_t(M_t^{(k)}) - F_t(\widetilde{M}_t^\lambda) \le \frac{2 L_f\|M_t^{(0)} - \widetilde{M}_t^\lambda\|_2^2}{(k+1)^2}.
  \end{equation}
  Note that \eqref{eq_algorithm} focuses on the convergence rate of object function $ F_t(M_t^{(k)})$ instead of $M_t^{(k)}$.from (\ref{eq_algorithm}) we know 
  \begin{equation*}
  \begin{aligned}
  & \sum_{j=1}^T \omega_j \mathcal{C}_j(M_t^{(k)},M_t^{(k)}) -2 \sum_{j=1}^T \omega_j \left\langle \frac{1}{n_j}\sum_{i=1}^{n_j} Y_{j i}X_{j i},M^{(k)}_{t} \right\rangle + \lambda \|M^{(k)}_{t} \|_1\\
  \le &
  \sum_{j=1}^T \omega_j \mathcal{C}_j(\widehat{M}_t^{\lambda},\widehat{M}_t^{\lambda})-2 \sum_{j=1}^T \omega_j \left\langle \frac{1}{n_j}\sum_{i=1}^{n_j} Y_{j i}X_{j i}, \widehat{M}_t^{\lambda} \right\rangle + \lambda \|\widehat{M}_t^{\lambda}\|_1 + \frac{2L_f\gamma_t^2}{(k+1)^2}
  \\\le & 
  \sum_{j=1}^T \omega_j \mathcal{C}_j(M,M) -2 \sum_{j=1}^T \omega_j \left\langle \frac{1}{n_j}\sum_{i=1}^{n_j} Y_{j i}X_{j i}, M\right\rangle + \lambda \|M\|_1 + \frac{2L_f\gamma_t^2}{(k+1)^2}
  \end{aligned}
  \end{equation*}
  for any $M\in \mathbb{M}$.
  Therefore, we have
  \begin{equation*}
  \begin{aligned}
  &\sum_{j=1}^T \omega_j \mathcal{C}_j(M_t^{(k)}- M_j^0,M_t^{(k)}- M_j^0)\\ \le & \sum_{j = 1}^T \omega_j \mathcal{C}_j(M-M_j^0,M-M_j^0) +\left\langle 2\sum_{j=1}^T \omega_j  \Delta_j,M_t^{(k)} - M\right \rangle \\+&\lambda (\|M\|_1-\|M_t^{(k)}\|_1) +\frac{2L_f\gamma_t^2}{(k+1)^2}.
  \end{aligned} 
  \end{equation*}\\
  With the assumption that $\lambda >2\|\sum_{j=1}^T \omega_j\Delta_j\|_{\infty}$, we use the similar methods in proof of Theorem 1 and immediately know that
  \begin{equation*}
  \begin{aligned}
  &\left\|M_t^{(k)} - M_t^0\right\|_2\\\le & \left\|M_t^0-\sum_{j=1}^T \omega_h(j-t) M_j^0\right\|_2 +\left(\left\|M_t^0-\sum_{j=1}^T \omega_h(j-t) M_j^0\right\|_2^2 +\frac{2\lambda}{\mu}\|M_t^0\|_1 + \frac{2L_f\gamma_t^2}{\mu(k+1)^2}\right)^{1/2}.
  \end{aligned}
  \end{equation*}
  Next, we consider $F_t(\widetilde{M}_t^\lambda +  M) - F_t(\widetilde{M}_t^\lambda)$ with the decomposition 
  \begin{equation*}
  \begin{aligned}
  M &=P_{S_1^\perp} M P_{S_2^\perp}\oplus P_{S_1}M P_{S_2}\oplus P_{S_1}MP_{S_2^\perp} \oplus P_{S_1^\perp}MP_{S_2}\\&:=
  M^{\perp}\oplus M^{//}\oplus M^{//1}\oplus M^{//2}
  \\&=  \oplus_{j=1}^{l_1} \sigma^{\perp}_j M^{\perp}_j \oplus_{j=1}^{l_2}\sigma_j^{//}M^{//}_j\oplus_{j=1}^{l_3} \sigma_j^{//1}M_j^{//_1} \oplus_{j=1}^{l_4}\sigma_j^{//2} M_j^{//2},
  \end{aligned}
  \end{equation*}
  where $\widetilde{M}_t^\lambda=\sum_{i=1}^r \sigma_i u_i v_i^\top$ with support $S_1,S_2$, $P_{S_1^\perp}MP_{S_2^{\perp}}=\sum_{i=1}^{l_1} \sigma^{\perp}_i u^{\perp}_i (v^{\perp}_j)^\top$ and 
  \begin{equation*}
  \begin{aligned}
  &\widetilde{M}_t^\lambda +\theta M^{\perp}_j = \sum_{i=1}^r \sigma_i u_i v_i^\top + \theta  u^{\perp}_j (v^{\perp}_j)^\top,\\
  & \widetilde{M}_t^\lambda+\theta M_j^{//} = \sum_{i=1}^r\sigma_i u_i v_i^\top + \theta u_{\alpha_j}v_{\beta_j}^\top,\\
  &  \widetilde{M}_t^\lambda+\theta M_j^{//1} = \sum_{i\not = \alpha_{1,j}}\sigma_i u_i v_i^\top + \sqrt{\sigma^2_{\alpha_{1,j}} + \theta^2}u_{\alpha_{1,j}}\left(\frac{\sigma_{\alpha_{1,j}}}{\sqrt{\sigma^2_{\alpha_{1,j}} + \theta^2}}v_{\alpha_{1,j}} +\frac{\theta}{\sqrt{\sigma^2_{\alpha_{1,j}} + \theta^2}}v^{\perp}_{\beta_{1,j}}\right)^\top ,\\
  &   \widetilde{M}_t^\lambda+\theta M_j^{//2} = \sum_{i\not = \alpha_{2,j}}\sigma_i u_i v_i^\top + \sqrt{\sigma^2_{\alpha_{2,j}} + \theta^2}\left(\frac{\sigma_{\alpha_{2,j}}}{\sqrt{\sigma^2_{\alpha_{2,j}} + \theta^2}}u_{\alpha_{2,j}} +\frac{\theta}{\sqrt{\sigma^2_{\alpha_{2,j}} + \theta^2}}u^{\perp}_{\beta_{2,j}}\right)v_{\alpha_{2,j}}^\top.
  \end{aligned}
  \end{equation*}
  Using the convexity of $F_t(\cdot)$, we have that for all $M,N\in \mathbb{M}$ and $B\in \partial F_t(M)$,
  \begin{equation}
  \label{convex_eq}
  F_t(N) \ge F_t(M)+ \langle B, N-M\rangle
  \end{equation}
  and
  \begin{equation}
  \label{decom_sum}
  F_t(\widetilde{M}_t^\lambda + M ) - F_t(\widetilde{M}_t^\lambda ) \ge  \sum_{\cdot \in \mathcal{A}}\left( F_t(\widetilde{M}_t^\lambda +  M^\cdot )- F_t(\widetilde{M}_t^\lambda )\right)
  \end{equation}
  where $\mathcal{A}=\left\{\perp,//,//1,//2\right\}$.
  Because $\widetilde{M}_t^\lambda$ is the minimizer of $F_t(M)$, there exists a matrix $\widehat{B} \in \partial F_t(\widetilde{M}_t^\lambda)$ satisfies that for all $M\in \mathbb{M}$,
  $$\langle \widehat{B}, \widetilde{M}_t^\lambda- M\rangle \le 0,$$
  which means that there exists 
  \begin{equation}
  \label{partial hat}
  \widehat{V}_t\in\partial \|\widetilde{M}_t^\lambda \|_1 = \sum_{j=1}^r u_j v_j^\top +P_{S_1^\perp}\widehat{W}P_{S_2^\perp},\quad \|\widehat{W}\|_{\infty}\le 1
  \end{equation}
  satisfying that
  \begin{equation}
  \label{mono_sub}
  2\sum_{j=1}^T \omega_j \left(\mathcal{C}_j( \widetilde{M}_t^\lambda, \widetilde{M}_t^\lambda-M) - \left\langle\frac{1}{n_j}\sum_{i=1}^{n_j}Y_{ji}X_{ji},\widetilde{M}_t^\lambda-M\right\rangle\right) +\lambda \langle \widehat{V}_t,\widetilde{M}_t^\lambda -M\rangle \le 0.
  \end{equation}
  Set $\varepsilon > \xi>0$.
  \begin{itemize}
  \item[a.] We know that for all $B_\xi \in \partial F_t(\widetilde{M}_t^\lambda +\xi M^{\perp})$,
  \begin{equation*}
  \begin{aligned}
  &F_t(\widetilde{M}_t^\lambda +\varepsilon M^{\perp})- F_t(\widetilde{M}_t^\lambda +\xi M^{\perp})  \\\ge &(\varepsilon-\xi )\langle B_\xi,M^{\perp}\rangle\\=&
  (\varepsilon-\xi)\left[2\sum_{j=1}^T \omega_j \left(\mathcal{C}_j( \widetilde{M}_t^\lambda + \xi M^{\perp}, M^{\perp})- \left\langle\frac{1}{n_j}\sum_{i=1}^{n_j}Y_{j i}X_{j i},M^{\perp}\right\rangle\right) +\lambda \langle V_\xi,M^{\perp}\rangle \right]
  \end{aligned}
  \end{equation*}
  holds for all
  \begin{align*}&V_\xi \in \partial \|\widetilde{M}_t^\lambda +\xi M^{\perp}\|_1 \\ =&
  \left\{\sum_{j=1}^r u_j v_j^\top +\sum_{j=1}^{l_1} u^{\perp}_j (v^{\perp}_j)^\top + P_{(S_1\oplus_{j=1}^{l_1} u_j)^{\perp}}W P_{(S_2\oplus_{j=1}^{l_1} v_j)^{\perp}},\|W\|_{\infty}\le 1\right\}.
  \end{align*}
  With (\ref{partial hat}) and (\ref{mono_sub}), setting $M= \widetilde{M}_t^\lambda +M^{\perp}$, we know that
  \begin{equation*}
  \begin{aligned}
  &2\sum_{j=1}^T \omega_j \left(\mathcal{C}_j( \widetilde{M}_t^\lambda + \xi M^{\perp}, M^{\perp}) - \left\langle\frac{1}{n_j}\sum_{i=1}^{n_j}Y_{j i}X_{j i},M^{\perp}\right\rangle\right) +\lambda \langle V_\xi,M^{\perp}\rangle
  \\\ge &
  2\sum_{j=1}^T \omega_j \mathcal{C}_j( \xi M^{\perp},M^{\perp}) + \lambda\langle V_\xi - \widehat{V},M^{\perp}\rangle \\=&
  2\xi \sum_{j=1}^T \omega_j \mathcal{C}_j(M^{\perp},M^{\perp})\\+&\lambda \left\langle  \sum_{j=1}^{l_1}u^{\perp}_j (v^{\perp}_j)^\top + P_{(S_1\oplus_{j=1}^{l_1} u_j)^{\perp}}W P_{(S_2\oplus_{j=1}^{l_1} v_j)^{\perp}} - P_{S_1^{\perp}} \widehat{W}P_{S_2^{\perp}},\sum_{j=1}^{l_1}\sigma^\perp_j u_j^\perp (v_j^\perp)^\top\right\rangle \\\ge &
  2\xi\mu\|M^\perp\|^2_2,
  \end{aligned}
  \end{equation*}
  in which we use that
  \begin{equation*}
  \begin{aligned}
  & \left\langle  \sum_{j=1}^{l_1}u^{\perp}_j (v^{\perp}_j)^\top + P_{(S_1\oplus_{j=1}^{l_1} u_j)^{\perp}}W P_{(S_2\oplus_{j=1}^{l_1} v_j)^{\perp}} - P_{S_1^{\perp}} \widehat{W}P_{S_2^{\perp}},\sum_{j=1}^{l_1}\sigma_j^\perp u_j^\perp (v_j^\perp)^\top\right\rangle \\
  =& \sum_{j=1}^{l_1}\sigma_j^\perp \left(1- P_{u_j^\perp}P_{S_1^\perp}\widehat{W}P_{S_2^\perp}P_{v_j^\perp}\right)\ge 0.
  \end{aligned}
  \end{equation*}
  So we have 
  \begin{equation*}
  F_t(\widetilde{M}_t^\lambda +\varepsilon M^{\perp}) - F_t(\widetilde{M}_t^\lambda + \xi M^{\perp})
  \ge 2(\varepsilon-\xi)\xi \mu\|M^\perp\|_2^2,
  \end{equation*}
  which means that
  \begin{equation}
  \label{re_a}
  F_t(\widetilde{M}_t^\lambda +\varepsilon M^{\perp}) - F_t(\widetilde{M}_t^\lambda)
  \ge \mu \|M^\perp\|_2^2\int_{0}^ \varepsilon 2\xi d\xi =\varepsilon^2\mu\|M^\perp\|_2^2.
  \end{equation}
  \item[b.] Similarly, we know that for all $B_\xi \in \partial F_t(\widetilde{M}_t^\lambda +\xi M^{//})$
  \begin{equation*}
  \begin{aligned}
  &F_t(\widetilde{M}_t^\lambda +\varepsilon M^{//})- F_t(\widetilde{M}_t^\lambda +\xi M^{//})  \\\ge &
  (\varepsilon-\xi)\left[2\sum_{j=1}^T \omega_j \left(\mathcal{C}_j( \widetilde{M}_t^\lambda + \xi M^{//}, M^{//}) - \left\langle\frac{1}{n_j}\sum_{i=1}^{n_j}Y_{j i}X_{j i},M^{//}\right\rangle\right) +\lambda \langle V_\xi,M^{//}\rangle \right]
  \end{aligned}
  \end{equation*}
  holds for all
  $$V_\xi \in \partial \|\widetilde{M}_t^\lambda +\xi M^{//}\|_1  \supset \partial \|\widetilde{M}_t^\lambda\|_1 .$$
  Choose $V_\xi=\widehat{V}$, so
  \begin{equation*}
  2\sum_{j=1}^T \omega_j \left(\mathcal{C}_j( \widetilde{M}_t^\lambda + \xi M^{//}, M^{//})- \left\langle\frac{1}{n_j}\sum_{i=1}^{n_j}Y_{j i}X_{j i},M^{//}\right\rangle\right) +\lambda \langle V_\xi,M^{//}\rangle  \ge 2\xi \mu\|M^{//}\|_2^2,
  \end{equation*}
  which means that
  \begin{equation}
  \label{re_b}
  F_t(\widetilde{M}_t^\lambda +\varepsilon M^{//})- F_t(\widetilde{M}_t^\lambda)\ge \varepsilon^2 \mu\|M^{//}\|_2^2.
  \end{equation}
  \item[c.] Similarly, we have that for all $B_\xi \in \partial F_t(\widetilde{M}_t^\lambda +\xi M^{//1})$
  \begin{equation*}
  \begin{aligned}
  &F_t(\widetilde{M}_t^\lambda +\varepsilon M^{//1})- F_t(\widetilde{M}_t^\lambda +\xi M^{//1})  \\\ge &
  (\varepsilon-\xi)\left[2\sum_{j=1}^T \omega_j \left(\mathcal{C}_j( \widetilde{M}_t^\lambda + \xi M^{//1}, M^{//1}) - \left\langle\frac{1}{n_j}\sum_{i=1}^{n_j}Y_{j i}X_{j i},M^{//1}\right\rangle\right) +\lambda \langle V_\xi,M^{//1}\rangle \right]
  \end{aligned}
  \end{equation*}
  holds for all
  \begin{align*}&V_\xi \in \partial \|\widetilde{M}_t^\lambda +\xi M^{//1}\|_1  \\=&\left\{\sum_{j\not\in \alpha_{1}}u_j v_j^\top +\sum_{j=1}^{l_3}u_{\alpha_{1,j}}\widetilde{v}_j^\top + P_{S_1^\perp}W P_{(S_2/_{j=1}^{l_3}v_{\alpha_{1,j}}\oplus_{j=1}^{l_3} \widetilde{v}_{j})^\perp},\|W\|_{\infty}\le 1\right\}\end{align*}
  with $$\widetilde{v}_j=\frac{\sigma_{\alpha_{1,j}}}{\sqrt{\sigma^2_{\alpha_{1,j}} + \xi^2}}v_{\alpha_{1,j}} +\frac{\xi}{\sqrt{\sigma^2_{\alpha_{1,j}} + \xi^2}}v^{\perp}_{\beta_{1,j}}.$$
  So we have 
  \begin{equation*}
  \begin{aligned}
  &2\sum_{j=1}^T \omega_j \left(\mathcal{C}_j( \widetilde{M}_t^\lambda + \xi M^{//1}, M^{//1})- \left\langle\frac{1}{n_j}\sum_{i=1}^{n_j}Y_{j i}X_{j i},M^{//1}\right\rangle\right) +\lambda \langle V_\xi,M^{//1}\rangle
  \\\ge &
  2\sum_{j=1}^T \omega_j \mathcal{C}_j( \xi M^{//1},M^{//1}) + \lambda\langle V_\xi - \widehat{V},M^{//1}\rangle \\=&
  2\xi \sum_{j=1}^T \omega_j \mathcal{C}_j(M^{//1},M^{//1}) \\+&\lambda \left\langle  \sum_{j=1}^{l_3}\left(u_{\alpha_{1,j}}\widetilde{v}_j^T - u_{\alpha_{1,j}}v_{\alpha_{1,j}}^T\right)+ P_{S_1^\perp} W P_{(S_2/_{j=1}^{l_3}v_{\alpha_{1,j}}\oplus_{j=1}^{l_3} \widetilde{v}_{j})^\perp} - P_{S_1^\perp}\widehat{W}P_{S_2^\perp}\right.,\\&\left. \sum_{j=1}^{l_3}\sigma_j^{//1} u_{\alpha_{1,j}}v_{\beta_{1,j}}^\perp \right\rangle \\\ge &
  2\xi\mu\|M^{//1}\|_2^2 + \lambda \sum_{j=1}^{l_3}\frac{\sigma_{j}^{//1}\xi }{\sqrt{\sigma_{\alpha_{1,j}}^2+ \xi^2}}.
  \end{aligned}
  \end{equation*}
  Then
  \begin{equation*}
  F_t(\widetilde{M}_t^\lambda +\varepsilon M^{//1})- F_t(\widetilde{M}_t^\lambda +\xi M^{//2})\ge 2(\varepsilon-\xi)\xi \mu\|M^{//1}\|_2^2 +(\varepsilon-\xi)\lambda \sum_{j=1}^{l_3}\frac{\sigma_{j}^{//1}\xi }{\sqrt{\sigma_{\alpha_{1,j}}^2+ \xi^2}},
  \end{equation*}
  which means that
  \begin{equation}
  \label{re_c}
  \begin{aligned}
  F_t(\widetilde{M}_t^\lambda +\varepsilon M^{//1})- F_t(\widetilde{M}_t^\lambda)&\ge \mu\|M^{//1}\|_2^2 \int_{0}^\varepsilon 2\xi d\xi  + \lambda \sum_{j=1}^{l_3}\int_{0}^\varepsilon \frac{\sigma_{j}^{//1}\xi }{\sqrt{\sigma_{\alpha_{1,j}}^2 +\xi^2}}d\xi
  \\&= \varepsilon^2 \mu\|M^{//1}\|_2^2 + \lambda\sum_{j=1}^{l_3}\sigma_{j}^{//1} \left(\sqrt{\sigma_{\alpha_{1,j}}^2+\varepsilon^2}-\sigma_{\alpha_{1,j}}\right).
  \end{aligned}
  \end{equation}
  \item[d.] The result is similar to (\ref{re_c}) as
  \begin{equation}
  \label{re_d}
  F_t(\widetilde{M}_t^\lambda +\varepsilon M^{//2})- F_t(\widetilde{M}_t^\lambda)\ge  \varepsilon^2 \mu\|M^{//2}\|_2^2 + \lambda\sum_{j=1}^{l_4}\sigma_{j}^{//2} \left(\sqrt{\sigma_{\alpha_{2,j}}^2+\varepsilon^2}-\sigma_{\alpha_{2,j}}\right).
  \end{equation}
  \end{itemize}
  Using (\ref{decom_sum}), (\ref{re_a}), (\ref{re_b}), (\ref{re_c}), (\ref{re_d}), we know that
  \begin{equation}
  \label{value_ineq}
  \begin{aligned}
  F_t(\widetilde{M}_t^\lambda +  M ) - F_t(\widetilde{M}_t^\lambda ) &\ge  \sum_{\cdot \in \mathcal{A}}\left( F_t(\widetilde{M}_t^\lambda + M^{\cdot} )- F_t(\widetilde{M}_t^\lambda )\right)\\& \ge 
  \mu \sum_{\cdot \in \mathcal{A}}\|M^\cdot\|_2^2 + \lambda\sum_{j=1}^{l_3}\sigma_{j}^{//1} \left(\sqrt{\sigma_{\alpha_{1,j}}^2+1}-\sigma_{\alpha_{1,j}}\right)\\&+\lambda\sum_{j=1}^{l_4}\sigma_{j}^{//2} \left(\sqrt{\sigma_{\alpha_{2,j}}^2+1}-\sigma_{\alpha_{2,j}}\right)
  \\& \ge \mu\sum_{\cdot\in \mathcal{A}}\|M^{\cdot}\|_2^2.
  \end{aligned}
  \end{equation}
  Meanwhile, we can easily check that $\langle M^{\cdot},M^{*}\rangle = 0$ hold  for all $\cdot,*\in \mathcal{A},\ \cdot\not = *$, so we have 
  \begin{equation}
  \label{norm_ineq}
  \|M\|_2^2 = \left\|\sum_{\cdot \in\mathcal{A}}M^{\cdot}\right\|_2^2=\sum_{\cdot\in \mathcal{A}}\|M^{\cdot}\|_2^2.
  \end{equation}
  With (\ref{eq_algorithm}), (\ref{value_ineq}), (\ref{norm_ineq}) and set $M=M^{(k)}_t-\widetilde{M}_t^\lambda$, we have that
  \begin{equation*}
  \|M^{(k)}_t-\widetilde{M}_t^\lambda\|_2^2 \le  \frac{2L_f \gamma_t^2}{(k+1)^2\mu}.
  \end{equation*}
  Finally, with Theorem 1, the proof is finished. 
\end{pf}

\section{Proof of Corollary 4}
\label{sec:iteration}
\begin{pf}{\it of Corollary 4:}
  From (18), under the conditions in Corollary 3, with probability at least $1-3/(m_1+m_2)$, 
   \begin{equation*}
  (m_1m_2)^{-1/2}\left\|\widetilde{M}_t^\lambda - M_t^0\right\|_2 \le \mathcal{C}_2\left(\frac{{\sigma^*}^2(\Phi_\mathcal{X}\vee \Phi_\mathcal{Y})^2r_t\log(m_1+m_2)}{\mu^2m_1m_2 n T}\right)^{2/5}.
  \end{equation*}
  When $k_1$ satisfies
  \begin{align*}
      k_1 \ge \mathcal{C}_2^{-1}\left(\frac{{\sigma^*}^2(\Phi_\mathcal{X}\vee \Phi_\mathcal{Y})^2r_t\log(m_1+m_2)}{\mu^2m_1m_2 n T}\right)^{-2/5} \left(\frac{2L_f\gamma_1^2}{\mu m_1m_2}\right)^{1/2}.
  \end{align*}
  then
  \begin{equation*}
      (m_1m_2)^{-1/2} \left(\frac{2 L_f\gamma_t^2}{\mu (k_1+1)^2}\right)^{1/2} \le \mathcal{C}_2\left(\frac{{\sigma^*}^2(\Phi_\mathcal{X}\vee \Phi_\mathcal{Y})^2r_t\log(m_1+m_2)}{\mu^2m_1m_2 n T}\right)^{2/5}
  \end{equation*}
  and thus
  \begin{equation*}
      (m_1m_2)^{-1/2} \left\|M_1^{(k_1)} - M_1^0\right\|_2 \le 2\mathcal{C}_2\left(\frac{{\sigma^*}^2(\Phi_\mathcal{X}\vee \Phi_\mathcal{Y})^2r_t\log(m_1+m_2)}{\mu^2m_1m_2 n T}\right)^{2/5}.
  \end{equation*}
  Similarly, for $t=2,3,\cdots,T$ in the random initial, we have 
  \begin{equation*}
      K_0 =T k_1 \ge \mathcal{C}_2^{-1}\left(\frac{{\sigma^*}^2(\Phi_\mathcal{X}\vee \Phi_\mathcal{Y})^2r_t\log(m_1+m_2)}{\mu^2m_1m_2 n T}\right)^{-2/5} \left(\frac{2L_f\gamma_1^2}{\mu m_1m_2}\right)^{1/2}T.
  \end{equation*}
  In the proposed initial strategy, for $t=2,3,\dots,T$,   when 
  \begin{equation*}
  k_t \ge \mathcal{C}_2^{-1}\left(\frac{{\sigma^*}^2(\Phi_\mathcal{X}\vee \Phi_\mathcal{Y})^2r_t\log(m_1+m_2)}{\mu^2m_1m_2 n T}\right)^{-2/5} \left(\frac{2L_f\gamma_t^2}{\mu m_1m_2}\right)^{1/2},
  \end{equation*}
  we have
  \begin{equation*}
  \gamma_t \le 3\mathcal{C}_2(m_1m_2)^{1/2}\left(\frac{{\sigma^*}^2(\Phi_\mathcal{X}\vee \Phi_\mathcal{Y})^2r_t\log(m_1+m_2)}{\mu^2m_1m_2 n T}\right)^{2/5}+ \frac{D_1}{T},
  \end{equation*}
  which gives
  \begin{equation*}
  (m_1m_2)^{-1/2}\left\|M_t^{(k_t)} - M_1^0\right\|_2\le2\mathcal{C}_2\left(\frac{{\sigma^*}^2(\Phi_\mathcal{X}\vee \Phi_\mathcal{Y})^2r_t\log(m_1+m_2)}{\mu^2m_1m_2 n T}\right)^{2/5}.
  \end{equation*}
  Choosing
  \begin{align*}
  K_1=\sum_{t=1}^T k_t &= \mathcal{C}_2^{-1}\left(\frac{{\sigma^*}^2(\Phi_\mathcal{X}\vee \Phi_\mathcal{Y})^2r_t\log(m_1+m_2)}{\mu^2m_1m_2 n T}\right)^{-2/5} \left(\frac{2L_f}{\mu m_1m_2}\right)^{1/2}\sum_{t=1}^T {\gamma_t}\\&\le 
  \mathcal{C}_2^{-1}\left(\frac{{\sigma^*}^2(\Phi_\mathcal{X}\vee \Phi_\mathcal{Y})^2r_t\log(m_1+m_2)}{\mu^2m_1m_2 n T}\right)^{-2/5} \left(\frac{2L_f}{\mu m_1m_2}\right)^{1/2} \times\\&
  \left(\gamma_1 + T\left(3\mathcal{C}_2(m_1m_2)^{1/2}\left(\frac{{\sigma^*}^2(\Phi_\mathcal{X}\vee \Phi_\mathcal{Y})^2r_t\log(m_1+m_2)}{\mu^2m_1m_2 n T}\right)^{2/5}+ \frac{D_1}{T}\right)\right)\\&=
  \mathcal{C}_2^{-1}\left(\frac{{\sigma^*}^2(\Phi_\mathcal{X}\vee \Phi_\mathcal{Y})^2r_t\log(m_1+m_2)}{\mu^2m_1m_2 n T}\right)^{-2/5} \left(\frac{2L_f}{\mu m_1m_2}\right)^{1/2}(\gamma_1 +D_1) \\&+ 3\sqrt{2}(\frac{L_f}{\mu})^{1/2}T,
  \end{align*}
  and the proof is finished.
\end{pf}
For independent case, we need to obtain the error bound
 \begin{equation}\label{eq_bound_indpt_alg}
  (m_1m_2)^{-1/2}\left\|M_t^{(k_t)} - M_t^0\right\|_2 \le 2C_2\left(\frac{\sigma_*^2r_t\log(m_1+m_2)}{\mu^2m_1m_2 n T}\right)^{2/5}
  \end{equation}
  and the parallel result is 
  \begin{cor}
  \label{cor_iteration_step_indpt}
  Under assumptions of Theorem 2, to attain \eqref{eq_bound_indpt_alg} for each $t=1,\ldots,T$,
  the total iteration step $K_0$ of random initial choice satisfies
  \begin{equation*}
  K_0 \ge  \frac{1}{2}C_2^{-1}\left(\frac{\sigma_*^2r_t\log(m_1+m_2)}{\mu^2m_1m_2}\right)^{-2/5} \left(\frac{2L_f}{\mu m_1m_2}\right)^{1/2}\gamma_1 n^{2/5}T^{7/5},
  \end{equation*}
  and total iteration step $K_1$ of the proposed initial strategy satisfies
  \begin{equation*}
  K_1 \ge  \frac{1}{2} C_2^{-1}\left(\frac{\sigma_*^2 r_t\log(m_1+m_2)}{\mu^2m_1m_2}\right)^{-2/5}\left(\frac{2L_f}{\mu m_1m_2}\right)^{1/2} (\gamma_1 +D_1) (nT)^{2/5}+ 3\sqrt{2}\left(\frac{L_f}{\mu}\right)^{1/2}T.
  \end{equation*}
  \end{cor}
  The proof of Corollary \ref{cor_iteration_step_indpt} and is similar to that of Corollary 4 and hence we omit it here.

\section{Additional Numerical Results}
\label{supp:data}
\subsection{Netflix Dataset with varying the choosing of number of time intervals $T$}

\begin{figure}[!ht]
    \centering
    \includegraphics[width=0.7\linewidth]{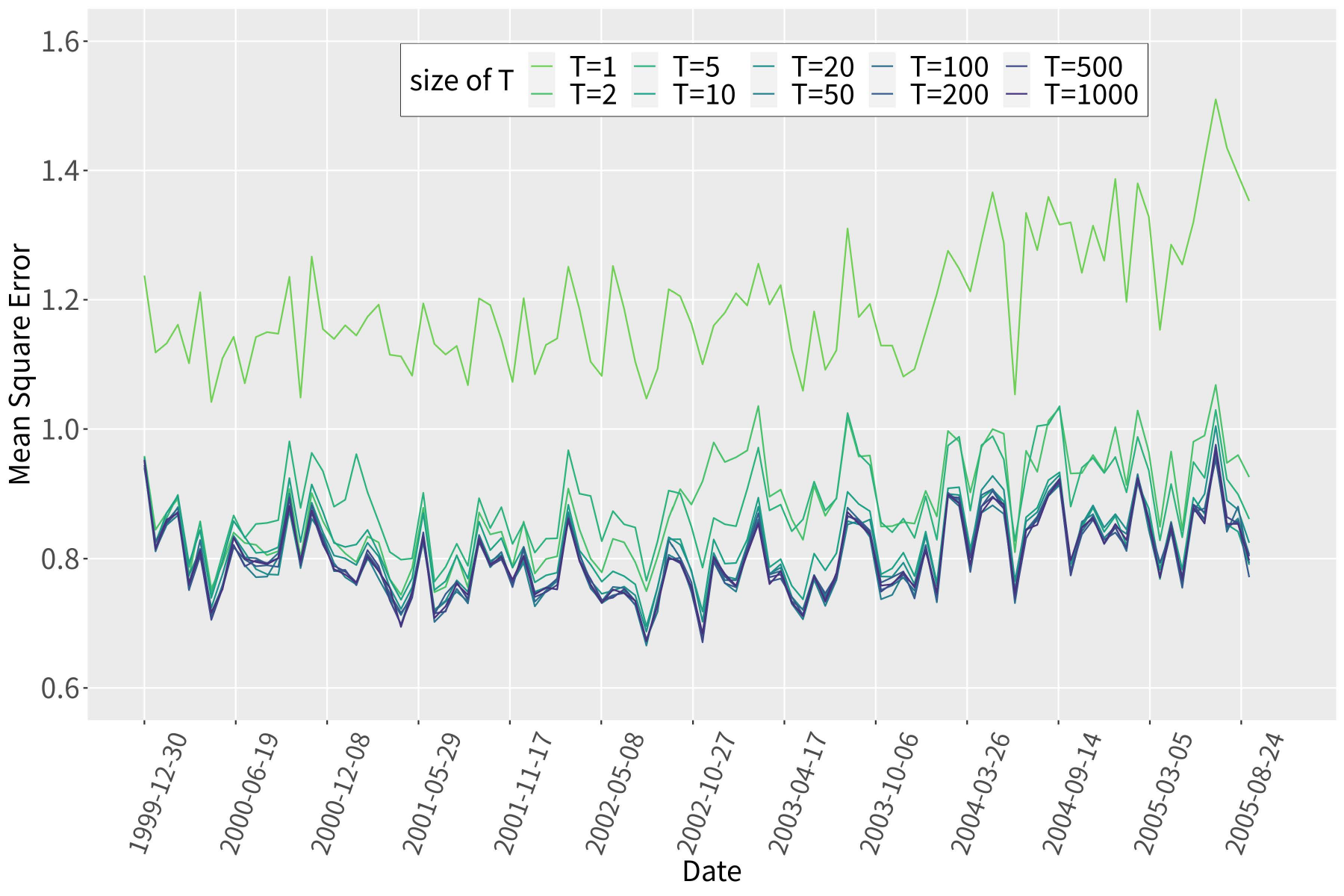}
    \caption{Mean square error for different $T$ in each time interval.}
    \label{fig:netflix_diff_t}
\end{figure}

We conduct a numerical analysis using Netflix data, encompassing 1034 movies that were viewed more than 25000 times. We sample 3000 users from those whose ratings occurred more than 30 times in the dataset. The analysis focus on the first 500000 ratings recorded from October 1998 to November 2005. These ratings are randomly divided, with $80\%$ assigned as training data and the remaining $20\%$ as test data. To explore the impact of varying $T$ while keeping $nT$ constant, we consider 10 different values of $T$ ranging from 1 to 1000 and split the observations into $T$ time intervals in chronological order. After applying our method, results are presented in Figure \ref{fig:netflix_diff_t} and Table \ref{tab:netflix_diff_t}.

For sufficiently large $T$ ($T\ge 50$), the variation in $T$ exerts a negligible influence on precision, but it results in an increase in time consumption. When $T$ is too small ($T< 50$), merging data across a substantial temporal range into a single time point can result in a loss of temporal information, rendering the utilization of data inefficient. This observation also corresponds to the dynamic nature inherent in the underlying low-rank matrix, signifying its dynamic essence rather than a static state. 

\begin{table}[!ht]
    \centering
    \caption{Average mean square error for different $T$.}
    \label{tab:netflix_diff_t}
    \begin{tabular}{c|c|c|c|c|c|c|c|c|c|c}
    \hline
         $T$& $1$ & $2$ & $5$ & $10$ & $20$  & $50$ & $100$  & $200$  & $500$  & $1000$ \\
         \hline
         MSE*& 1.194 &0.881 & 0.886 & 0.825 &0.812 & 0.800&0.803 &0.802 & 0.801 & 0.802\\
         \hline
    \end{tabular}
\end{table}

\subsection{Netflix Dataset with Link Function}
\label{supp:9.1}
\begin{figure}[!ht]
  \centering 
  \includegraphics[width=0.7\linewidth]{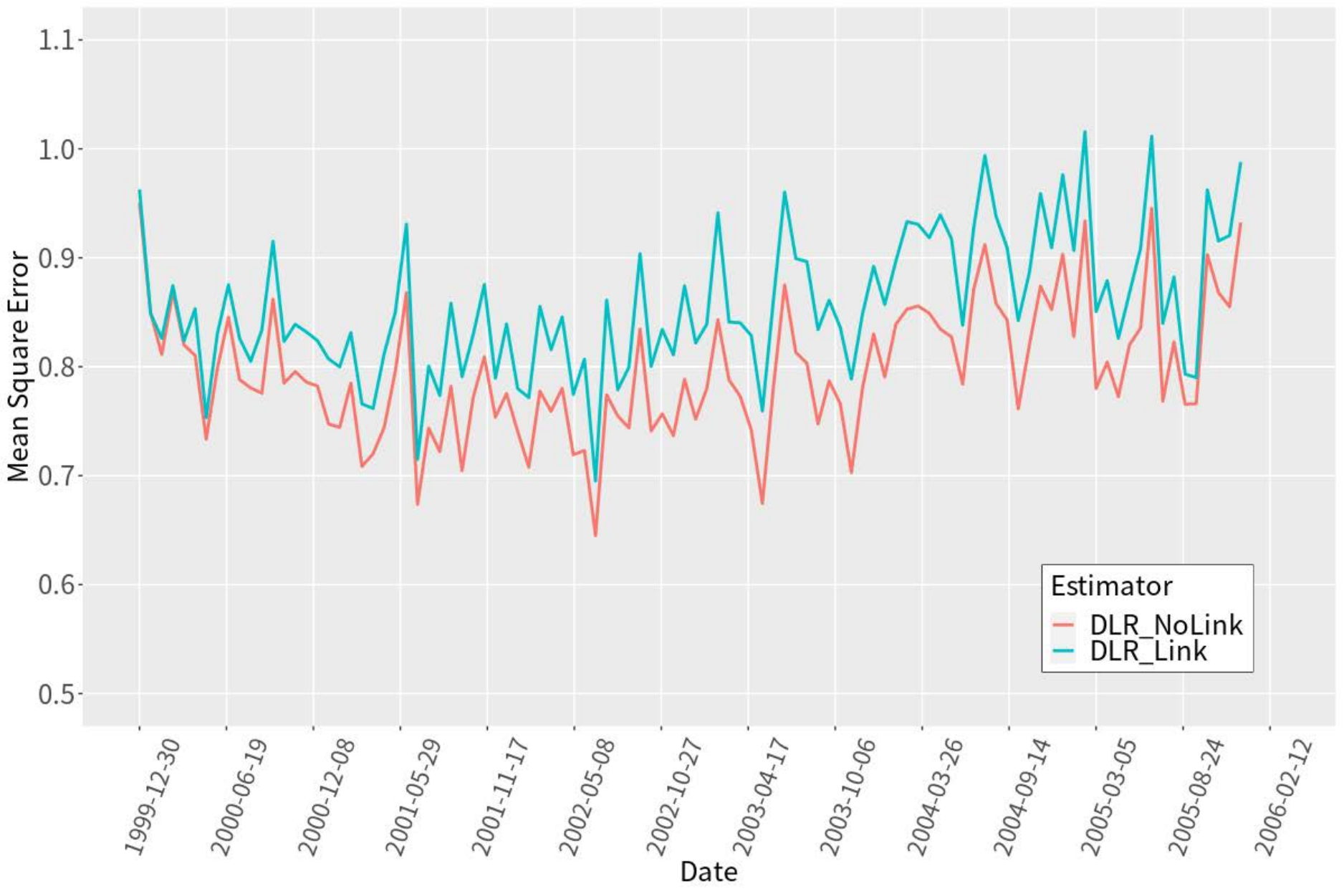}
  \caption{MSE with and without link functions}
  \label{fig:link_version}
  \end{figure}

\change{In the real data experiment with Netflix dataset, each rating $Y_i$ is a discrete integer from $1$ to $5$. 
Here we add experiment results with link function for the Netflix dataset. We assume that each user possesses a latent rating to each movie. Given the rating system's constraint of integer values ranging from 1 to 5, we consider the observed rating $Y_i$ derived from a transformation $\varphi$ on the corresponding latent rating $Z_i$, and assume that users typically tend to choose integer values in proximity to their latent ratings with a certain probability as their final ratings, i.e.,}
  \begin{align*}
      Y_i = \varphi(Z_i) = \left\{\begin{array}{ll}
      1 & Z_i < 0.5\\
           \lfloor Z_i +0.5\rfloor  + \delta_i  & 0.5\le Z_i < 5.5\\
           5 & Z_i \ge 5.5
      \end{array}
      \right.
  \end{align*}
  where $\delta_i,i=1,2,\cdots,n$ follow independent Bernoulli distributions with probability $Z_i - \lfloor Z_i +0.5\rfloor$ to take value 1 and $\lfloor\cdot \rfloor$ is the floor operator.
  And we estimate $M_t^0$ by
  \begin{align*}
       \widetilde{M}_t^{\lambda}=\argmin_{M\in\mathbb{M}}\mathbb{E}_{\delta} \sum_{j=1}^T\frac{\omega_h(j-t)}{n_j}\sum_{i=1}^{n_j}\left[\mathbb{E}(\varphi(\left\langle M,X_{ji}\right\rangle))^2 - 2Y_{ji} \varphi\left(\left\langle X_{ji}, M \right\rangle\right)\right] +\lambda \|M\|_1. 
  \end{align*}
  \change{ We followed a similar procedure and conducted new experiments with results  presented in Figure \ref{fig:link_version}. It is interesting to note  that the inclusion of this link function performs worse. One possible reason might be that the link function may not accurately reflect the true mechanism of users' ratings. Thus, in this example, we opted to not use such a link function to illustrate the proposed method. However, this does not necessarily imply a link function is not needed in general, e.g., a more flexible form of link function by nonparametric techniques can be considered  for future research.}
\subsection{Image Classification}
\label{supp:cifar10}
\change{
 An experiment was conducted utilizing the CIFAR-10 dataset to assess the efficacy of the trace regression framework in improving subsequent image-related tasks, such as classification or regression. The CIFAR-10 dataset is publicly available at \href{https://www.cs.toronto.edu/~kriz/cifar.html}{https://www.cs.toronto.edu/$\sim$ kriz/cifar.html}. This evaluation is particularly relevant when the original dataset is plagued by incomplete information and noise. Employing a similar approach as detailed in the real-world example involving video data processing, we treated each image's individual channel as a matrix denoted by $M$. Subsequently, we employed the Robust Principal Component Analysis (RPCA) method to acquire the sparse ($S$) and low-rank ($L$) components, respectively. For the purpose of dataset generation, we randomly preserved a subset of elements from the low-rank component ($L$), subjecting these elements to perturbation using i.i.d. normal noises. The recovery process involved applying our method to retrieve the low-rank component from the saved subset of elements in $L$ and subsequently adding the sparse component $S$. We compared the performance of directly using the compressed data with utilizing the data obtained after image recovery based on the trace regression framework. For the classification tasks, we utilized LeNet and ResNet18 models. The corresponding results are presented in Table \ref{table:classfication}. In the table, $\rho$ represents the compression rate, and Acc denotes the classification accuracy in the test dataset. The findings suggest that prior image recovery before classification can lead to an enhancement in classification accuracy by approximately $3\%$ to $5\%$ for the aforementioned compressed datasets.
}

\begin{table}[]
    \centering
    \caption{Comparison of classification results}
    \label{table:classfication}
    \begin{tabular}{c|c|c|c}
    \hline
       $\rho$ & Network  &No recovery Acc & Recovery Acc\\
       \hline
        $18.5\%$ & LeNet & $50.59\%$ & $55.94\%$\\
        \hline
         $18.5\%$ & ResNet & $74.57\%$& $75.75\%$\\
        \hline
        $39.1\%$ & LeNet & $52.86\%$ & $58.7\%$\\
        \hline
         $39.1\%$ & ResNet & $76.83\%$ & $80.40\%$\\
        \hline
        $58.6\%$ & LeNet & $54.64\%$ & $61.03\%$\\
        \hline
         $58.6\%$ & ResNet & $77.24\%$ & $82.1\%$\\
        \hline
        $100\%$ & LeNet & $59.09\%$ & $61.56\%$\\
        \hline
        $100\%$ & ResNet & $80.69\%$ & $83.41\%$\\
        \hline
    \end{tabular}
\end{table}
\subsection{The selection of $\lambda$ and other parameters in the FISTA algorithm}  
\label{secS:lambda}

\change{
We first remark that the optimal tuning parameter $\lambda$ can be chosen for each time points $t$. The penalty term can be relaxed to depend on $t$, and the theoretical analysis is still valid with slight modification. In practice, based on our experience with  extensive numerical studies, the universal chosen $\lambda$ would not produce similarity in the low-rank structure in terms of ranks and estimated eigen-space, and can substantially reduce the computational cost associated with tuning hyper-parameters. Theoretically, considering the heterogeneous case in Assumption \ref{assump_hete} as an example and assuming that the change of $\nu_t$ over $t$ is not very large, i.e. there exist $\nu$ to bound $\nu_t$ such that $(1-c)\nu <\nu_t< (1+C)\nu$ with constants $c,C>0$. With the same order assumption for sample size $n_t\asymp n$, $\lambda$ and $h$ has the optimal selection from Theorem 2 and (17) for each $t$ by
\begin{align}
\label{eq:tuning}
h_t=C_{h,t}\left(\frac{\sigma_*^2 r_t \log \left(m_1+m_2\right)}{\nu_t^2 m_1 m_2 n T}\right)^{1 / 5},\quad 
\lambda_t=2 C_{1,t} \sigma_* \sqrt{\frac{\log \left(m_1+m_2\right)}{n T h}},
\end{align}
where $C_{h,t}$ and $C_{1,t}$ are related to $M_t^0$. With regularity and smoothness assumption 2, there exist $C_h,C_1>0$ to bound $C_{h,t},c_{1,t}$ as their change is also not very large. So we can choose a universal bandwidth $h$ and tuning parameter $\lambda$ with the error bound for $\widetilde{M}^\lambda_t$ optimal up to a constant scalar.}

\change{
Next, we present the  detailed parameters in the FISTA algorithm of the simulation.
In our proposed method, we set the tolerance (tor) to around $1\times 10^{-3}$, which is a commonly used for the termination of optimization. For the initial time point $t=1$, we employ the following strategy,
{\small\begin{align*}
    [M_{1}^{(0)}]_{ij} = \left\{\begin{array}{ll}
        0, & \text{if the $(i,j)$ element is not observed at time 1,} \\
        Y_{1k}, & \text{if the $(i,j)$ element is observed at time 1, using the observation $Y_{1k}.$}
    \end{array}\right.
\end{align*} }
Then the output of last time $(t-1)$ is used as the initial matrix for current time $t$, i.e., 
\begin{align*}
    M_{t}^{(0)} = M_{t-1}^{(k_{t-1})}.
\end{align*}
The Lipchitz constant $L_f$ here has an upper bound which can be calculated directly as
\begin{align*}
    L_f = 2\left\|\sum_{j=1}^T \frac{\omega_h(j-t)}{n_j}\sum_{i=1}^{n_j} (\|X_{ji}\|_2X_{ji})\right\|_2,
\end{align*}
and the detail proof can be found in Supplementary S.6. 
These parameters are set in the same way for the benchmark, independent and dependent cases.
}

\change{
We employ a 5-fold cross-validation (CV) approach to select the tuning parameter $\lambda$ in both independent and dependent settings. In our empirical findings, we observed that the results exhibit minimal differences when $\lambda$ falls within a certain suitable range. To mitigate computational complexity in tuning the parameter, we suggest employing a coarse grid to estimate a rough range for $\lambda$ and subsequently using a finer grid to make the final selection. When altering the matrix size, sample size, or the number of simulation points, it is not necessary to re-select $\lambda$ using CV. Instead, we can rely on the theoretical relationship
\begin{align*}
    \lambda = 2C_1\sigma_*\sqrt{\frac{\log(m_1+m_2)}{nTh}}
\end{align*}
to adjust the new $\lambda$ accordingly.
}

\subsection{Rank Changes in Simulation and Real Data Experiment}
\label{supp:rank}
For the simulation, we show the rank of recovered matrices in independent setting with $\rho = 0.2$ and $\sigma= 1$ in Table \ref{table:simulation} after truncating those dimensions with single values smaller than $\sigma$. For the Netflix dataset, the rank for Filter 1 is in Table \ref{table:netflix_filter1} and for Filter 2 is in Table \ref{table:netflix_filter2} after truncating those dimensions with single values smaller than $3\sqrt{1035}$. For the lions video dataset, the rank for low rank part $L$ is in Table \ref{table:lions} after truncating those dimensions with single values smaller than $0.5\sqrt{480}$.

\begin{table}[!ht]
    \centering
    \caption{Rank for Simulation in independent setting with $\rho = 0.2 $ and $\sigma= 1$}
    \label{table:simulation}
    \begin{tabular}{c|c|c|c|c|c|c|c|c|c|c|c|c|c|c|c|c|c|c|c|c}
    \hline
     T & 1&2 &3 &4 &5 & 6 & 7& 8 & 9 & 10&  11  & 12&   13&   14&   15&   16&   17 &  18 &  19  & 20\\
     Rank & 12 & 10 & 9 & 9 & 9 & 9 & 9 & 9 & 9 & 9& 9 & 9& 9 & 9 & 9 & 9 & 9 & 9 & 9 & 9\\
     \hline
    T & 21 &  22  & 23&   24 &  25 &  26 &  27 &  28  & 29 &  30  & 31 &  32 &  33  & 34 &  35&   36&   37 & 38  & 39 &  40 \\
     Rank  & 9 & 9& 9 & 9 & 9 & 9 & 9 & 9 & 9 & 9& 9 & 9& 9 & 9 & 9 & 9 & 9 & 9 & 9 & 9 \\
     \hline
    T &41  & 42&   43 &  44 &  45  & 46  & 47  & 48 &  49 &  50 &  51  & 52& 
 53 &  54  & 55  & 56 &  57 &  58 &  59 &  60  \\
 Rank  & 9 & 9& 9 & 9 & 9 & 9 & 9 & 9 & 9 & 9& 9 & 9& 9 & 9 & 9 & 9 & 9 & 9 & 9 & 9 \\
  \hline
T & 61 &  62  & 63 &  64 &  65 &  66 &  67 &  68  & 69&   70  & 71  & 72 &  73 &  74 &  75 &  76  & 77 &  78 &  79  & 80 \\
  Rank & 9 & 9& 9 & 9 & 9 & 9 & 9 & 9 & 9 & 9& 9 & 9& 9 & 9 & 9 & 9 & 9 & 9 & 9 & 9\\
 \hline
T &  81 &  82&   83 &  84  & 85 &  86 &  87  & 88 &  89 &  90 &  91 &  92 &  93&   94 &  95 &  96&   97 &  98 &  99 & 100\\
  Rank& 9 & 9& 9 & 9 & 9 & 9 & 9 & 9 & 9 & 9  & 9 & 9& 9 & 9 & 9 & 9 & 9 & 9 & 9& 11\\
   \hline
    \end{tabular}
\end{table}

\begin{table}[!ht]
    \centering
    \caption{Rank for Netflix dataset using Filter 1}
    \label{table:netflix_filter1}
    \begin{tabular}{c|c|c|c|c|c|c|c|c|c|c|c|c|c|c|c|c|c|c|c|c}
    \hline
     T & 1&2 &3 &4 &5 & 6 & 7& 8 & 9 & 10&  11  & 12&   13&   14&   15&   16&   17 &  18 &  19  & 20\\
     Rank & 3& 3 &3 &3 &3 &3 &3& 3& 3 &3 &3& 3 &3& 3& 3& 3& 3 &3& 3& 3 \\
     \hline
    T & 21 &  22  & 23&   24 &  25 &  26 &  27 &  28  & 29 &  30  & 31 &  32 &  33  & 34 &  35&   36&   37 & 38  & 39 &  40 \\
     Rank  & 2& 2&2 &2 &2 &2 &2& 2& 2 &2 &2& 2 &2& 2& 2& 2&2 &2& 2& 2  \\
     \hline
    T &41  & 42&   43 &  44 &  45  & 46  & 47  & 48 &  49 &  50 &  51  & 52& 
 53 &  54  & 55  & 56 &  57 &  58 &  59 &  60  \\
 Rank& 2& 2&2 &2 &2 &2 &2& 2& 2 &2 &2& 2 &2& 2& 2& 2&2 &2& 2& 2  \\
  \hline
T & 61 &  62  & 63 &  64 &  65 &  66 &  67 &  68  & 69&   70  & 71  & 72 &  73 &  74 &  75 &  76  & 77 &  78 &  79  & 80 \\
  Rank& 2& 2&2 &2 &2 &2 &2& 2& 2 &2 &2& 2 &2& 2& 2& 2&2 &2& 2& 2 \\
 \hline
T &  81 &  82&   83 &  84  & 85 &  86 &  87  & 88 &  89 &  90 &  91 &  92 &  93&   94 &  95 &  96&   97 &  98 &  99 & 100\\
  Rank & 2& 3 &3 &3 &3 &3 &3& 3& 3 &3 &3& 3 &3& 3& 3& 3& 3 &3& 3& 3\\
   \hline
    \end{tabular}
\end{table}

\begin{table}[!ht]
    \centering
    \caption{Rank for Netflix dataset using Filter 2}
    \label{table:netflix_filter2}
    \begin{tabular}{c|c|c|c|c|c|c|c|c|c|c|c|c|c|c|c|c|c|c|c|c}
    \hline
     T & 1&2 &3 &4 &5 & 6 & 7& 8 & 9 & 10&  11  & 12&   13&   14&   15&   16&   17 &  18 &  19  & 20\\
     Rank & 3&  3&  3 & 3&  4 & 4 & 3&  3&  4&  4&  4&  4 & 4 & 3 & 4&  4&  3&  3&  4 & 3 \\
     \hline
    T & 21 &  22  & 23&   24 &  25 &  26 &  27 &  28  & 29 &  30  & 31 &  32 &  33  & 34 &  35&   36&   37 & 38  & 39 &  40 \\
     Rank   & 3 & 3&  3&  3 & 4 & 3 & 3&  4&  3&  3 & 3 & 3&  3 & 3&  3 & 3&  3&  3 & 3 & 3 \\
     \hline
    T &41  & 42&   43 &  44 &  45  & 46  & 47  & 48 &  49 &  50 &  51  & 52& 
 53 &  54  & 55  & 56 &  57 &  58 &  59 &  60  \\
 Rank & 4 & 3&  3 & 3 & 3 & 3 & 4 & 3 & 3 & 3 & 3&  3&  3 & 3 & 4&  4 & 3&  4 & 3 & 3 \\
  \hline
T & 61 &  62  & 63 &  64 &  65 &  66 &  67 &  68  & 69&   70  & 71  & 72 &  73 &  74 &  75 &  76  & 77 &  78 &  79  & 80 \\
  Rank & 3&  3&  4&  3 & 4 & 3 & 4&  4&  4 & 3 & 3&  3&  3 & 3 & 3 & 4&  4&  4  & 4&  3\\
 \hline
T &  81 &  82&   83 &  84  & 85 &  86 &  87  & 88 &  89 &  90 &  91 &  92 &  93&   94 &  95 &  96&   97 &  98 &  99 & 100\\
  Rank &  4 & 4 & 3 & 4&  4&  4&  3&  3 & 4&  3&  4&  4&  4&  4 & 4 & 4 & 4&  3 & 3 & 3\\
   \hline
    \end{tabular}
\end{table}

\begin{table}[!ht]
    \centering
    \caption{Rank for lions video dataset}
    \label{table:lions}
    \begin{tabular}{c|c|c|c|c|c}
    \hline
         T &  5 & 25 & 45 & 65 & 85 \\
         \hline
         Rank (Channel Red)& 6&5&6&5&6\\
         Rank (Channel Green) & 6&4&5&4&6\\
         Rank (Channel Blue) & 4 & 4 & 4 & 4& 4\\
         \hline
    \end{tabular}
\end{table}

\end{document}